\documentclass[12pt]{article}
\usepackage[top = 2cm, bottom = 2cm, left = 2cm, right = 2cm]{geometry}

\usepackage{amsmath}
\usepackage{amsthm}
\usepackage{amssymb}
\usepackage{amsfonts}
\usepackage{mathtools}
\usepackage{graphicx,psfrag,epsf}
\usepackage{enumerate}
\usepackage[numbers, sort&compress]{natbib}
\usepackage{algorithm}
\usepackage{algorithmicx}
\usepackage{algpseudocode}
\usepackage{multirow}
\usepackage{url}
\usepackage{booktabs}

\newtheorem{theorem}{Theorem}
\newtheorem{lemma}{Lemma}
\newtheorem{corollary}{Corollary}

\usepackage{newtxtext}
\usepackage[subscriptcorrection]{newtxmath}

\begin{document}

\def\spacingset#1{\renewcommand{\baselinestretch}%
{#1}\small\normalsize} 
\spacingset{1.5}

\title{Estimation of time-varying treatment effects using marginal structural models dependent on partial treatment history}
\author{Nodoka Seya$^1$, Masataka Taguri$^1$, Takeo Ishii$^2$\\
$^{1}$Department of Health Data Science, Tokyo Medical University\\
$^{2}$Department of Medical Science and Cardiorenal Medicine,\\
Yokohama City University Graduate School of Medicine}
\date{}
\maketitle

\begin{abstract}
Inverse probability (IP) weighting of marginal structural models (MSMs) can provide consistent estimators of time-varying treatment effects under correct model specifications and identifiability assumptions, even in the presence of time-varying confounding. However, this method has two problems: (i) inefficiency due to IP-weights cumulating all time points and (ii) bias and inefficiency due to the MSM misspecification. To address these problems, we propose (i) new IP-weights for estimating parameters of the MSM that depends on partial treatment history and (ii) closed testing procedures for selecting partial treatment history (how far back in time the MSM depends on past treatments). We derive the theoretical properties of our proposed methods under known IP-weights and discuss their extension to estimated IP-weights. Although some of our theoretical results are derived under additional assumptions beyond standard identifiability assumptions, some of which can be checked empirically from the data. In simulation studies, our proposed methods outperformed existing methods both in terms of performance in estimating time-varying treatment effects and in selecting partial treatment history. Our proposed methods have also been applied to real data of hemodialysis patients with reasonable results. 
\end{abstract}


\spacingset{2.0}
\section{Introduction}
\label{sec: intro}
In real-world clinical practice, especially for chronic diseases, individuals do not always remain in the same treatment state, but may initiate or discontinue treatment midway based on their response to past treatment states. 
When the treatment state is time-varying in this way, several estimands may be considered.
In recent years, methodologies for treatment strategies based on responses to past treatments, known as dynamic treatment regime \citep{murphy2001marginal}, have been developed. 
In practice, however, there are cases where the interest is in the effect of the basic ``treatment itself'' rather than the ``treatment strategy''. 
This is especially important in situations where the primary goal is to understand the fundamental efficacy of the treatment.
Therefore, this study defines time-varying treatment effects of interest as the contrast between always treated versus never treated, and aims to improve performance in estimating these effects.
Inverse probability (IP) weighting of marginal structural models (MSMs) proposed by \cite{MSMvsSNM} can provide consistent estimators of time-varying treatment effects under correct model specifications and identifiability assumptions, specifically, \textbf{(A1)} consistency, \textbf{(A2)} sequential exchangeability, and \textbf{(A3)} positivity, even in the presence of time-varying confounding.
However, IP-weighting of MSMs has two problems.

The first problem is inefficiency due to IP-weighting.
This problem also occurs in the context of a point treatment, but it is more severe in the context of time-varying treatments (especially when the number of time points is large) because IP-weights for MSMs, which targets the effect of the entire treatment history, are multiplied over all time points. 
In contrast, IP-weighting of history-restricted MSMs (HRMSMs), proposed by \cite{HRMSM}, which targets the effect of recent partial treatment history, can overcome inefficiency caused by the large number of time points, because IP-weights for HRMSMs are multiplied only over recent time points.
However, as we discuss later, IP-weights for HRMSMs treat past treatments as confounders, so IP-weighting of HRMSMs may be more inefficient than that of MSMs if the association between treatments at different time points is strong, which is a situation similar to the poor overlap of the propensity score in the context of a point treatment. 
Furthermore, depending on the choice of partial treatment history in the HRMSM, there may be a serious difference between the estimand based on the HRMSM and time-varying treatment effects of interest, leading to a misunderstanding of the overall treatment effect and wrong decision-making.

The second problem is the MSM misspecification. 
Specifying the MSM which does not encompass the true MSM leads to bias, while specifying the MSM which is larger than the true MSM leads to inefficiency, just as in ordinary regression problems.
In most applications, the MSM is specified by a priori knowledge.
Alternatively, information criteria for MSMs have been proposed, to select the MSM from the data.
The first information criterion for the MSM is QICw \cite{QICw}.
\cite{cQICw} noted that the penalty term in QICw is not valid and proposed cQICw which corrects it.
\cite{wCp} proposed $\text{w}C_p$ which is equivalent to cQICw if IP-weights are treated as known.
The typical model selection by the information criterion aims to select the model with minimum risk.
However, as the information criterion is a point estimator of risk, inefficiency in its IP-weighted estimation may lead to poor selection performance.
Furthermore, cQICw or $\text{w}C_p$ is a measure of the goodness of fit of the MSM overall (average across all treatment histories), so the MSM selected by cQICw or $\text{w}C_p$ not always have good properties for estimating time-varying treatment effects (the contrast of two specific treatment histories).

To address the first problem, we propose new IP-weights for estimating parameters of the MSM dependent on partial treatment history, which are expected to provide more efficient estimators than existing IP-weights, even when the number of time points is large and the association between treatments at different time points is strong, as is the case in most real-world data.
The key idea of this method is to use different IP-weights according to how far back in time the MSM depends on past treatments (partial treatment history). 
Then, to avoid the second problem, we also propose the closed testing procedure based on comparing two IP-weighted estimators (one for the MSM and one for the HRMSM), which select partial treatment history.
The MSM parameterizes counterfactual means for the entire treatment history, while the HRMSM parameterizes them for recent partial treatment history.
However, as shown in Section \ref{subsec: mselect}, the two parameterizations can share the same estimand if the counterfactual means for the entire treatment history depend only on recent partial treatment history.
This equivalence is fundamental to our testing procedure.
This method can be viewed as selecting variables in the MSM from a different perspective than information criteria.

This article is structured as follows. 
After describing the data structure and estimand (Section \ref{sec: setup}), we review MSMs and HRMSMs and discuss the link between them (Section \ref{sec: exist}).
We then describe our proposed methods and these theoretical results (Section \ref{sec: prop}).
We also conduct simulation studies to evaluate the performance of our proposed methods (Section \ref{sec: sim}).
Furthermore, we apply our proposed methods to real data while checking whether our assumptions are reasonable for that data (Section \ref{sec: ana}).
Finally, we give concluding remarks and discuss future challenges (Section \ref{sec: discuss}).

\section{The data structure and estimand}
\label{sec: setup}
Suppose that $n$ independent and identically distributed copies of 
\begin{equation*}
    O_i\coloneqq\left(L_{i}(0), A_{i}(0),L_{i}(1), A_{i}(1),\ldots\,L_{i}(K-1), A_{i}(K-1), Y_i\right)
\end{equation*}  
are observed in this order, where $L_{i}(t)$ and $A_{i}(t)\in \mathcal{A}$ are a covariate vector and a treatment variable at time $t={0,\ldots,K-1}$, and $Y_{i}\in\mathbb{R}$ is an outcome at time $K$.
Here, $L_i(0)\coloneqq (B_i, Z_i(0))$ and $L_i(t)\coloneqq Z_i(t)$ for $t= {1,\ldots, K-1}$, where $B_i\in\mathbb{R}^{p}$ is a $p$-dimensional time-fixed covariate vector and $Z_i(t)\in\mathbb{R}^{q}$ is a $q$-dimensional time-varying covariate vector for $t={0,\ldots, K-1}$.
We consider $\mathcal{A}=\{0,1\}$ with $A_{i}(t)=1$ if $i$ received treatment at time $t$ and $A_{i}(t)=0$ otherwise.
Let $\bar{L}_i(t)\coloneqq\{L_i(k); 0\leq k\leq t\}$ and $\bar{A}_i(t)\coloneqq\{A_i(k); 0\leq k\leq t\}$ denote the covariate and treatment history up to time $t$. 
We denote the treatment history from time $t^{'}$ up to time $t$ by $\underline{A}_i(t^{'}, t)\coloneqq\{A_i(k); t^{'}\leq k\leq t\}$ for $t^{'}= {0,\ldots,t}$.
In particular, $\bar{L}_i\coloneqq\bar{L}_i(K-1)$, $\bar{A}_i\coloneqq\bar{A}_i(K-1)$, and $\underline{A}_i(t^{'})\coloneqq\underline{A}_i(t^{'}, K-1)$.
Then, the observed data can also be written as $O_i=(\bar{L}_i, \bar{A}_i, Y_i)$.
For convenience, we denote $\bar{L}_i(-1)\equiv\bar{A}_i(-1)\equiv\underline{A}_i(t^{'}, t)\equiv\emptyset$ for $t^{'}> t$ and omit the subscript $i$ unless necessary.

Let $\bar{\mathcal{A}}$ be the support of $\bar{A}$ and introduce the potential outcome $Y^{\bar{a}}$ under each $\bar{a}\in \bar{\mathcal{A}}$  (i.e., the outcome if, possibly contrary to fact, treatment regime $\bar{a}$ is followed).
We also denote $Y^{\underline{a}(K-m)}\coloneqq Y^{\bar{A}(K-m-1),\underline{a}(K-m)}$ for $m ={1,\ldots, K}$ and $\underline{a}(K-m)\in \mathcal{\underline{A}}(K-m)$, where $\mathcal{\underline{A}}(K-m)$ is the support of $\underline{A}(K-m)$.
Then, the average causal effect of continuing treatment of the last $m$ time points can be expressed as $\theta^{(m)}\coloneqq \mathbb{E}[Y^{\underline{a}(K-m)={1}_m}]-\mathbb{E}[Y^{\underline{a}(K-m)={0}_m}]$, where $a_m$ is a vector of length $m$ with all elements of $a\in\{0,1\}$.
While it is possible to formulate $\theta^{(m)}$ for any $m$ as above, our estimand is the effect of continuing treatment of the last $K$ time points (i.e., from the beginning to the end), i.e., $\theta^{(K)}=\mathbb{E}[Y^{\bar{a}={1}_K}]-\mathbb{E}[Y^{\bar{a}={0}_K}]$.

\section{Review of IP-weighted estimation of marginal structural models}
\label{sec: exist}
In this section, we briefly review MSMs (Section \ref{subsec: MSMs}) and HRMSMs (Section \ref{subsec: HRMSMs}), and then discuss the link between them (Section \ref{subsec: MSM HRMSM link}) as preparation for Section \ref{sec: prop}.
For more details of MSMs, see \cite{MSMvsSNM, MSMepi, MSMjasa}.

\subsection{IP-weighted estimation of marginal structural models}
\label{subsec: MSMs}
Since there are $2^K$ possible values of $\bar{a}$ and the number of patients who exactly received the treatment history of interest is small, inference is often conducted under the MSM:
\begin{equation*}
   \label{eq: MSM general}
   \mathbb{E}[Y^{\bar{a}}]=\gamma\left(\bar{a};\psi\right),
\end{equation*}
where $\gamma\left(\bar{a};\psi\right)$ is a known function of $\bar{a}$ and $\psi$ is a vector of unknown parameters. 
If $\gamma\left(\bar{a};\psi\right)$  is correctly specified, $\psi^*$ can characterize $\theta^{(K)}$ in the form of $\theta^{(K)}=\gamma\left(1_K ;\psi^*\right)-\gamma\left(0_K ;\psi^*\right)$, where $\psi^*$ is a true value of $\psi$.
For example, $\theta^{(K)}=\sum_{S\subseteq \{1,...,K\}}\psi_S^*$ under the following saturated MSM:
\begin{equation}
\label{eq: MSM saturated}
\mathbb{E}[Y^{\bar a}]
= \sum_{S\subseteq \{1,...,K\}}\psi_S\prod_{j\in S}a(K-j),
\end{equation}
or $\theta^{(K)}=\sum_{j=1}^{K}\psi_j^*$ under the following main effect MSM:
\begin{equation}
    \label{eq: MSM main}
    \mathbb{E}[Y^{\bar{a}}]=\psi_0 + \sum_{j=1}^{K}\psi_j a(K-j).
\end{equation}

As shown by \cite{MSMvsSNM}, under the correctly specified MSM and identifiability assumptions (see Appendix A.1), $\theta^{(K)}$ can be consistently estimated using the regression model:
\begin{equation*} 
    \mathbb{E}[Y_i\mid \bar{A}_i]=\gamma\left(\bar{A}_i;\psi\right),
\end{equation*}
and the following IP-weights:
\begin{equation*}
    W_{sw,i}\coloneqq\prod_{k=0}^{K-1}\frac{f[A_i(k)\mid \bar{A}_i(k-1)]}{f[A_i(k)\mid \bar{L}_i(k),\bar{A}_i(k-1)]},
\end{equation*}
called stabilized weights (SW).
For example, under the MSM (\ref{eq: MSM main}) and identifiability assumptions, $\sum_{j=1}^{K}\hat{\psi}_{j}$ is consistent for $\theta^{(K)}$, where $(\hat{\psi}_{0},\ldots, \hat{\psi}_{K})^T=(X^TWX)^{-1}X^TWY$, $Y=(Y_1, \ldots, Y_n)^T$, $W= diag(W_{sw,1},\ldots, W_{sw,n})$, $X=(X_1,\ldots, X_n)^T$, and $X_i=(1, A_i(K-1),\ldots, A_i(0))^T$.
In a broader sense, the model for $\mathbb{E}[Y^{\bar{a}}\mid V(0)]$ is also called MSM, where $V(0) \subset L(0)$.
The estimation procedure in this case is the same as above, except for conditioning $V(0)$ on the outcome regression model and the numerator of SW.

\subsection{IP-weighted estimation of history-restricted marginal structural models}
\label{subsec: HRMSMs}
\cite{HRMSM} proposed inference based on the HRMSM:
\begin{equation*}
    \label{eq: HRMSM general}
    \mathbb{E}[Y^{\underline{a}(K-m)}]=\delta\left(\underline{a}(K-m);\phi\right),
\end{equation*}
where $\delta\left(\underline{a}(K-m);\phi\right)$ is a known function of $\underline{a}(K-m)$ and $\phi$ is a vector of unknown parameters for $m$ specified by the analyst. 
If $\delta\left(\underline{a}(K-m);\phi\right)$ is correctly specified, $\phi^{*}$ can characterize $\theta^{(m)}$ in the form of $\theta^{(m)}=\delta(1_m;\phi^{*})-\delta(0_m;\phi^{*})$, where $\phi^{*}$ is a true value of $\phi$.
For example, $\theta^{(m)}=\sum_{S\subseteq \{1,...,m\}}\phi_S^*$ under the following saturated HRMSM:
\begin{equation*}
\label{eq: HRMSM saturated}
\mathbb{E}[Y^{\underline{a}(K-m)}]
= \sum_{S\subseteq \{1,...,m\}}\phi_S\prod_{j\in S}a(K-j),
\end{equation*}
or $\theta^{(m)}=\sum_{j=1}^m\phi^*_j$ under the following main effect HRMSM:
\begin{equation*}
    \label{eq: HRMSM main}
    \mathbb{E}[Y^{\underline{a}(K-m)}]=\phi_0 + \sum_{j=1}^{m}\phi_ja(K-j).
\end{equation*}

As shown by \cite{HRMSM}, under correctly specified HRMSM and identifiability assumptions (see Appendix A.2), $\theta^{(m)}$ can be consistently estimated using the following model:
\begin{equation*} 
\label{eq: HRMSM reg}
    \mathbb{E}[Y_i\mid \underline{A}_i(K-m)]=\delta\left(\underline{A}_i(K-m);\phi\right),
\end{equation*}
and the following IP-weights:
\begin{equation*}
    W_{rsw, i}^{(m)}\coloneqq\prod_{k=K-m}^{K-1}\frac{f[A_i(k)\mid \underline{A}_i(K-m, k-1)]}{f[A_i(k)\mid \bar{L}_i(k),\bar{A}_i(k-1)]},
\end{equation*}
which we call restricted stabilized weights (RSW).
Note that identifiability assumptions for HRMSMs are necessary conditions of that for MSMs. 
In a broader sense, the model for $\mathbb{E}[Y^{\underline{a}(K-m)}\mid V(K-m)]$ is also called HRMSM, where $V(K-m) \subset (\bar{L}(K-m), \bar{A}(K-m-1))$.
The estimation procedure in this case is the same as above, except for conditioning $V(K-m)$ on the outcome regression model and the numerator of RSW.

\subsection{The link between MSMs and HRMSMs}
\label{subsec: MSM HRMSM link}
For ${\theta}^{(m)}$ and the parameters of MSMs, the following lemma holds.
The proof is given in Appendix C.1.
\begin{lemma}
\label{lem: MSM HRMSM link}
For $m=1,...,K$, the following statements hold:
\begin{itemize}
    \item[(i)] Under the MSM (\ref{eq: MSM saturated}), the following equation holds:
    \begin{equation*}
\theta^{(m)}={\sum_{T\subseteq \{1,...,m\},T\neq\emptyset}}\psi_{T}\ +\sum_{S\subseteq \{m+1,...,K\},S\neq \emptyset}\sum_{T\subseteq \{1,...,m\},T\neq\emptyset}\psi_{S\cup T}\ \mathbb{P}\left[\prod_{j\in S }{A}(K-j)=1\right].
\end{equation*}
    \item[(ii)] Especially, under the MSM (\ref{eq: MSM main}), the following equation holds:
\begin{equation*}
    \theta^{(m)}=\sum_{j=1}^{m}\psi_j.
\end{equation*}
\end{itemize}
\end{lemma}
Thus, $\theta^{(m)}$, the estimand of the HRMSM, can be expressed using the parameters of the MSM and the treatment probabilities.

\section{The proposed methodology}
\label{sec: prop}
In this section, we propose alternative methods to address the problems of existing methods in the following steps.
First, we propose the closed testing procedure based on comparing the estimator weighted by SW and RSW to select partial treatment history (Section \ref{subsec: mselect}).
Second, we propose alternative IP-weights to allow for more efficient estimation than existing IP-weights (Section \ref{subsec: PSW}).
Third, we also propose the closed testing procedure based on the comparison of the estimator weighted by IP-weights proposed in Section \ref{subsec: PSW} and by RSW (Section \ref{subsec: mselectPSW}).
Finally, we provide some remarks on estimation using our proposed methods (Section \ref{subsec: procedure}).

\subsection{Closed testing procedure for selecting partial treatment history}
\label{subsec: mselect}
In this section, we set the problem of selecting up to which time point the treatment variable should be included in the MSM back in time, i.e., selecting $m$ such that the following equation holds:
\begin{equation*}
   \label{eq: MSM m}
   \mathbb{E}[Y^{\bar{a}(K-m-1),\underline{a}(K-m)=1_m}]-\mathbb{E}[Y^{\bar{a}(K-m-1),\underline{a}(K-m)=0_m}]=\theta^{(K)}.
\end{equation*}
In constructing selection methods, we focus on two IP-weighted estimators (differing only in IP-weights) based on the following saturated model of $Y$ on $ \underline{A}(K-m)$:
\begin{equation} 
\label{eq: reg m saturated}
    \mathbb{E}[Y_i\mid \underline{A}_i(K-m)]=\sum_{S\subseteq \{1,...,m\}}\psi_S\prod_{j\in S}A(K-j),
\end{equation}
for each $m$.
One is the SW estimator:
\begin{equation*}
    \hat{\theta}_{sw}^{(m)}\coloneqq\frac{{\sum_{i=1}^n\prod_{k=K-m}^{K-1}I(A_i(k)=1) W_{sw,i} Y_i}}{{\sum_{i=1}^n\prod_{k=K-m}^{K-1}I(A_i(k)=1)  W_{sw,i}}}-\frac{{\sum_{i=1}^n\prod_{k=K-m}^{K-1}I(A_i(k)=0) W_{sw,i} Y_i}}{{\sum_{i=1}^n\prod_{k=K-m}^{K-1}I(A_i(k)=0)W_{sw,i}}},
\end{equation*}
and the other is the RSW estimator: 
\begin{equation*}
    \hat{\theta}_{rsw}^{(m)}\coloneqq\frac{{\sum_{i=1}^n\prod_{k=K-m}^{K-1}I(A_i(k)=1) W_{rsw,i}^{(m)} Y_i}}{{\sum_{i=1}^n\prod_{k=K-m}^{K-1}I(A_i(k)=1)  W_{rsw,i}^{(m)}}}-\frac{{\sum_{i=1}^n\prod_{k=K-m}^{K-1}I(A_i(k)=0) W_{rsw,i}^{(m)} Y_i}}{{\sum_{i=1}^n\prod_{k=K-m}^{K-1}I(A_i(k)=0)W_{rsw,i}^{(m)}}}.
\end{equation*}
Clearly $\hat{\theta}_{sw}^{(m)}$ and $\hat{\theta}_{rsw}^{(m)}$ are regular and asymptotically linear (RAL) estimators, so $\hat{\theta}_{sw}^{(m)}$ converges in probability to 
\begin{equation*}
    {\theta}_{sw}^{(m)}\coloneqq\frac{\mathbb{E}\left[\prod_{k=K-m}^{K-1}I(A(k)=1) W_{sw} Y\right]}{\mathbb{E}\left[\prod_{k=K-m}^{K-1}I(A(k)=1) W_{sw}\right]}-\frac{\mathbb{E}\left[\prod_{k=K-m}^{K-1}I(A(k)=0) W_{sw} Y\right]}{\mathbb{E}\left[\prod_{k=K-m}^{K-1}I(A(k)=0)W_{sw}\right]},
\end{equation*}
and $\hat{\theta}_{rsw}^{(m)}$ converges in probability to 
\begin{equation*}
    {\theta}_{rsw}^{(m)}\coloneqq\frac{\mathbb{E}\left[\prod_{k=K-m}^{K-1}I(A(k)=1) W_{rsw}^{(m)} Y\right]}{\mathbb{E}\left[\prod_{k=K-m}^{K-1}I(A(k)=1) W_{rsw}^{(m)}\right]}-\frac{\mathbb{E}\left[\prod_{k=K-m}^{K-1}I(A(k)=0) W_{rsw}^{(m)} Y\right]}{\mathbb{E}\left[\prod_{k=K-m}^{K-1}I(A(k)=0)W_{rsw}^{(m)}\right]},
\end{equation*}
under suitable regularity conditions. 

For ${\theta}_{sw}^{(m)}$ and ${\theta}_{rsw}^{(m)}$, the following lemma holds.
The proof is given in Appendix C.2.
\begin{lemma}
\label{lem: converge diff}
Assume (A1)--(A3). Then, for $m=1,...,K$, the following statements hold:
\begin{itemize}
    \item[(i)] Under the MSM (\ref{eq: MSM saturated}), the following equations hold:
    \begin{equation*}
\begin{split}
    &\theta^{(K)}-\theta_{sw}^{(m)}=\sum_{S\subseteq \{m+1,...,K\},S\neq\emptyset}\left[\sum_{T\subseteq \{1,...,m\},T\neq\emptyset}\psi_{S\cup T}\left(1-p_{1,S}^{(m)}\right)+\psi_{S}\left\{1-\left(p_{1,S}^{(m)}-p_{0,S}^{(m)}\right)\right\}\right],\\
    &\theta^{(K)}-\theta_{rsw}^{(m)}=\sum_{S\subseteq \{m+1,...,K\},S\neq\emptyset}\left[\sum_{T\subseteq \{1,...,m\},T\neq\emptyset}\psi_{S\cup T}\left(1-\mathbb{P}\left[\prod_{j\in S }{A}(K-j)=1\right]\right)+\psi_S\right],\\
    &\theta_{sw}^{(m)}-\theta_{rsw}^{(m)}=\sum_{S\subseteq \{m+1,...,K\},S\neq\emptyset}\left[\sum_{T\subseteq \{1,...,m\},T\neq\emptyset}\psi_{S\cup T}\left(p_{1,S}^{(m)}- \mathbb{P}\left[\prod_{j\in S }{A}(K-j)=1\right]\right)+\psi_{S}\left\{p_{1,S}^{(m)}-p_{0,S}^{(m)}\right\}\right],
\end{split}
\end{equation*}
    where $p_{a,S}^{(m)}\coloneqq\mathbb{P}[\prod_{j\in S} A(K-j)=1\mid \underline{A}(K-m)=a_m]$ for $a\in\{0,1\}$ .
    \item[(ii)] Especially, under the MSM (\ref{eq: MSM main}), the following equations hold:
\begin{equation*}
    \theta^{(K)}-\theta_{sw}^{(m)}=\sum_{j=m+1}^{K}\psi_j \left\{1-q_j^{(m)}\right\},\quad \theta^{(K)}-\theta_{rsw}^{(m)}=\sum_{j=m+1}^{K}\psi_j,\quad \theta_{sw}^{(m)}-\theta_{rsw}^{(m)}=\sum_{j=m+1}^{K}\psi_jq_j^{(m)},
\end{equation*}
where $q_j^{(m)}\coloneqq\mathbb{P}[A(K-j)=1\mid \underline{A}(K-m)=1_m]-\mathbb{P}[A(K-j)=1\mid \underline{A}(K-m)=0_m]$.
\end{itemize}
\end{lemma}

Now the following corollary immediately follows.
\begin{corollary}
\label{col: SWvsRSW}
Assume (A1)--(A3). Then, for $m=1,...,K$, the following statements hold:
\begin{itemize}
    \item[(i)] Assume the MSM (\ref{eq: MSM saturated}). Further assume\\
\rm{\textbf{(A4)} If $\{\psi_{S\cup T}\mid S\subseteq\{m+1,...,K\},T\subseteq\{1,...,m\},S\neq\emptyset\}$ includes non-zero component, then
$\sum_{S\subseteq \{m+1,...,K\},S\neq\emptyset}\left[\sum_{T\subseteq \{1,...,m\},T\neq\emptyset}\psi_{S\cup T}\left(1-p_{1,S}^{(m)}\right)+\psi_{S}\left\{1-\left(p_{1,S}^{(m)}-p_{0,S}^{(m)}\right)\right\}\right]\neq 0$, \\
$\sum_{S\subseteq \{m+1,...,K\},S\neq\emptyset}\left[\sum_{T\subseteq \{1,...,m\},T\neq\emptyset}\psi_{S\cup T}\left(1-\mathbb{P}\left[\prod_{j\in S }{A}(K-j)=1\right]\right)+\psi_S\right]\neq 0$, and\\
$\sum_{S\subseteq \{m+1,...,K\},S\neq\emptyset}\left[\sum_{T\subseteq \{1,...,m\},T\neq\emptyset}\psi_{S\cup T}\left(p_{1,S}^{(m)}- \mathbb{P}\left[\prod_{j\in S }{A}(K-j)=1\right]\right)+\psi_{S}\left\{p_{1,S}^{(m)}-p_{0,S}^{(m)}\right\}\right]\neq0$.}\\
\textit{Then, the following statement holds:}
\begin{equation}
    \label{eq: SWvsRSW}
    \theta_{sw}^{(m)}=\theta^{(K)} \Leftrightarrow \theta_{rsw}^{(m)}=\theta^{(K)}\Leftrightarrow \theta_{sw}^{(m)}=\theta_{rsw}^{(m)}.
\end{equation}
    \item[(ii)] \textit{Especially, assume the MSM (\ref{eq: MSM main}). Further assume}\\
\rm{\textbf{(A4)'} If $(\psi_{m+1},...,\psi_K)\neq 0_{K-m}$, then
$\sum_{m+1}^{K}\psi_j\neq 0$, $\sum_{m+1}^{K}\psi_j q_j^{(m)}\neq 0$, and
$\sum_{m+1}^{K}\psi_j\left\{1-q_j^{(m)}\right\}\neq 0$.}\\
\textit{Then, the statement (\ref{eq: SWvsRSW}) holds.}
\end{itemize}
\end{corollary}
Although we construct proposed methods based on the statement (\ref{eq: SWvsRSW}), before explaining this, we discuss the assumptions.
(A4) or (A4)' is an assumption that eliminates situations where, if there are non-zero elements in the parameter vector for $\bar{a}(K-m-1)$, they cancel each other out and result in zero overall.
Therefore, (A4) or (A4)' will hold except for specific parameter values. 
In Section \ref{subsec: ana results}, we empirically check this assumption for our applied data.

To help with understanding, we also give the corollary under sufficient conditions of (A4) or (A4)', which are more intuitively interpretable as follows.
\begin{corollary}
\label{col: SWvsRSW2}
Assume (A1)--(A3). Then, for $m=1,...,K$, the following statements hold:
\begin{itemize}
    \item[(i)] Assume the MSM (\ref{eq: MSM saturated}). Further assume\\ 
    \rm{\textbf{(A5)} Elements of $\{\psi_{S\cup T}\mid S\subseteq\{m+1,...,K\},T\subseteq\{1,...,m\},S\neq\emptyset\}$ have same sign.}\\
    \rm{\textbf{(A6)} $0<p_{1,S}^{(m)}-p_{0,S}^{(m)}<1$ and $0<p_{1,S}^{(m)}-\mathbb{P}\left[\prod_{j\in S }{A}(K-j)=1\right]<1$ for $S\neq \emptyset\subseteq\{m+1,...,K\}$.}\\
    \textit{Then, the statement (\ref{eq: SWvsRSW}) holds.}
    \item[(ii)] \textit{Especially, assume the MSM (\ref{eq: MSM main}). Further assume}\\
\rm{\textbf{(A5)'} Elements of $\{\psi_{m+1},...,\psi_K\}$ have same sign.}\\
\rm{\textbf{(A6)'} $0<q_j^{(m)}<1$} for $j=m+1,...,K$.\\
\textit{Then, the statement (\ref{eq: SWvsRSW}) holds.}
\end{itemize}
\end{corollary}
It is important to note that the additional assumptions required for Corollary 2 are generally not satisfied in real data. However, as discussed below, they may be reasonable under certain circumstances. As $\psi_{S}$ in the MSM (\ref{eq: MSM saturated}) or $\psi_j$ in the MSM (\ref{eq: MSM main}) are the parameters representing the effect of the same treatment received at different time points, there are some cases where assuming that they have the same sign, i.e., (A5) or (A5)' is reasonable.
In some real-world data, (A6) or (A6)' would hold because people who have received treatment at the last $m$ time points are more likely to have received treatment at the past time point than those who have not received treatment at the last $m$ time points.
In Section \ref{subsec: ana results}, we empirically check these assumptions for our applied data.

The statement (\ref{eq: SWvsRSW}) implies that the following three statements are equivalent: (i) $\hat{\theta}_{sw}^{(m)}$ can consistently estimate $\theta^{(K)}$, (ii) $\hat{\theta}_{rsw}^{(m)}$ can consistently estimate $\theta^{(K)}$, and (iii) the limits of convergence in probability of $\hat{\theta}_{sw}^{(m)}$ and $\hat{\theta}_{rsw}^{(m)}$ are the same.
Thus, Corollary \ref{col: SWvsRSW} can be seen as replacing problems depending on potential outcomes (selecting $m$ such that $\theta_{sw}^{(m)}=\theta^{(K)}$ holds and selecting $m$ such that $\theta_{rsw}^{(m)}=\theta^{(K)}$ holds) with the verifiable problem from the data (selecting $m$ such that $ \theta_{sw}^{(m)}=\theta_{rsw}^{(m)}$ holds).
Although obviously $\theta_{sw}^{(K)}=\theta_{rsw}^{(K)}$ holds, in terms of efficiency, $m$ should be as small as possible in satisfying $\theta_{sw}^{(m)}=\theta_{rsw}^{(m)}$.
Therefore, based on Corollary \ref{col: SWvsRSW}, we propose the method for selecting $m^* \coloneqq \min\{m \mid \theta_{sw}^{(m)}=\theta_{rsw}^{(m)}, 1\leq m\leq K\}$ by comparing $\hat{\theta}_{sw}^{(m)}$ and $\hat{\theta}_{rsw}^{(m)}$.

Let us now describe the proposed method. 
We set the problem of testing the null hypothesis $H_0^{(m)}: {\theta}_{sw}^{(m)}={\theta}_{rsw}^{(m)}$ against the alternative hypothesis $H_1^{(m)}: {\theta}_{sw}^{(m)}\neq{\theta}_{rsw}^{(m)}$, for $m\in\{1,\ldots,K\}$.
We define the test statistic as $D^{(m)}\coloneqq (\hat{\theta}_{sw}^{(m)}-\hat{\theta}_{rsw}^{(m)})^{2}/\widehat{\mathbb{V}}[\hat{\theta}_{sw}^{(m)}-\hat{\theta}_{rsw}^{(m)}]$, where $\widehat{\mathbb{V}}[\hat{\theta}_{sw}^{(m)}-\hat{\theta}_{rsw}^{(m)}]$ is an estimator of ${\mathbb{V}}[\hat{\theta}_{sw}^{(m)}-\hat{\theta}_{rsw}^{(m)}]$ and then define the indicator function for rejecting $H_0^{(m)}$ (test function) as $ h_{\alpha}(D^{(m)})\coloneqq I(D^{(m)}>\chi^2_\alpha(1))$,
where $\alpha$ is a significance level and $\chi^2_\alpha(1)$ is the upper $100\alpha$ percentile of the chi-squared distribution with 1 degree of freedom.
The elements of $\{H_0^{(m)}\mid 1\leq m\leq K\}$ are tested in ascending order from $m=1$, and let $\tilde{m}_\alpha$ be $m$ when it is accepted $H_0^{(m)}$, i.e., $h_{\alpha}(D^{(m)})=0$ for the first time.
That is, as an estimator of $m^*$, $\tilde{m}_\alpha$ is obtained according to the following algorithm.

\spacingset{1.2}
\begin{algorithm}[H]
\caption{Selecting $m$}
\begin{algorithmic}[0]
\Function {}{$D^{(1)},\ldots, D^{(K)}$}
    \State$\tilde{m}_{\alpha}\leftarrow 0$ and $\tilde{h} \leftarrow 1$
    \While {$\tilde{h}=1$}
        $\tilde{m}_{\alpha}\leftarrow \tilde{m}_{\alpha} + 1$
        \If {$\tilde{m}_{\alpha}\leq K-1$}
        $\tilde{h} \leftarrow h_{\alpha}(D^{(\tilde{m}_{\alpha})})$
        \Else 
        \ $\tilde{h} \leftarrow 0$
        \EndIf
    \EndWhile
    \State \Return $\tilde{m}_{\alpha}$
\EndFunction
\end{algorithmic}
\end{algorithm}

\spacingset{2.0}

Now the following theorem holds for $\tilde{m}_\alpha$.
The proof is given in Appendix C.3.
\begin{theorem}
\label{theorem: select sw rsw}
Assume regularity conditions for the asymptotic normality of $\hat{\theta}_{sw}^{(m)}-\hat{\theta}_{rsw}^{(m)}$ and convergence in probability of $\widehat{\mathbb{V}}[\hat{\theta}_{sw}^{(m)}-\hat{\theta}_{rsw}^{(m)}]$ to ${\mathbb{V}}[\hat{\theta}_{sw}^{(m)}-\hat{\theta}_{rsw}^{(m)}]$ for $m=1,\ldots, K$. 
Then, the following statements hold:
\begin{itemize}
    \item[(i)]\(\displaystyle\lim_{n\rightarrow \infty}\mathbb{P}[\tilde{m}_\alpha>m^*]\leq\alpha\).
    \item[(ii)] \(\displaystyle\lim_{n\rightarrow \infty}\mathbb{P}[h_{\alpha}(D^{(m)})=1]=1-F_{D^{(m)}}\left(\chi^2_\alpha(1)\right)\), where $F_{D^{(m)}}(\cdot)$ is the cumulative distribution function of the noncentral chi-squared distribution with 1 degree of freedom and noncentrality parameter $({\theta}_{sw}^{(m)}-{\theta}_{rsw}^{(m)})^2/\mathbb{V}[\hat{\theta}_{sw}^{(m)}-\hat{\theta}_{rsw}^{(m)}]$, for $m={1,\ldots, K}$.
\end{itemize}
\end{theorem}

The statement (i) of Theorem \ref{theorem: select sw rsw} implies that the probability of selecting $m$ larger than $m^*$ is asymptotically controlled to be less than $\alpha$.
The statement (ii) of Theorem \ref{theorem: select sw rsw} implies that the marginal power of each test depends on the absolute value of the difference in the limit of convergence in probability of the two IP-weighted estimators $|\theta_{sw}^{(m)}-\theta_{rsw}^{(m)}|$ and the variance of the difference between two estimators $\mathbb{V}[\hat{\theta}_{sw}^{(m)}-\hat{\theta}_{rsw}^{(m)}]$.
By the statement (ii) of Lemma \ref{lem: converge diff}, if (A5)' and (A6)' hold, the larger the absolute value of $\psi_j$ and $q_j^{(m)}$, the larger $|\theta_{sw}^{(m)}-\theta_{rsw}^{(m)}|$. 
Therefore, our proposed method is expected to have a higher probability of correctly selecting $m^*$, i.e., $\mathbb{P}[\tilde{m}_{\alpha}=m^*]$, as the stronger the treatment effect before the last $m$ time points and the stronger the association between the treatment variables.

Figure 1 shows the transition of the selection probability for each $m$ in the simulation data of Section \ref{subsec: sim set} by changing (a) effect of past treatment or (b) association between time-varying treatments, and the result is in line with this expectation.
On the other hand, for the existing information criteria, QICw and cQICw, the selection probability of $m^*$ did not increase as the association between time-varying treatments became stronger.
Thus, if a non-negligible treatment effect exists before the last $m$ time points, it would be well detected, as the association between treatment variables is often strong in real-world data.

\spacingset{1.1}
\begin{figure}
   \centering
    \includegraphics[width=\linewidth]{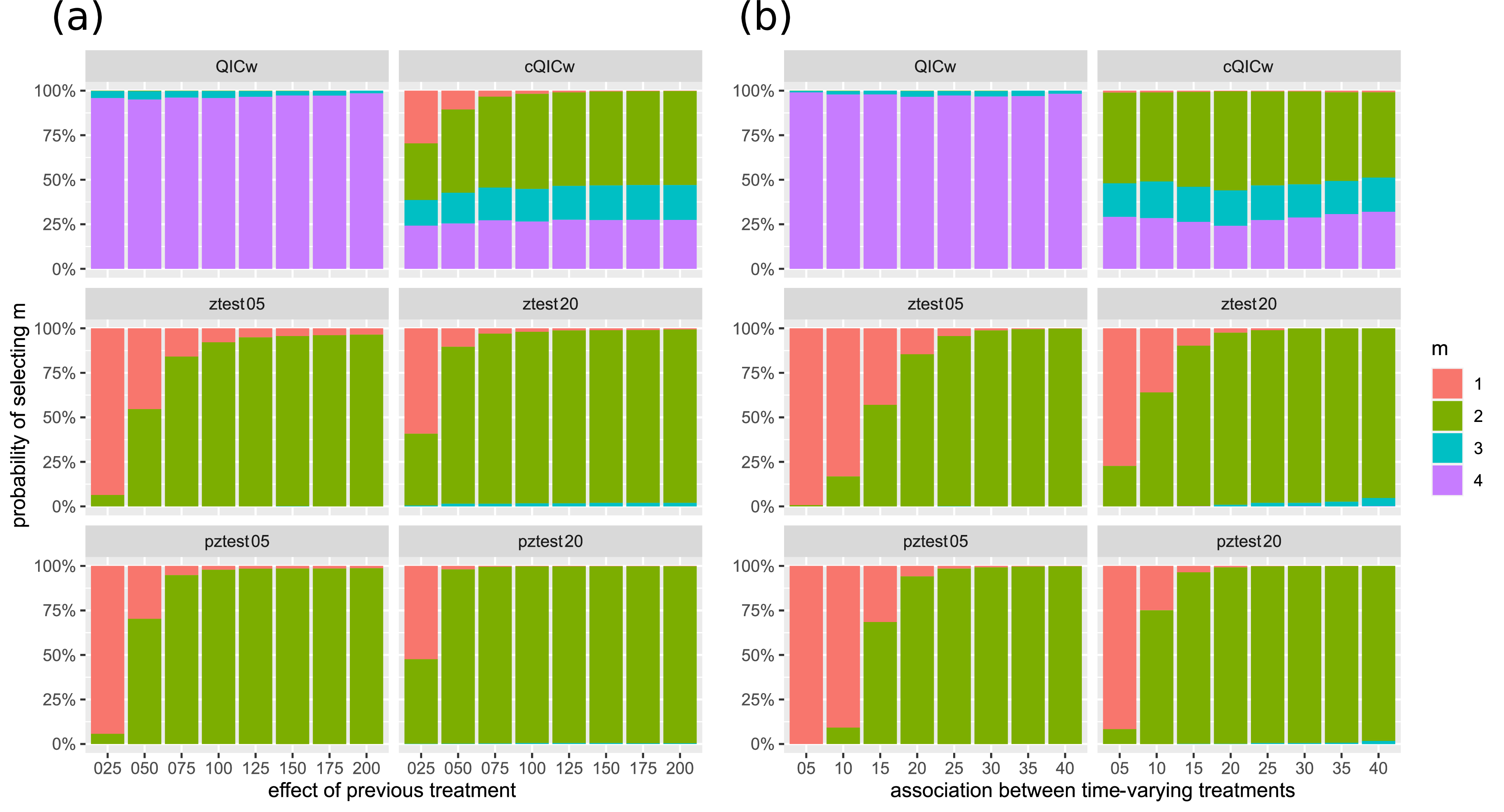}
    \caption{Plots of the selection probability of $m\in\{1,2,3,4\}$ corresponding to the main effect model over 1000 simulation runs based on the data generation process described in Section \ref{subsec: sim set} with $(\alpha_0,\alpha_1, \alpha_2, \pi_1, \delta_0, \delta_1, \delta_2, \delta_3)=(0,0,1,\pi_1,0,\delta_1,2,0)$, (a) setting $\pi_1=2.5$ and changing $\delta_1\in\{0.25, 0.50, 0.75, 1.00, 1.25, 1.50, 1.75, 2.00\}$ and (b) setting $\delta_1=1.5$ and changing $\pi_1\in\{0.5, 1.0, 1.5, 2.0, 2.5, 3.0,3.5,4.0\}$.
    In (a), the x-axis represents $\delta_1$ multiplied by 100, whose change is corresponding to the change of the effect of past treatment $\delta_1\alpha_2$.
    In (b), the x-axis represents $\pi_1$ multiplied by 10, whose change is corresponding to the change of the association between time-varying treatments.
    The first row is existing selection methods, where QICw is $\tilde{m}_{{\text{QICw}}}$ and cQICw is $\tilde{m}_{{\text{cQICw}}}$.
    The bottom two rows are proposed selection methods, where ztest05, ztest20, pztest05, pztest20 is $\tilde{m}_{0.05}$, $\tilde{m}_{0.20}$,  $\hat{m}_{0.05}$, $\hat{m}_{0.20}$, respectively.
    True is $m^*=2$.}
\end{figure}

\spacingset{2.0}

The test proposed by \cite{MSMtest} is also similar to each test in our proposed selection method in the sense that it is based on comparing different IP-weighted estimators, specifically, two or all three of the estimators weighted by SW, unstabilized weights \citep{MSMvsSNM}, basic/marginal stabilized weights \citep{MSW}.
Although \cite{MSMtest} did not discuss the testing procedure and the mapping between the limit of convergence of differences in estimators and the distribution of the potential outcome, the test proposed by \cite{MSMtest} could also be used in our framework, and then similar assumptions would be required to derive similar theoretical properties as our Corollaries 1 and 2 and Theorem 1.
Note that the test proposed by \cite{MSMtest} is expected to have lower power than each test in our proposed selection method because unstabilized weights and basic/marginal stabilized weights are generally more inefficient than RSW.

\subsection{IP-weights for marginal structural models dependent on partial treatment history}
\label{subsec: PSW}
Using $\tilde{m}_{\alpha}$ obtained by the closed testing procedure proposed in Section \ref{subsec: mselect}, we can construct the SW estimator $\hat{\theta}_{sw}^{(\tilde{m}_{\alpha})}$ or the RSW estimator $\hat{\theta}_{rsw}^{(\tilde{m}_{\alpha})}$ for $\theta^{(K)}$.
In this section, we propose an alternative IP-weighted estimator which is expected to be more efficient than these.

Here, we revisit the problem of existing IP-weights.
Since SW are cumulative weights for all $K$ time points, they become more inefficient as the number of time points $K$ increases.
On RSW, as the numerator part of the weights is $f[A_i(k)\mid \underline{A}_i(K-m, k-1)]$ rather than $f[A_i(k)\mid \bar{A}_i(k-1)]$, especially the higher association between $\underline{A}_i(k-m)$ and $\bar{A}_i(k-m-1)$, the less control the variability of the denominator part $f[A_i(k)\mid \bar{L}_i(k),\bar{A}_i(k-1)]$ has, resulting in efficiency loss. 

To address these problems, we propose the following partial SW (PSW):
\begin{equation*}
    W_{psw,i}^{(m)}\coloneqq\prod_{k=K-m}^{K-1}\frac{f[A_i(k)\mid \bar{A}_i(k-1)]}{f[A_i(k)\mid \bar{L}_i(k),\bar{A}_i(k-1)]},
\end{equation*}
and the corresponding PSW estimator:
\begin{equation*}
\hat{\theta}_{psw}^{(m)}\coloneqq\frac{{\sum_{i=1}^n\prod_{k=K-m}^{K-1}I(A_i(k)=1) W_{psw,i}^{(m)} Y_i}}{{\sum_{i=1}^n\prod_{k=K-m}^{K-1}I(A_i(k)=1)  W_{psw,i}^{(m)}}}-\frac{{\sum_{i=1}^n\prod_{k=K-m}^{K-1}I(A_i(k)=0) W_{psw,i}^{(m)} Y_i}}{{\sum_{i=1}^n\prod_{k=K-m}^{K-1}I(A_i(k)=0)W_{psw,i}^{(m)}}},
\end{equation*}
for $m={1,\ldots, K}$.
Clearly $\hat{\theta}_{psw}^{(m)}$ is an RAL estimator, so $\hat{\theta}_{psw}^{(m)}$ converges in probability to 
\begin{equation*}{\theta}_{psw}^{(m)}\coloneqq\frac{\mathbb{E}\left[\prod_{k=K-m}^{K-1}I(A(k)=1) W_{psw}^{(m)} Y\right]}{\mathbb{E}\left[\prod_{k=K-m}^{K-1}I(A(k)=1) W_{psw}^{(m)}\right]}-\frac{\mathbb{E}\left[\prod_{k=K-m}^{K-1}I(A(k)=0) W_{psw}^{(m)} Y\right]}{\mathbb{E}\left[\prod_{k=K-m}^{K-1}I(A(k)=0)W_{psw}^{(m)}\right]},
\end{equation*}
under suitable regularity conditions. 

We make the additional assumption \textbf{(A7)} $Y^{\bar{a}}\perp \bar{A}(K-m-1)$.
One may wonder whether (A7) holds, as it is generally interpreted as a situation where $\bar{A}(K-m-1)$ are randomized.
However, as we discuss later, when combined with a situation where $\bar{A}(K-m-1)$ have no effects, it is possible to state that (A7) holds under more realistic situations.

Now the following theorem holds for $\theta_{psw}^{(m)}$ for $m=1,...,K$.
The proof is given in Appendix C.4.
\begin{theorem}
\label{theorem: psw}
Assume (A1)--(A3) and (A7). Then, ${\theta}_{psw}^{(m)}={\theta}_{sw}^{(m)}$ holds.
\end{theorem}
Theorem \ref{theorem: psw} implies that under (A7), if $\theta_{sw}^{(m)}=\theta^{(K)}$ holds, then $\theta_{psw}^{(m)}=\theta^{(K)}$ also holds in general. 
Thus, under (A7), using $\hat{\theta}_{psw}^{(\tilde{m}_{\alpha})}$ instead of $\hat{\theta}_{sw}^{(\tilde{m}_{\alpha})}$ as an estimator of $\theta^{(K)}$ would also be justified.

Further, for $m=1,...,K$, the following theorem holds for the asymptotic variance of $\hat{\theta}_{w}^{(m)}$: 
\begin{equation*}
\begin{split}
        {asyvar_{w}^{(m)}}\coloneqq&\frac{\mathbb{E}\left[\prod_{k=K-m}^{K-1}I(A(k)=1)\{W_{w}^{(m)}(Y-\mu_{1,w}^{(m)})\}^2\right]}{\mathbb{E}\left[\prod_{k=K-m}^{K-1}I(A(k)=1) W_{w}^{(m)}\right]^2}\\
        &+\frac{\mathbb{E}\left[\prod_{k=K-m}^{K-1}I(A(k)=0)\{W_{w}^{(m)}(Y-\mu_{0,w}^{(m)})\}^2\right]}{\mathbb{E}\left[\prod_{k=K-m}^{K-1}I(A(k)=0) W_{w}^{(m)}\right]^2},
\end{split}
\end{equation*}
where $W_{sw}$ is denoted as $W_{sw}^{(m)}$ for convenience and
\begin{equation*}
     \mu_{a,w}^{(m)}=\frac{\mathbb{E}\left[\prod_{k=K-m}^{K-1}I(A(k)=a) W_{w}^{(m)} Y\right]}{\mathbb{E}\left[\prod_{k=K-m}^{K-1}I(A(k)=a) W_{w}^{(m)}\right]},
\end{equation*}
for $w\in \{sw, rsw, psw\}$.
The proof is given in Appendix C.5.

\begin{theorem}
\label{theorem: asyvar}
For $w\in\{sw,rsw, psw\}$ and $a\in\{0,1\}$, assume $\mu_{a,w}^{(m)}=\mathbb{E}[Y^{\bar{a}={a}_K}]$.
Then, the following statements hold:
\begin{itemize}
    \item[(i)] $asyvar_{sw}^{(m)}=\{1+\mathbb{V}[W_{sw}/W_{psw}^{(m)}]\} asyvar_{psw}^{(m)}+c_1$, where
    \begin{equation*}
    \begin{split}
         c_1=&\frac{\mathbb{COV}[\{W_{sw}/W_{psw}^{(m)}\}^2, I(\underline{A}(K-m)=1_m)\{W_{psw}^{(m)}(Y-\mathbb{E}[Y^{\bar{a}={1}_K}])\}^2]}{\mathbb{P}[\underline{A}(K-m)=1_m]^2}\\
         &+\frac{\mathbb{COV}[\{W_{sw}/W_{psw}^{(m)}\}^2, I(\underline{A}(K-m)=0_m)\{W_{psw}^{(m)}(Y-\mathbb{E}[Y^{\bar{a}={0}_K}])\}^2]}{\mathbb{P}[\underline{A}(K-m)=0_m]^2}.
    \end{split}
    \end{equation*}
    \item[(ii)] $asyvar_{rsw}^{(m)}=\{1+\mathbb{V}[W_{rsw}^{(m)}/W_{psw}^{(m)}]\} asyvar_{psw}^{(m)}+c_2$, where
    \begin{equation*}
    \begin{split}
         c_2=&\frac{\mathbb{COV}[\{W_{rsw}^{(m)}/W_{psw}^{(m)}\}^2, I(\underline{A}(K-m)=1_m)\{W_{psw}^{(m)}(Y-\mathbb{E}[Y^{\bar{a}={1}_K}])\}^2]}{\mathbb{P}[\underline{A}(K-m)=1_m]^2}\\
         &+\frac{\mathbb{COV}[\{W_{rsw}^{(m)}/W_{psw}^{(m)}\}^2, I(\underline{A}(K-m)=0_m)\{W_{psw}^{(m)}(Y-\mathbb{E}[Y^{\bar{a}={0}_K}])\}^2]}{\mathbb{P}[\underline{A}(K-m)=0_m]^2}.
    \end{split}
    \end{equation*}
\end{itemize}
\end{theorem}
By Theorem \ref{theorem: asyvar}, especially if $c_1=0$ and $c_2=0$, then the following statements hold:
\begin{equation*}
        \frac{asyvar_{psw}^{(m)}}{asyvar_{sw}^{(m)}}=\frac{1}{1+\mathbb{V}[W_{sw}/W_{psw}^{(m)}]} \leq1\text{\quad and\quad } \frac{asyvar_{psw}^{(m)}}{asyvar_{rsw}^{(m)}}=\frac{1}{1+\mathbb{V}[W_{rsw}^{(m)}/W_{psw}^{(m)}]}\leq 1.
\end{equation*}
The above inequalities imply $asyvar_{psw}^{(m)}\leq asyvar_{sw}^{(m)}$ and $asyvar_{psw}^{(m)}\leq asyvar_{rsw}^{(m)}$.
In practice, although $c_1=0$ and $c_2=0$ may rarely be exactly satisfied, $c_1$ and $c_2$ are not expected to have enough influence to change the direction of the above inequalities.
In fact, $\hat\theta_{psw}^{(m)}$ had smaller Monte Carlo standard errors than $\hat\theta_{sw}^{(m)}$ and $\hat\theta_{rsw}^{(m)}$ in our simulations of Section \ref{sec: sim} (see column 8 of Table 1 and Appendix Tables D.1 - D.6), and smaller estimated standard errors than $\hat\theta_{sw}^{(m)}$ and $\hat\theta_{rsw}^{(m)}$ in an empirical application of  Section \ref{sec: ana} (see column 5 of Table 2).

We now discuss (A7), which is the key assumption for the validity of our PSW estimator for $\mathbb{E}[Y^{\bar{a}}]$.
On the PSW estimator for $\mathbb{E}[Y^{\bar{a}}\mid L(0)]$, (A7) can be relaxed to another assumption \textbf{(A7)'} $Y^{\bar{a}}\perp \bar{A}(K-m-1)\mid L(0)$.
The following theorem holds for (A7) and (A7)'.
The proof is given in Appendix C.6.
\begin{theorem}
\label{theorem: SCMV}
Assume the following structural causal models \citep{SCM}:
  \begin{equation}
  \label{eq: SMCV}
  \begin{split}
   L(k)&=f_{L(k)}\left(\bar{L}(k-1),\bar{A}(k-1),\varepsilon_{L(k)}\right) ,\quad 0\leq k\leq K-1, \\
    A(k)&=f_{A(k)}\left(\bar{L}(k),\bar{A}(k-1),\varepsilon_{A(k)}\right) ,\quad 0\leq k\leq K-1, \\
   Y &= f_{Y}\left(\bar{L}(K-1),\bar{A}(K-1),\varepsilon_{Y}\right),
  \end{split}
\end{equation}
where error terms $\{\varepsilon_{L(0)},\ldots,\varepsilon_{L(K-1)}, \varepsilon_{A(0)},\ldots, \varepsilon_{A(K-1)},\varepsilon_{Y}\}$ are independent of each other.
Furthermore, assume the following two assumptions hold:\\
\rm{\textbf{(A8)}} \rm{There is a directed path from $A(k-1)$ to $L(k)$ for $1\leq k\leq K-m$.}\\
\rm{\textbf{(A9)}} \rm{There is no directed path from $\bar{A}(K-m-1)$ to $Y$ that is not through $\underline{A}(K-m)$.}\\
\textit{Then (A7)' holds.
In addition, if the following assumption holds, then (A7) holds:}\\
\rm{\textbf{(A10)}} \rm{There is no directed path from $L(0)$ to $Y$ that is not through $\underline{A}(K-m)$.}\\
\textit{Note that a directed path is defined as a sequence of nodes connected by directed edges, where each edge points from one node to the next in the sequence.}
\end{theorem}

Essentially, under the assumed structural causal model, (A8) and (A9) together imply that all directed paths from $L(k)$ for $1 \leq k \leq K-m$ to $Y$ are through $\underline{A}(K-m)$, and thus (A7)' holds.
If $L(k)$ is a time-varying confounder, then (A8) generally holds.
Further, (A9) implies $Y^{\bar{a}}=Y^{\underline{a}(K-m)}$.
Therefore, for $m$ such that $\theta_{sw}^{(m)}=\theta_{rsw}^{(m)}$, it may be reasonable to assume (A7)' holds and then the PSW estimator based on $\mathbb{E}[Y^{\bar{a}} \mid L(0)]$ can be consistent for $\theta^{(K)}$.
In practice, it may be sufficient to condition on $B$ rather than $L(0)=(B, Z(0))$, since $\underline{Z}(K-m)$ is likely to affect $Y$ more than $Z(0)$.
Furthermore, there may be some situations where it is reasonable to assume (A7) holds and then the PSW estimator based on $\mathbb{E}[Y^{\bar{a}}]$ can be consistent for $\theta^{(K)}$ for $m$ such that $\theta_{sw}^{(m)}=\theta_{rsw}^{(m)}$.
A typical situation is (A10).
In practice, if $\underline{L}(K-m)$ rather than $B$ more strongly influences $Y$, then (A10) may be roughly valid.

Since (A7) holds under specific conditions, we also propose directly checking whether $\theta_{psw}^{(m)}=\theta_{sw}^{(m)}$ holds when $m=\tilde{m}_{\alpha}$ and choosing IP-weights to be used accordingly.
Specifically, we propose to use $\hat{\theta}_{sw}^{(\tilde{m}_{\alpha})}$ if the null hypothesis $H_0^{(\tilde{m}_{\alpha})}: {\theta}_{psw}^{(\tilde{m}_{\alpha})}={\theta}_{sw}^{(\tilde{m}_{\alpha})}$ is rejected and to use $\hat{\theta}_{psw}^{(\tilde{m}_{\alpha})}$ otherwise, i.e., $\hat{\theta}_{sw/psw}^{(\tilde{m}_{\alpha})}\coloneqq I((\hat{\theta}_{psw}^{(\tilde{m}_{\alpha})}-\hat{\theta}_{sw}^{(\tilde{m}_{\alpha})})^2/{\widehat{\mathbb{V}}}[\hat{\theta}_{psw}^{(\tilde{m}_{\alpha})}-\hat{\theta}_{sw}^{(\tilde{m}_{\alpha})}]>\chi^2_\alpha(1))(\hat{\theta}_{sw}^{(\tilde{m}_{\alpha})}-\hat{\theta}_{psw}^{(\tilde{m}_{\alpha})}) + \hat{\theta}_{psw}^{(\tilde{m}_{\alpha})}$.
$\hat{\theta}_{sw}^{(\tilde{m}_{\alpha})}$ can be replaced by $\hat{\theta}_{rsw}^{(\tilde{m}_{\alpha})}$, and denote this estimator as $\hat{\theta}_{rsw/psw}^{(\tilde{m}_{\alpha})}$.
However, it is expected that $\hat{\theta}_{sw}^{(\tilde{m}_{\alpha})}$ is more efficient than $\hat{\theta}_{rsw}^{(\tilde{m}_{\alpha})}$, even with a large number of time points, as the association between treatment variables at different time points is quite strong in most real-world data.
Furthermore, $\hat{\theta}_{sw}^{(\tilde{m}_{\alpha})}$ is expected to be more robust than $\hat{\theta}_{rsw}^{(\tilde{m}_{\alpha})}$ in the sense that the bias due to misselection of $m^*$ is smaller.
In fact, by Lemma \ref{lem: converge diff}, under (A5)' and (A6)', $|{\theta}_{rsw}^{(m)}-{\theta}^{(K)}|\geq  |{\theta}_{sw}^{(m)}-{\theta}^{(K)}|$ holds.
Thus, $\hat{\theta}_{sw/psw}^{(\tilde{m}_{\alpha})}$ would be better than $\hat{\theta}_{rsw/psw}^{(\tilde{m}_{\alpha})}$.

\subsection{Selecting partial treatment history using proposed inverse probability weights}
\label{subsec: mselectPSW}
We also propose to replace $\hat{\theta}_{sw}^{(m)}$ in the variable selection method proposed in Section \ref{subsec: mselect} with $\hat{\theta}_{psw}^{(m)}$ proposed in Section \ref{subsec: PSW}.
Let $\hat{m}_{\alpha}$ be $m$ selected by this method.
By Theorems \ref{theorem: select sw rsw} and \ref{theorem: psw}, it is expected that $\hat{m}_{\alpha}$ will have a higher probability of correctly selecting $m^*$ than $\tilde{m}_{\alpha}$ under (A7).

\subsection{Remarks}
\label{subsec: procedure}
Based on the discussion in previous sections, we recommend using $\hat{\theta}_{psw}^{(\tilde{m}_{\alpha})}$, $\hat{\theta}_{psw}^{(\hat{m}_{\alpha})}$ or $\hat{\theta}_{sw/psw}^{(\tilde{m}_{\alpha})}$ as an estimator of $\theta^{(K)}$.
Of course, one could also use $\hat{\theta}_{sw}^{(\tilde{m}_{\alpha})}$, $\hat{\theta}_{rsw}^{(\tilde{m}_{\alpha})}$, $\hat{\theta}_{sw}^{(\hat{m}_{\alpha})}$, $\hat{\theta}_{rsw}^{(\hat{m}_{\alpha})}$, or $\hat{\theta}_{rsw/psw}^{(\tilde{m}_{\alpha})}$.

For simplicity, we have considered the saturated model (i.e., including the interaction term) for each $m$ as a candidate model. 
However, the other model could also be used to select $m^*$ and/or to estimate $\theta^{(K)}$.
For example, using the following main effect model:
\begin{equation} 
\label{eq: reg m main}
    \mathbb{E}[Y_i\mid \underline{A}_i(K-m)]=\psi_0+\sum_{j=1}^{m}\psi_jA(K-j),
\end{equation}
replace $\hat{\theta}_{w}^{(m)}$ by $\hat{\theta}_{w,main}^{(m)}\coloneqq \sum_{j=1}^{m}\hat{\psi}_j$, where $(\hat{\psi}_{0},\ldots, \hat{\psi}_{m})^T=(X^TWX)^{-1}X^TWY$, $Y=(Y_1, \ldots, Y_n)^T$, $W= diag(W_{w,1},\ldots, W_{w,n})$, $X=(X_1,\ldots, X_n)^T$, and $X_i=(1, A_i(K-1),\ldots, A_i(K-m))^T$, for $w\in\{sw,rsw, psw\}$.
Note that our testing procedures do not deal with functional form selection.
Our testing procedures are a framework for selecting $m$ (a certain type of variable selection) given a functional form (e.g., saturated model or main effect model). 
If the functional form is misspecified, the theoretical properties would not be guaranteed.
After selecting $m$ by our testing procedure using the saturated model as a candidate model, one could also consider approaches such as selecting the functional form for that $m$ by other methods.
We have also considered testing procedures that start at $m=1$, but if, for example, a priori knowledge suggests that up to $m=4$ is affected, then one could start at $m=5$.

In addition, although we have treated IP-weights as known, IP-weights are unknown and must be estimated in practice.
Typically, pooled logistic regression models are used to estimate IP-weights \citep{MSMepi}.
Nevertheless, even in this case, (statistical) consistency is ensured if models for estimating IP-weights are correctly specified \citep{MSMvsSNM}, and thus, all theoretical results provided in Section \ref{sec: prop} are still valid, except for Theorem \ref{theorem: asyvar}.
Theorem \ref{theorem: asyvar} cannot be applied directly, as deriving the asymptotic variance for each IP-weighted estimator requires considering variability due to estimating IP-weights.

\section{Simulation studies}
\label{sec: sim}
In this section, we conduct simulation studies to assess the empirical performance of our proposed methods.
For each of the six simulations, we run 1000 simulations and evaluate performance from two perspectives: (i) selecting $m^*$ and (ii) estimating $\theta^{(K)}$.

\subsection{Simulation setting}
\label{subsec: sim set}
In all six simulations, we generate the data in the following steps based on \cite{QICw, MSMtest}:
\begin{itemize}
    \item $L_i(0) {\sim} N(\alpha_0+\alpha_1, 1)$ and $A_i(0) {\sim} Bin\left(1,\text{expit}(-3 + L_i(0))\right)$
    \item $L_i(k)\mid \bar{L}_i(k-1),\bar{A}_i(k-1) {\sim} N(\alpha_0 L_i(0) + \alpha_1 L_i(k-1) + \alpha_2 A_i(k-1), 1)$, for $k=1,2,3$
    \item $A_i(k)\mid \bar{L}_i(k),\bar{A}_i(k-1) {\sim} Bin\left(1,\text{expit}(-3 + L_i(k) + \pi_1 A_i(k-1))\right)$, for $k=1,2,3$
    \item $Y_i\mid \bar{L}_i(3),\bar{A}_i(3) {\sim} N(\delta_0L_i(0) + \delta_1L_i(3) + \delta_2 A_i(3) + \delta_3  A_i(3) L_i(3), 1)$,
\end{itemize}
for $i=1,\ldots, n$.
The true MSM is as follows:
\begin{equation*}
    \mathbb{E}[Y^{\bar{a}}]=\mathbb{E}[Y^{a(2),a(3)}]=\delta_2 a(3)+\delta_1\alpha_2 a(2)+\delta_3\alpha_2 a(3) a(2).
\end{equation*}
Thus, $K=4$, $m^*=2$ and $\theta^{(K)}= \delta_2 + \delta_1\alpha_2 + \delta_3\alpha_2$.
Except for the fifth simulation, $n=5000$. For the fifth simulation, $n=500$.

In the first simulation, we set $(\alpha_0,\alpha_1, \alpha_2, \pi_1, \delta_0, \delta_1, \delta_2, \delta_3)=(0,0,1,4,0,1,2,1)$ and use the saturated model (\ref{eq: reg m saturated}) as the candidate model.
The purpose is to confirm that our proposed methods work as theory suggests when (A7) and the MSM with an interaction term hold.
In the second simulation, we set $(\alpha_0,\alpha_1, \alpha_2, \pi_1,\delta_0, \delta_1, \delta_2, \delta_3)=(0,0,1,4,0,1,2,0)$ and use the main effect model (\ref{eq: reg m main}) as the candidate model.
The purpose is to confirm that our proposed methods also work as theory suggests when the main effect MSM holds.
In the third simulation, we set $(\alpha_0,\alpha_1, \alpha_2, \pi_1, \delta_0, \delta_1, \delta_2, \delta_3)=(0.5,0,1,4,0.5,1,2,0)$ and use the main effect model (\ref{eq: reg m main}) as the candidate model.
The purpose is to investigate the performance of our proposed methods when (A7) does not hold.
The settings for the fourth simulation are the same as the first simulation, except that it uses the main effect model (\ref{eq: reg m main}) as the candidate model. 
The purpose is to investigate the performance of our proposed methods when the functional form of the MSM is misspecified.
The settings for the fifth simulation are the same as those for the first simulation, except that $n=500$.
The purpose is to investigate the performance of our proposed methods with a smaller sample size.
The settings for the sixth simulation are the same as those for the first simulation, except that $\pi_1=40$.
The purpose is to investigate the performance of our proposed methods when the association between treatment variables at different time points is larger.

On selecting $m$, we compare six methods: QIC minimization (denoted as $\tilde{m}_{{\text{QICw}}}$) and cQICw minimization (denoted as $\tilde{m}_{{\text{cQICw}}}$) as two existing methods, and $\tilde{m}_{0.05}$, $\tilde{m}_{0.20}$,  $\hat{m}_{0.05}$, and $\hat{m}_{0.20}$ as four proposed methods.
On estimating $\theta^{(K)}$, we compare twenty-two methods with combinations of selection methods and IP-weights: $\hat{\theta}_{sw}^{(m)}$, $\hat{\theta}_{rsw}^{(m)}$ , and $\hat{\theta}_{psw}^{(m)}$ for $m\in \{\tilde{m}_{{\text{QICw}}}, \tilde{m}_{\text{cQICw}}, \tilde{m}_{0.05}, \tilde{m}_{0.20}, \hat{m}_{0.05}, \hat{m}_{0.20}\}$, and $\hat{\theta}_{sw/psw}^{(m)}$ and $\hat{\theta}_{rsw/psw}^{(m)}$ for $m\in \{\tilde{m}_{0.05}, \tilde{m}_{0.20}\}$.
$\hat{\theta}_{sw}^{(m)}$ and $\hat{\theta}_{rsw}^{(m)}$ are using only existing IP-weights and $\hat{\theta}_{psw}^{(m)}$, $\hat{\theta}_{sw/psw}^{(m)}$ and $\hat{\theta}_{rsw/psw}^{(m)}$ are using proposed IP-weights.
For all comparison methods, we fit pooled logistic regression models as correct treatment assignment models to estimate IP-weights and use naïve sandwich variance estimators that do not take into account uncertainty due to estimating IP-weights and selecting MSMs.
In this case, $\text{w}C_p$ is equivalent to cQICw, thus omitted from comparison.

\subsection{Simulation results}
\label{subsec: sim results}
Figure 2 and Table 1 show results of the first simulation.
On the selection probability of $m$ as shown in (a) of Table 1, all four proposed selection methods had a higher probability of correctly selecting $m^*=2$ than two existing selection methods. 
Existing selection methods tended to select a larger $m$ than $m^*=2$, i.e., $m=3,4$, whereas the probability of selecting $m=3,4$ in proposed methods was generally controlled to be less than $\alpha$, as expected.
We then discuss the estimation performance of $\theta^{(K)}$ as shown in (b) of Table 1 and Figure 2.
As a premise, for any selection method, the probability of selecting $m=1$ was low, so bias was quite small.
Comparing by selection methods, estimators based on four proposed selection methods had a smaller variability than estimators based on two existing selection methods.
Comparing by IP-weights, estimators using three proposed IP-weights, i.e., $\hat{\theta}_{psw}^{(m)}$, $\hat{\theta}_{sw/psw}^{(m)}$ and $\hat{\theta}_{rsw/psw}^{(m)}$ had a smaller variability than estimators using two existing IP-weights, i.e., $\hat{\theta}_{sw}^{(m)}$ and $\hat{\theta}_{rsw}^{(m)}$.
Furthermore, in this scenario where (A7) holds, $\hat{\theta}_{sw/psw}^{(m)}$ and $\hat{\theta}_{rsw/psw}^{(m)}$ tended to select PSW as expected and showed similar performance to $\hat{\theta}_{psw}^{(m)}$.
The second simulation showed similar results to the first simulation (see Appendix D.1).

\spacingset{1.1}
\begin{figure}
    \centering
    \includegraphics[width=\linewidth]{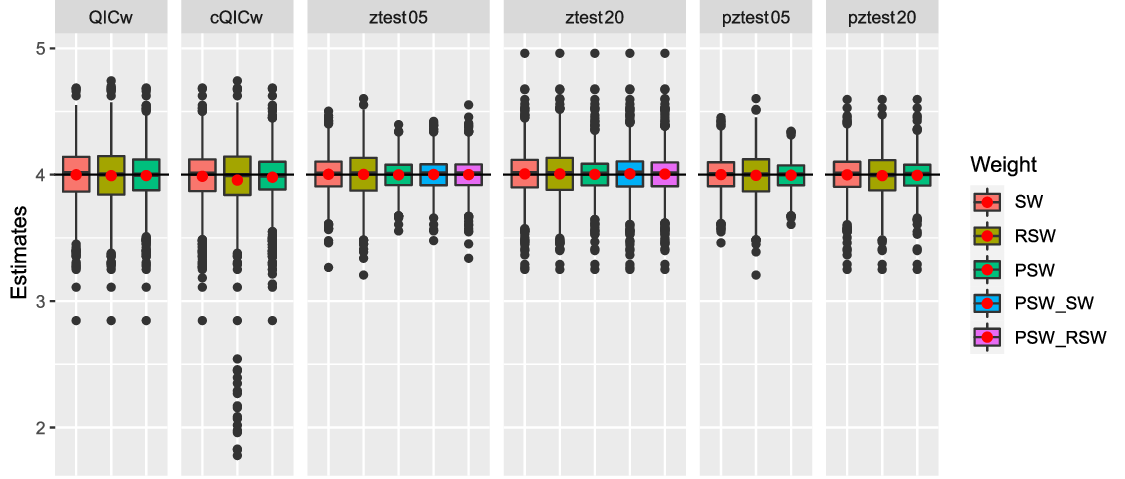}
    \caption{
    Box-plots of estimates of $\theta^{(K)}$ over 1000 runs of the first simulation with $(\alpha_0,\alpha_1, \alpha_2, \pi_1, \delta_0, \delta_1, \delta_2, \delta_3)=(0,0,1,4,0,1,2,1)$.
    The horizontal line is drawn at true value $\theta^{(K)}=4$.
    Twenty-two methods for estimating $\theta^{(K)}$ with combinations of selection methods and IP-weights are compared.
    Six gray blocks represent selection methods, where QICw, cQICw, ztest05, ztest20, pztest05, pztest20 is $\tilde{m}_{{\text{QICw}}}$, $\tilde{m}_{{\text{cQICw}}}$, $\tilde{m}_{0.05}$, $\tilde{m}_{0.20}$, $\hat{m}_{0.05}$, $\hat{m}_{0.20}$, respectively.
    For $m\in \{\tilde{m}_{\text{QICw}}, \tilde{m}_{\text{cQICw}}, \tilde{m}_{0.05}, \tilde{m}_{0.20}, \hat{m}_{0.05}, \hat{m}_{0.20}\}$, SW, RSW, PSW is $\hat{\theta}_{sw}^{(m)}$, $\hat{\theta}_{rsw}^{(m)}$, $\hat{\theta}_{psw}^{(m)}$, respectively.
    For $m\in \{\tilde{m}_{0.05}, \tilde{m}_{0.20}\}$, PSW\_SW, PSW\_RSW is $\hat{\theta}_{sw/psw}^{(m)}$, $\hat{\theta}_{rsw/psw}^{(m)}$, respectively.}
\end{figure}

\spacingset{2.0}

\spacingset{1.1}
\begin{table}
\caption{(a) Selection probability of each $m\in \{1,2,3,4\}$ and (b) Estimation performance for $\theta^{(K)}$ over 1000 runs of the first simulation with $(\alpha_0,\alpha_1, \alpha_2, \pi_1, \delta_0, \delta_1, \delta_2, \delta_3)=(0,0,1,4,0,1,2,1)$.
In (a), six methods for selecting $m^*$ are compared, where QICw, cQICw, ztest05, ztest20, pztest05, pztest20 is $\tilde{m}_{{\text{QICw}}}$, $\tilde{m}_{{\text{cQICw}}}$, $\tilde{m}_{0.05}$, $\tilde{m}_{0.20}$, $\hat{m}_{0.05}$, $\hat{m}_{0.20}$, respectively.
Bold letter represents the selection probability of true $m^*=2$.
In (b), twenty-two methods for estimating $\theta^{(K)}$ with combinations of selection methods and IP-weights are compared. 
For $m\in \{\tilde{m}_{\text{QICw}}, \tilde{m}_{\text{cQICw}}, \tilde{m}_{0.05}, \tilde{m}_{0.20}, \hat{m}_{0.05}, \hat{m}_{0.20}\}$, SW, RSW, PSW is $\hat{\theta}_{sw}^{(m)}$, $\hat{\theta}_{rsw}^{(m)}$, $\hat{\theta}_{psw}^{(m)}$, respectively.
For $m\in \{\tilde{m}_{0.05}, \tilde{m}_{0.20}\}$, PSW\_SW, PSW\_RSW is $\hat{\theta}_{sw/psw}^{(m)}$, $\hat{\theta}_{rsw/psw}^{(m)}$, respectively.
Bias is the average of the estimates over 1000 simulations minus the true value $\theta^{(K)}=4$.
SE is the Monte Carlo standard error over 1000 simulations.
RMSE is the root mean squared error of the estimates over 1000 simulations.
CP is the proportion out of 1000 simulations for which the 95 percent confidence interval using the naïve sandwich variance estimator, that does not take into account uncertainty due to estimating IP-weights and selecting MSMs, includes the true value $\theta^{(K)}=4$.\\
}
\centering
\resizebox{0.95\textwidth}{!}{%
\begin{tabular}{lcccclcccc}
\hline
\multirow{2}{*}{Selection method} & \multicolumn{4}{c}{(a) Selection probability}                                                                  & \multirow{2}{*}{Weight} & \multicolumn{4}{c}{(b) Estimation performance} \\ \cline{2-5} \cline{7-10} 
                                    & $m=1$                   & \textbf{$m=2$}                    & $m=3$                    & $m=4$                    &                         & Bias         & SE          & RMSE      & CP \\ \hline
\multirow{3}{*}{QICw}               & \multirow{3}{*}{0.000} & \multirow{3}{*}{\textbf{0.001}} & \multirow{3}{*}{0.596} & \multirow{3}{*}{0.403} & SW                      & 0.000        & 0.225       & 0.225      & 0.932         \\
                                    &                        &                                 &                        &                        & RSW                     & -0.008       & 0.250       & 0.250      & 0.926         \\
                                    &                        &                                 &                        &                        & PSW                     & -0.007       & 0.211       & 0.211      & 0.935         \\ \hline  
\multirow{3}{*}{cQICw}              & \multirow{3}{*}{0.017} & \multirow{3}{*}{\textbf{0.309}} & \multirow{3}{*}{0.348} & \multirow{3}{*}{0.326} & SW                      & -0.013       & 0.222       & 0.222      & 0.923         \\
                                    &                        &                                 &                        &                        & RSW                     & -0.043       & 0.336       & 0.338      & 0.920         \\
                                    &                        &                                 &                        &                        & PSW                     & -0.020       & 0.210       & 0.211      & 0.926         \\ \hline  
\multirow{5}{*}{ztest05}            & \multirow{5}{*}{0.000} & \multirow{5}{*}{\textbf{0.943}} & \multirow{5}{*}{0.055} & \multirow{5}{*}{0.002} & SW                      & 0.003        & 0.155       & 0.155      & 0.943         \\
                                    &                        &                                 &                        &                        & RSW                     & 0.002        & 0.193       & 0.193      & 0.958         \\
                                    &                        &                                 &                        &                        & PSW                     & -0.001       & 0.120       & 0.120      & 0.950         \\
                                    &                        &                                 &                        &                        & PSW\_SW                 & 0.002        & 0.129       & 0.130      & 0.937         \\
                                    &                        &                                 &                        &                        & PSW\_RSW                & 0.001        & 0.132       & 0.132     & 0.941         \\ \hline  
\multirow{5}{*}{ztest20}            & \multirow{5}{*}{0.000} & \multirow{5}{*}{\textbf{0.775}} & \multirow{5}{*}{0.173} & \multirow{5}{*}{0.052} & SW                      & 0.006        & 0.189       & 0.189      & 0.915         \\
                                    &                        &                                 &                        &                        & RSW                     & 0.006        & 0.200       & 0.201      & 0.958         \\
                                    &                        &                                 &                        &                        & PSW                     & 0.003        & 0.157       & 0.157      & 0.929         \\
                                    &                        &                                 &                        &                        & PSW\_SW                 & 0.007        & 0.178       & 0.178      & 0.906         \\
                                    &                        &                                 &                        &                        & PSW\_RSW                & 0.005        & 0.174       & 0.174      & 0.928         \\ \hline  
\multirow{3}{*}{pztest05}           & \multirow{3}{*}{0.000} & \multirow{3}{*}{\textbf{0.945}} & \multirow{3}{*}{0.053} & \multirow{3}{*}{0.002} & SW                      & 0.000        & 0.148       & 0.148      & 0.949         \\
                                    &                        &                                 &                        &                        & RSW                     & -0.005       & 0.186       & 0.186      & 0.969         \\
                                    &                        &                                 &                        &                        & PSW                     & -0.004       & 0.117       & 0.117      & 0.957         \\ \hline  
\multirow{3}{*}{pztest20}           & \multirow{3}{*}{0.000} & \multirow{3}{*}{\textbf{0.793}} & \multirow{3}{*}{0.160} & \multirow{3}{*}{0.047} & SW                      & -0.001       & 0.164       & 0.164      & 0.944         \\
                                    &                        &                                 &                        &                        & RSW                     & -0.007       & 0.174       & 0.174      & 0.982         \\
                                    &                        &                                 &                        &                        & PSW                     & -0.005       & 0.136       & 0.136      & 0.957  \\ \hline      
\end{tabular}%
}
\end{table}

\spacingset{2.0}

Results of the third simulation were roughly similar to the first simulation, except for estimators using PSW (see Appendix D.2).
In this simulation where (A7) does not hold, a non-negligible bias occurred in $\hat{\theta}_{psw}^{(m)}$.
However, $\hat{\theta}_{sw/psw}^{(m)}$ and $\hat{\theta}_{rsw/psw}^{(m)}$ tended to select $\hat{\theta}_{sw}^{(m)}$ and $\hat{\theta}_{rsw}^{(m)}$, respectively, so the bias was quite small, as expected.
Although $\hat{\theta}_{rsw/psw}^{(m)}$ showed a large variability, influenced by the inefficiency of $\hat{\theta}_{rsw}^{(m)}$, the performance of $\hat{\theta}_{sw/psw}^{(m)}$ was comparable to that of $\hat{\theta}_{sw}^{(m)}$.
In addition, the PSW estimator for $\mathbb{E}[Y^{\bar{a}}\mid L(0)]$, i.e., replacing the model (\ref{eq: reg m main}) with 
\begin{equation}
\label{eq: reg m main L0}
    \mathbb{E}[Y_i\mid \underline{A}_i(K-m),L_i(0)]
=\psi_0 + \sum_{j=1}^{m}\psi_j A_i(K-j)+\psi_{m+1}L_i(0),
\end{equation}
and conditioning $L(0)$ on the numerator of IP-weights, has a quite small bias, as expected (see Appendix D.3).
 The above results suggest that $\hat{\theta}_{sw/psw}^{(m)}$ tends to select $\hat{\theta}_{sw}^{(m)}$ when (A7) does not hold and can estimate with small bias, and selects $\hat{\theta}_{psw}^{(m)}$ when (A7) holds and can improve efficiency with small bias.
Furthermore, it was confirmed that the PSW estimator conditional on $L(0)$ is valid under (A7)', which is weaker than (A7).

Results of the fourth simulation are shown in Appendix D.4.
Despite the candidate model class being misspecified in this simulation, the probability of correctly selecting $m^*=2$ exceeded 0.95 for all four of the proposed selection methods.
In contrast, this probability for existing selection methods was less than 0.5.
Of course, because of the misspecification of the model class, all methods introduced bias in estimating $\theta^{(K)}$.
The magnitude of the bias was similar across all methods.
Proposed selection methods yielded smaller variability than existing selection methods, and combining them with proposed IP-weights further reduced variability.

Results of the fifth simulation ($n=500$) are shown in Appendix D.5.
The overall trend was similar to the first simulation, but the coverage probability was generally lower. 
The probability of correctly selecting $m^*=2$ by the proposed selection methods was lower than that in the first simulation ($n=5000$).
The results suggest that $n=500$ may be insufficient for asymptotic theory to work in time-varying treatment settings like this.
Even though, $\hat{m}_{0.05}$ and $\hat{m}_{0.20}$ showed relatively good performance. 

Results of the sixth simulation are shown in Appendix D.6.
Proposed selection methods performed as well as or better than in the first simulation, while existing selection methods selected wrongly $m=1$ in all 1000 simulation runs.
Comparing the results of the three IP-weights in these existing selection methods, RSW exhibited greater bias than SW or PSW.
RSW would have the problem not only with instability but also with bias when $m$ is underselected.
The results suggest that both our testing procedure and PSW show greater advantages over existing methods when the association between treatment variables at different time points is stronger.

\section{An empirical application}
\label{sec: ana}
In this section, we apply our proposed methods to real data and check whether our assumptions are reasonable for that data. 

\subsection{Data and analysis methods}
\label{subsec: ana plan}
We used the same data as a previous clinical study \cite{AlloFeb}, which performed IP-weighted estimation of Cox MSMs \cite{CoxMSM} to investigate effects of the xanthine oxidase inhibitor (allopurinol or febuxostat) on survival or cardiovascular events in hemodialysis patients in Japan.
We used no history of xanthine oxidoreductase inhibitors as of March 2016 as the eligibility criterion, resulting in 5194 patients being included in our target subjects.
Xanthine oxidoreductase inhibitors are medications generally used to lower uric acid levels.

Time-varying variables were measured in months from March 2016 ($t=0$) to February 2017 ($t=11$) and $K=12$.
For $t=0,\ldots, 11$, $A_i(t)\in\{0,1\}$ is an indicator of the prescription of the xanthine oxidoreductase inhibitor in month $t$. 
Covariates used in the analysis are the same as in \cite{AlloFeb}.
For $t=0,\ldots, 11$, the time-varying covariate vector $Z_i(t)\in \mathbb{R}^{45}$ includes laboratory, concomitant medication, and vital sign data, and the time-fixed covariate vector $B_i\in\mathbb{R}^{26}$ includes age, sex, diabetes mellitus, and comorbidity data.
$Y_i$ is the uric acid level (mg/dL) at March 2017 ($t=12$).

We excluded subjects who died or were censored by $t=12$ from our analysis, leaving $n=4640$ subjects.
To handle missing data on baseline covariates $L_i(0)$, we perform multiple imputation with a fully conditional specification method \citep{MICE}. 
To handle missing data on time-varying covariates $L_i(t),t\geq1$, we use the last observation carried forward method.
We consider the model (\ref{eq: reg m main}) with $m\in\{1,\ldots,11\}$ as the candidate model.

\subsection{Analysis results}
\label{subsec: ana results}
In all four proposed selection methods ($\tilde{m}_{0.05}$,  $\hat{m}_{0.05}$, $\tilde{m}_{0.20}$, $\hat{m}_{0.20}$), $m=2$ was selected.
Table 2 shows the analysis results of using each IP-weights for the model (\ref{eq: reg m main}) with $m=2$.
Regardless of which IP-weights were used, the results suggest that continued use of the xanthine oxidoreductase inhibitor leads to a significant reduction in uric acid levels after one year.
Although the PSW estimator had the smallest estimated standard error (SE), it was almost indistinguishable from that of the SW estimator. 
The SE of the RSW estimator was considerably larger than those of the other two estimators.
A possible reason for this is that the association between treatment variables at different time points in our applied data is quite large (see Appendix E.5).
In both $\hat{\theta}_{sw/psw}^{(m)}$ and $\hat{\theta}_{rsw/psw}^{(m)}$,
PSW was selected and thus $\hat{\theta}_{psw}^{(m)}$, $\hat{\theta}_{sw/psw}^{(m)}$ and $\hat{\theta}_{rsw/psw}^{(m)}$ had the same results of $\hat{\theta}_{psw}^{(m)}$.
When SW was used, results were similar to those obtained with PSW, whereas results obtained with RSW differed largely from them.
This may be due to the instability of the RSW estimator.

\spacingset{1.1}
\begin{table}
\caption{Results of analyzing hemodialysis patients' data using various types of IP-weights for the model (\ref{eq: reg m main}) with $m=2$.
The 1st column gives the type of IP-weights $w$ for estimating the treatment effect $\theta^{(K)}$.
The 2nd column gives $\hat{\psi}_0$, i.e., point estimates of the mean of potential outcome under never treated ${\mathbb{E}}[Y^{\bar{a}=0_K}]$.
The 3rd column gives $\sum_{j=0}^m\hat{\psi}_j$, i.e., point estimates of the mean of potential outcome under always treated ${\mathbb{E}}[Y^{\bar{a}=1_K}]$.
The 4th column gives $\hat{\theta}_{w}^{(m)}$, i.e., point estimates of $\theta^{(K)}$.
The 5th column gives their estimated standard errors (SE) calculated by naïve sandwich variance estimators, and the 6th and 7th columns give their 95 percent lower confidence limits (LCL), i.e., $\hat{\theta}_w^{(m)}-1.96\times\text{SE}$ and 95 percent upper confidence limits (UCL), i.e., $\hat{\theta}_w^{(m)}+1.96\times\text{SE}$, respectively.
The 8th column gives two-sided $p$-value ($\alpha=0.05$) calculated using SE for the null hypothesis $\theta^{(K)}=0$.\\
}
\centering
\resizebox{0.85\textwidth}{!}{%
\begin{tabular}{lclclccccc}
\hline
\multirow{2}{*}{IP-weights $w$} & \multicolumn{1}{c}{$\mathbb{E}[Y^{\bar{a}=0_K}]$} &  & \multicolumn{1}{c}{$\mathbb{E}[Y^{\bar{a}=1_K}]$} &  & \multicolumn{5}{c}{$\theta^{(K)}=\mathbb{E}[Y^{\bar{a}=1_K}]-\mathbb{E}[Y^{\bar{a}=0_K}]$} \\ 
\cline{2-2} \cline{4-4} \cline{6-10}
 & $\hat{\psi}_{0}$ &  & $\sum_{j=0}^{m}\hat{\psi}_{j}$ &  & $\hat{\theta}_{w}^{(m)}$ & SE & LCL & UCL & $p$-value \\ \hline
$sw$     & 7.480 &  & 5.989 &  & -1.491 & 0.128 & -1.742 & -1.240 & \textless{}0.001 \\ 
$rsw$    & 8.597 &  & 6.639 &  & -1.958 & 0.587 & -3.109 & -0.807 & 0.0009 \\ 
$psw$    & 7.470 &  & 6.033 &  & -1.437 & 0.126 & -1.683 & -1.191 & \textless{}0.001 \\ 
$sw/psw$ & 7.470 &  & 6.033 &  & -1.437 & 0.126 & -1.683 & -1.191 & \textless{}0.001 \\ 
$rsw/psw$& 7.470 &  & 6.033 &  & -1.437 & 0.126 & -1.683 & -1.191 & \textless{}0.001 \\ \hline
\end{tabular}%
}
\end{table}

\spacingset{2.0}

We also performed a sensitivity analysis using the model (\ref{eq: reg m main}) for each IP-weights when varying $m\in\{1,...,11\}$ (see Appendix E.1). 
When $m=1$, the estimate for $\theta^{(K)}$ was -1.214, whereas for $m\geq2$, the estimates were all around -1.5 according to $\hat{\theta}_{psw}^{(m)}$ and $\hat{\theta}_{sw}^{(m)}$.
In addition, as $m$ increased, the confidence interval tended to widen.
Therefore, in our applied data, $m=2$ can be seen as one of the good selections.

In addition, we empirically checked whether several assumptions made in Section \ref{sec: prop} hold for our applied data.
Appendix E.4 shows the results of calculating $q_j^{(m)}$ using the empirical distribution.
Calculated $q_j^{(m)}$ falls within the interval $(0,1)$ for all $m$ and $j$, suggesting that (A6') holds.
Appendix E.3 shows the estimation results for $(\psi_1,...,\psi_{11})$ in the main effect MSM with $m=11$ using PSW. 
Since most of $(\psi_1,...,\psi_{11})$ had confidence intervals that included zero, it is difficult to draw any conclusions from our applied data about whether (A5') holds.
However, by combining information about $q_j^{(1)}$ and $(\psi_1,...,\psi_{11})$, and calculating and comparing the three equations in statement (ii) of Lemma \ref{lem: converge diff} from the data (see Appendix E.2), it was suggested that situations where the conditions for these three equations to equal zero do not coincide are not common, i.e., indicating (A4'). 

\section{Concluding remarks}
\label{sec: discuss}
In this article, we proposed new methods to address two problems with IP-weighting of MSMs: (i) inefficiency due to IP-weights cumulating all time points and (ii) bias and inefficiency due to the MSM misspecification.
Specifically, we proposed new IP-weights to estimate parameters of the MSM dependent on partial treatment history more efficiently than existing IP-weights, and closed testing procedures based on comparing two IP-weighted estimators to select partial treatment history.
The former is for addressing the problem (i), and the latter is for avoiding the problem (ii).
The simulation results showed our proposed methods outperformed existing methods in terms of both performance in selecting the correct MSM and in estimating time-varying treatment effects.
Overall, the simulation results suggest that PSW is a promising method in terms of statistical efficiency and bias.

Note that there is no guarantee that the additional assumptions we made when asserting theoretical properties hold in real data, although it is expected to be reasonable in some cases as discussed in Section 4.1.
As shown in Section \ref{subsec: ana results} and the corresponding Appendix E, while there are ways to empirically check our assumptions, this does not necessarily yield meaningful information.
In most applications, especially when associations between treatments at different time points are very strong, as in our application, parameters in (A6') would be estimated stably, whereas parameters in (A5') would be estimated unstably. The latter is due to problems analogous to multicollinearity when estimating each parameter.
Therefore, in such cases, it is preferable to assess whether (A5') holds based on a priori clinical knowledge.

Another discussion point in our proposed MSM selection methods is how to determine $\alpha$.
In general, there is a trade-off that setting $\alpha$ large (small) decreases (increases) the probability of incorrectly selecting $m< m^*$, but increases (decreases) the probability of incorrectly selecting $m> m^*$. 
One guideline is to set $\alpha$ large if bias is important and to set it small if efficiency is important.
Another guideline would be to set $\alpha$ larger when the number of candidate models is large.
Instead of selecting a single value for $\alpha$, one could vary it across several values, as in a sensitivity analysis, to check the robustness of the results.

On variance estimation, we have constructed confidence intervals using naive sandwich variance estimators that do not take into account uncertainties due to (i) estimating IP-weights and (ii) selecting MSMs.
These confidence intervals achieved nominal coverage probability in the first and second simulations, but they were below in the third simulation of Section \ref{sec: sim}, so it is desirable to construct confidence intervals that take into account uncertainties due to (i) and (ii).
The challenge for (ii) is so-called post-selection inference \citep{seleinf}.

Furthermore, it may be possible to construct even better estimators than our proposed IP-weighted estimators by (i) extending to double robust estimators for parameters of MSMs, e.g., target maximum likelihood estimators \citep{MSMTMLE} and iterated conditional expectation or multiple robust estimators \citep{MSMICE1,MSMICE2,MultiRobust}, and/or (ii) combining with covariate balancing propensity score \citep{CBPSMSM}. 
These considerations are also future research projects.

\vspace{10mm}

\noindent
{\textbf{Acknowledgements}:} We thank the associate editor and the two reviewers for their constructive comments that helped improve the article.\\
{\textbf{Conflict of interest}:} Authors state no conflict of interest.\\
{\textbf{Author contribution}:} Nodoka Seya: Conceptualization; Methodology; Simulation; Data analysis; Writing – original draft; Writing – review \& editing; Project administration. 
Masataka Taguri: Methodology; Supervision; Writing – review \& editing; Funding acquisition. Takeo Ishii: Data curation; Writing – review \& editing.\\
{\textbf{Funding information}:} This work was partially supported by a Grant-in-Aid for Scientific Research (No. 24K14862) from the Ministry of Education, Culture, Sports, Science, and Technology of Japan.\\
{\textbf{Data availability statement}:} The data that support the findings of this study are available from Zenjinkai-Group but restrictions apply to the availability of these data, which were used under license for the current study, and so are not publicly available. Data are however available from the authors upon reasonable request and with permission of Zenjinkai-Group.

\bibliographystyle{vancouver}
\bibliography{PMSM}

\renewcommand\thesection{\Alph{section}}
\renewcommand{\thetable}{\Alph{section}.\arabic{table}}
\renewcommand{\thefigure}{\Alph{section}.\arabic{figure}}

\newpage

\setcounter{section}{0}

\section{Identifiability assumptions}

\subsection{Identifiability assumptions of $\mathbb{E}[Y^{\bar{a}}]$ for $\bar{a}\in \mathcal{\bar{A}}$}
\label{subsec: IdenA MSM}
\begin{itemize}
    \item[\textbf{(A1)}] consistency
    \begin{equation*}
      \text{If }\bar{A}=\bar{a}, \text{ then } Y=Y^{\bar{a}}, \text{ for } \bar{a} \in \mathcal{\bar{A}}.
    \end{equation*}
    \item[\textbf{(A2)}] sequential exchangeability
  \begin{equation*}
      Y^{\bar{a}}\perp A(t) \mid \bar{L}(t),\bar{A}(t-1), \text{ for } t\in \{0,\ldots,K-1\} \text{ and }\bar{a} \in \mathcal{\bar{A}}.
    \end{equation*}
   \item[\textbf{(A3)}] positivity
   \begin{equation*}
   \begin{split}
      &\text{ If }f\left[\bar{L}(t),\bar{A}(t-1)\right]\neq 0, \text{ then } \mathbb{P}\left[A(t)=a\mid \bar{L}(t),\bar{A}(t-1)\right]>0 \text{ w.p.1.},\\ &\text{ for }t\in \{0,\ldots,K-1\} \text{ and } a \in \mathcal{A}.
   \end{split}
   \end{equation*}
\end{itemize}

\subsection{Identifiability assumptions of $\mathbb{E}[Y^{\underline{a}(K-m)}]$ for $\underline{a}(K-m)\in \mathcal{\underline{A}}(K-m)$}
\label{subsec: IdenA HRMSM}
\begin{itemize}
    \item[\textbf{(A1)'}] consistency
    \begin{equation*}
        \text{if }\underline{A}(K-m)=\underline{a}(K-m), \text{ then } Y=Y^{\underline{a}(K-m)}, \text{ for } \underline{a}(K-m) \in \mathcal{\underline{A}}(K-m).
    \end{equation*}
    \item[\textbf{(A2)'}] sequential exchangeability
  \begin{equation*}
      Y^{\underline{a}(K-m)}\perp A(t) \mid \bar{L}(t),\bar{A}(t-1), \text{ for } t\in \{K-m,\ldots,K-1\} \text{ and }\underline{a}(K-m) \in \mathcal{\underline{A}}(K-m).
    \end{equation*}
   \item[\textbf{(A3)'}] positivity
   \begin{equation*}
   \begin{split}
     &\text{ if }f\left[\bar{L}(t),\bar{A}(t-1)\right]\neq 0, \text{ then } \mathbb{P}\left[A(t)=a\mid \bar{L}(t),\bar{A}(t-1)\right]>0 \text{ w.p.1.},\\ &\text{ for }t\in \{K-m,\ldots,K-1\} \text{ and } a \in \mathcal{A}.
   \end{split}
   \end{equation*}
\end{itemize}

\newpage
\section{Preparation of proofs}
In this section, we derive how $\theta_{sw}^{(m)}$, $\theta_{rsw}^{(m)}$, and $\theta_{psw}^{(m)}$ can be expressed under (A1)--(A3) in preparation for proofs in Section \ref{sec: proof}.

\subsection{Additional notation}
According to \cite{MSMvsSNM}, we introduce the pseudo-population distribution (i.e., the distribution after weighting by $W_w^{(m)}$) of 
$\tilde{O}\coloneqq(\{Y^{\underline{a}(K-m)};1\leq m\leq K\},Y,\bar{A},\bar{L})$:
\begin{equation}
\label{eq: pp dist}
    f_{w}^{(m)}[\tilde{O}]\coloneqq\frac{W_{w}^{(m)}f[\tilde{O}]}{\int W_{w}^{(m)}dF[\tilde{O}]}={W_{w}^{(m)}f[\tilde{O}]}.\tag{B.1.1}
\end{equation}
for $w\in\{sw, rsw, psw\}$.
The last equation holds since $\int W_{w}^{(m)}dF[\tilde{O}]=1$. 
By equation (\ref{eq: pp dist}), the following equation holds:
\begin{equation}
   \label{eq: Ew}
    \mathbb{E}_{w}^{(m)}[X_1]\coloneqq\int X_1dF_{w}^{(m)}[\tilde{O}]=\int X_1W_w^{(m)}dF[\tilde{O}]=\mathbb{E}[X_1W_w^{(m)}],\tag{B.1.2}
\end{equation}
where $X_1\subset \tilde{O}$.
We also denote the marginal and conditional distribution derived from the joint distribution (\ref{eq: pp dist}) as $f_{w}^{(m)}[\cdot]$ and $f_{w}^{(m)}[\cdot\mid \cdot]$, and denote the corresponding expectation as $\mathbb{E}_{w}^{(m)}[\cdot]$ and $\mathbb{E}_{w}^{(m)}[\cdot\mid \cdot]$.
For $sw$, we omit the superscript ${(m)}$.

Using the above notation, $\theta_{w}^{(m)}$ can be written as follows:
\begin{equation*}
\begin{split}
    \theta_{w}^{(m)}&=\frac{\mathbb{E}\left[\prod_{k=K-m}^{K-1}I(A(k)=1) W_{w}^{(m)} Y\right]}{\mathbb{E}\left[\prod_{k=K-m}^{K-1}I(A(k)=1) W_{w}^{(m)}\right]}-\frac{\mathbb{E}\left[\prod_{k=K-m}^{K-1}I(A(k)=0) W_{w}^{(m)} Y\right]}{\mathbb{E}\left[\prod_{k=K-m}^{K-1}I(A(k)=0)W_{w}^{(m)}\right]}\\
    &=\frac{\mathbb{E}_{w}^{(m)}[I(\underline{A}(K-m)=1_m)Y]}{\mathbb{E}_{w}^{(m)}[I(\underline{A}(K-m)=1_m)]}-\frac{\mathbb{E}_{w}^{(m)}[I(\underline{A}(K-m)=0_m)Y]}{\mathbb{E}_{w}^{(m)}[I(\underline{A}(K-m)=0_m)]}\because \text{(\ref{eq: Ew})}\\
    &=\mathbb{E}_{w}^{(m)}[Y\mid \underline{A}(K-m)=1_m]-\mathbb{E}_{w}^{(m)}[Y\mid \underline{A}(K-m)=0_m],
\end{split}
\end{equation*}
for $w\in\{sw, rsw, psw\}$.

\subsection{$\theta^{(m)}_{sw}$ under identifiability assumptions}
Under (A2) and (A3), $f_{sw}[Y^{\bar{a}}, \bar{A}, \bar{L}]$ can be expressed as follows:
\begin{equation*}
\begin{split}
    f_{sw}[Y^{\bar{a}}, \bar{A}, \bar{L}]&=f[Y^{\bar{a}}]\prod_{k=0}^{K-1}f[L(k)\mid \bar{A}(k-1),\bar{L}(k-1),Y^{\bar{a}}]\prod_{k=0}^{K-1}f[A(k)\mid \bar{A}(k-1),\bar{L}(k),Y^{\bar{a}}]\\
    &\quad \ \times \prod_{k=0}^{K-1}\frac{f[A(k)\mid \bar{A}(k-1)]}{f[A(k)\mid \bar{A}(k-1),\bar{L}(k)]}\\
    &=f[Y^{\bar{a}}]\prod_{k=0}^{K-1}f[L(k)\mid \bar{A}(k-1),\bar{L}(k-1),Y^{\bar{a}}]\prod_{k=0}^{K-1}f[A(k)\mid \bar{A}(k-1)].
\end{split}
\end{equation*}
The above equation implies the following equation holds:
\begin{equation}
   \label{eq: sw dist}
    f_{sw}[Y^{\bar{a}},\bar{A}]=f_{sw}[Y^{\bar{a}}]f_{sw}[\bar{A}]=f[Y^{\bar{a}}]f[\bar{A}].\tag{B.2.1}
\end{equation}
Thus, under (A1)--(A3), ${\theta}_{sw}^{(m)}$ can be expressed as follows:
\begin{equation}
    \label{eq: SW converge}
    \begin{split}
        {\theta}_{sw}^{(m)}&=\mathbb{E}_{sw}[Y\mid \underline{A}(K-m)=1_m]-\mathbb{E}_{sw}[Y\mid \underline{A}(K-m)=0_m]\\
        &=\mathbb{E}_{sw}[Y^{\bar{A}(K-m-1),\underline{a}(K-m)=1_m}\mid \underline{A}(K-m)=1_m]\\
        &\quad-\mathbb{E}_{sw}[Y^{\bar{A}(K-m-1),\underline{a}(K-m)=0_m}\mid \underline{A}(K-m)=0_m]\because \text{(A1)}\\
        &=\sum_{\bar{a}(K-m-1)\in\bar{\mathcal{A}}(K-m-1)}\mathbb{E}_{sw}[Y^{\bar{a}(K-m-1),\underline{a}(K-m)=1_m}\mid \bar{A}(K-m-1)=\bar{a}(K-m-1),\\
        &\quad \quad\underline{A}(K-m)=1_m]\times \mathbb{P}_{sw}[\bar{A}(K-m-1)=\bar{a}(K-m-1)\mid \underline{A}(K-m)=1_m]\\
        &\quad-\sum_{\bar{a}(K-m-1)\in\bar{\mathcal{A}}(K-m-1)}\mathbb{E}_{sw}[Y^{\bar{a}(K-m-1),\underline{a}(K-m)=0_m}\mid \bar{A}(K-m-1)=\bar{a}(K-m-1),\\
        &\quad \quad \quad\underline{A}(K-m)=0_m]\times \mathbb{P}_{sw}[\bar{A}(K-m-1)=\bar{a}(K-m-1)\mid \underline{A}(K-m)=0_m]\\
        &\quad \because \text{iterated expectation}\\
        &=\sum_{\bar{a}(K-m-1)\in\bar{\mathcal{A}}(K-m-1)}\mathbb{E}[Y^{\bar{a}(K-m-1),\underline{a}(K-m)=1_m}]\\
        &\quad \quad \times \mathbb{P}[\bar{A}(K-m-1)=\bar{a}(K-m-1)\mid \underline{A}(K-m)=1_m]\\
        &\quad -\sum_{\bar{a}(K-m-1)\in\bar{\mathcal{A}}(K-m-1)}\mathbb{E}[Y^{\bar{a}(K-m-1),\underline{a}(K-m)=0_m}]\\
        &\quad \quad \quad \times \mathbb{P}[\bar{A}(K-m-1)=\bar{a}(K-m-1)\mid \underline{A}(K-m)=0_m]. \because \text{(\ref{eq: sw dist})}
    \end{split}
    \tag{B.2.2}
\end{equation}

\subsection{$\theta^{(m)}_{rsw}$ under identifiability assumptions}
Under (A2) and (A3), $f_{rsw}^{(m)}[Y^{\underline{a}(K-m)},\bar{A},\bar{L}]$ can be expressed as follows:
\begin{equation*}
\begin{split}
    &f_{rsw}^{(m)}[Y^{\underline{a}(K-m)},\bar{A},\bar{L}]\\
    &=f[Y^{\underline{a}(K-m)}]\prod_{k=0}^{K-1}f[L(k)\mid \bar{A}(k-1),\bar{L}(k-1),Y^{\underline{a}(K-m)}]\prod_{k=0}^{K-1}f[A(k)\mid \bar{A}(k-1),\bar{L}(k),Y^{\underline{a}(K-m)}]\\
    &\quad \ \times \prod_{k=K-m}^{K-1}\frac{f[A(k)\mid \underline{A}(K-m,k-1)]}{f[A(k)\mid \bar{A}(k-1),\bar{L}(k)]}\\
    &=f[Y^{\underline{a}(K-m)}]\prod_{k=0}^{K-1}f[L(k)\mid \bar{A}(k-1),\bar{L}(k-1),Y^{\underline{a}(K-m)}]\prod_{k=0}^{K-m-1}f[A(k)\mid \bar{A}(k-1),\bar{L}(k),Y^{\underline{a}(K-m)}]\\
    &\quad \ \times \prod_{k=K-m}^{K-1}f[A(k)\mid \underline{A}(K-m,k-1)]
\end{split}
\end{equation*}
The above equation implies the following equation holds:
\begin{equation}
 \label{eq: rsw dist}
    f_{rsw}^{(m)}[Y^{\underline{a}(K-m)},\underline{A}(K-m)] =f_{rsw}^{(m)}[Y^{\underline{a}(K-m)}]f_{rsw}^{(m)}[\underline{A}(K-m)]= f[Y^{\underline{a}(K-m)}]f[\underline{A}(K-m)].\tag{B.3.1}
\end{equation}
Thus, under (A1)--(A3), ${\theta}_{rsw}^{(m)}$ can be expressed as follows:
\begin{equation}
    \label{eq: RSW converge}
    \begin{split}
        {\theta}_{rsw}^{(m)}&=\mathbb{E}_{rsw}^{(m)}[Y\mid \underline{A}(K-m)=1_m]-\mathbb{E}_{rsw}^{(m)}[Y\mid \underline{A}(K-m)=0_m]\\
        &=\mathbb{E}_{rsw}^{(m)}[Y^{\underline{a}(K-m)=1_m}\mid \underline{A}(K-m)=1_m]-\mathbb{E}_{rsw}^{(m)}[Y^{\underline{a}(K-m)=0_m}\mid \underline{A}(K-m)=0_m] \because \text{(A1)}\\
        &=\mathbb{E}[Y^{\underline{a}(K-m)=1_m}]-\mathbb{E}[Y^{\underline{a}(K-m)=0_m}]=\theta^{(m)} \because \text{(\ref{eq: rsw dist})}
    \end{split}
    \tag{B.3.2}
\end{equation}

\subsection{$\theta^{(m)}_{psw}$ under identifiability assumptions}
Under (A2) and (A3), $f_{psw}^{(m)}[Y^{\underline{a}(K-m)},\bar{A},\bar{L}]$ can be expressed as follows:
\begin{equation*}
\begin{split}
    &f_{psw}^{(m)}[Y^{\underline{a}(K-m)},\bar{A},\bar{L}]\\
    &=f[Y^{\underline{a}(K-m)}]\prod_{k=0}^{K-1}f[L(k)\mid \bar{A}(k-1),\bar{L}(k-1),Y^{\underline{a}(K-m)}]\prod_{k=0}^{K-1}f[A(k)\mid \bar{A}(k-1),\bar{L}(k),Y^{\underline{a}(K-m)}]\\
    &\quad \ \times \prod_{k=K-m}^{K-1}\frac{f[A(k)\mid \bar{A}(k-1)]}{f[A(k)\mid \bar{A}(k-1),\bar{L}(k)]}\\
    &=f[Y^{\underline{a}(K-m)}]\prod_{k=0}^{K-1}f[L(k)\mid \bar{A}(k-1),\bar{L}(k-1),Y^{\underline{a}(K-m)}]\prod_{k=0}^{K-m-1}f[A(k)\mid \bar{A}(k-1),\bar{L}(k),Y^{\underline{a}(K-m)}]\\
    &\quad \ \times \prod_{k=K-m}^{K-1}f[A(k)\mid \bar{A}(k-1)]
\end{split}
\end{equation*}
The above equation implies the following equation holds:
\begin{equation*}
 \begin{split}
      &f_{psw}^{(m)}[Y^{\underline{a}(K-m)},\underline{A}(K-m)\mid \bar{A}(K-m-1)] \\
      &=f_{psw}^{(m)}[Y^{\underline{a}(K-m)}\mid \bar{A}(K-m-1)]f_{psw}^{(m)}[\underline{A}(K-m)\mid \bar{A}(K-m-1)]\\
      &= f[Y^{\underline{a}(K-m)}\mid \bar{A}(K-m-1)]f[\underline{A}(K-m)\mid \bar{A}(K-m-1)],
 \end{split}
\end{equation*}
and then the following equation holds by (A1):
\begin{equation}
 \label{eq: psw dist}
 \begin{split}
      &f_{psw}^{(m)}[Y^{\bar{a}},\underline{A}(K-m)\mid \bar{A}(K-m-1)=\bar{a}(K-m-1)] \\
      &=f_{psw}^{(m)}[Y^{\bar{a}}\mid \bar{A}(K-m-1)=\bar{a}(K-m-1)]\\
      &\quad \times f_{psw}^{(m)}[\underline{A}(K-m)\mid \bar{A}(K-m-1)=\bar{a}(K-m-1)]\\
      &= f[Y^{\bar{a}}\mid \bar{A}(K-m-1)=\bar{a}(K-m-1)]f[\underline{A}(K-m)\mid \bar{A}(K-m-1)=\bar{a}(K-m-1)]. \end{split}
   \tag{B.4.1}
\end{equation}
Thus, under (A1)--(A3), ${\theta}_{psw}^{(m)}$ can be expressed as follows:
\begin{equation}
    \label{eq: PSW converge}
    \begin{split}
        {\theta}_{psw}^{(m)}&=\mathbb{E}_{psw}^{(m)}[Y\mid \underline{A}(K-m)=1_m]-\mathbb{E}_{psw}^{(m)}[Y\mid \underline{A}(K-m)=0_m]\\
        &=\mathbb{E}_{psw}^{(m)}[Y^{\bar{A}(K-m-1),\underline{a}(K-m)=1_m}\mid \underline{A}(K-m)=1_m]\\
        &\quad-\mathbb{E}_{psw}^{(m)}[Y^{\bar{A}(K-m-1),\underline{a}(K-m)=0_m}\mid \underline{A}(K-m)=0_m]\because \text{(A1)}\\
        &=\sum_{\bar{a}(K-m-1)\in\bar{\mathcal{A}}(K-m-1)}\mathbb{E}_{psw}^{(m)}[Y^{\bar{a}(K-m-1),\underline{a}(K-m)=1_m}\mid \bar{A}(K-m-1)=\bar{a}(K-m-1),\\
        &\quad \quad\underline{A}(K-m)=1_m]\times \mathbb{P}_{psw}^{(m)}[\bar{A}(K-m-1)=\bar{a}(K-m-1)\mid \underline{A}(K-m)=1_m]\\
        &\quad-\sum_{\bar{a}(K-m-1)\in\bar{\mathcal{A}}(K-m-1)}\mathbb{E}_{psw}^{(m)}[Y^{\bar{a}(K-m-1),\underline{a}(K-m)=0_m}\mid \bar{A}(K-m-1)=\bar{a}(K-m-1),\\
        &\quad \quad \quad\underline{A}(K-m)=0_m]\times \mathbb{P}_{psw}^{(m)}[\bar{A}(K-m-1)=\bar{a}(K-m-1)\mid \underline{A}(K-m)=0_m]\\
        &\quad \because \text{iterated expectation}\\
        &=\sum_{\bar{a}(K-m-1)\in\bar{\mathcal{A}}(K-m-1)}\mathbb{E}[Y^{\bar{a}(K-m-1),\underline{a}(K-m)=1_m}\mid  \bar{A}(K-m-1)=\bar{a}(K-m-1)]\\
        &\quad \quad \times \mathbb{P}[\bar{A}(K-m-1)=\bar{a}(K-m-1)\mid \underline{A}(K-m)=1_m]\\
        &\quad -\sum_{\bar{a}(K-m-1)\in\bar{\mathcal{A}}(K-m-1)}\mathbb{E}[Y^{\bar{a}(K-m-1),\underline{a}(K-m)=0_m}\mid  \bar{A}(K-m-1)=\bar{a}(K-m-1)]\\
        &\quad \quad \quad \times \mathbb{P}[\bar{A}(K-m-1)=\bar{a}(K-m-1)\mid \underline{A}(K-m)=0_m].\because \text{(\ref{eq: psw dist})}
    \end{split}
    \tag{B.4.2}
\end{equation}

\section{Proofs}
\label{sec: proof}

\subsection{Proof of Lemma 1}
\label{subsec: lemma1}

\begin{proof}
By iterated expectation, ${\theta}^{(m)}$ can be expressed as follows:
\begin{equation}
\label{eq: C1 prepare}
\begin{split}
    \theta^{(m)}&=\mathbb{E}[Y^{\underline{a}(K-m)=1_m}]-\mathbb{E}[Y^{\underline{a}(K-m)=0_m}]\\&=\sum_{\bar{a}(K-m-1)\in\bar{\mathcal{A}}(K-m-1)}\mathbb{E}[Y^{\bar{a}(K-m-1),\underline{a}(K-m)=1_m}-Y^{\bar{a}(K-m-1),\underline{a}(K-m)=0_m}\mid \bar{A}(K-m-1)=\bar{a}(K-m-1)]\\
    &\quad\quad\times \mathbb{P}[\bar{A}(K-m-1)=\bar{a}(K-m-1)].
\end{split}
\tag{C.1.1}
\end{equation}

First, we prove statement (i).
Under the MSM (1), $Y^{\bar a}$ can be expressed as follows:
\begin{equation*}
   Y^{\bar a}
= \sum_{S\subseteq \{1,...,K\}}\psi_S\prod_{j\in S}a(K-j) + \varepsilon,
\end{equation*}
where $\mathbb{E}[\varepsilon]=0$.
Specifically, $Y^{\bar{a}(K-m-1),\underline{a}(K-m)=1_m}$ and $Y^{\bar{a}(K-m-1),\underline{a}(K-m)=0_m}$ can be expressed as follows:
\begin{equation*}
   Y^{\bar{a}(K-m-1),\underline{a}(K-m)=1_m}
= \sum_{S\subseteq \{m+1,...,K\}}\left\{\left(\sum_{T\subseteq \{1,...,m\}}\psi_{S\cup T}\right)\prod_{j\in S}a(K-j)\right\}+ \varepsilon,
\end{equation*}
\begin{equation*}
   Y^{\bar{a}(K-m-1),\underline{a}(K-m)=0_m}
= \sum_{S\subseteq \{m+1,...,K\}}\left\{\psi_{S}\prod_{j\in S}a(K-j)\right\}+ \varepsilon.
\end{equation*}
Thus, under the MSM (1), the equation (\ref{eq: C1 prepare}) can be expressed as follows:
\begin{equation}
\label{eq: MSM HRMSM link saturated}
\begin{split}
    \theta^{(m)}
    &=\sum_{\bar{a}(K-m-1)\in\bar{\mathcal{A}}(K-m-1)}\sum_{S\subseteq \{m+1,...,K\}}\left\{\left(\sum_{T\subseteq \{1,...,m\}}\psi_{S\cup T}\right)\prod_{j\in S}a(K-j) -\psi_S\prod_{j\in S }a(K-j)\right\}\\
    &\quad\quad\times \mathbb{P}[\bar{A}(K-m-1)=\bar{a}(K-m-1)]\\
    &=\sum_{\bar{a}(K-m-1)\in\bar{\mathcal{A}}(K-m-1)}\sum_{S\subseteq \{m+1,...,K\}}\left\{\sum_{T\subseteq \{1,...,m\},T\neq\emptyset}\psi_{S\cup T}\right\}\prod_{j\in S }a(K-j)\\
    &\quad\quad\times \mathbb{P}[\bar{A}(K-m-1)=\bar{a}(K-m-1)]\\
    &={\sum_{T\subseteq \{1,...,m\},T\neq\emptyset}}\psi_{T}\ +\sum_{S\subseteq \{m+1,...,K\},S\neq \emptyset}\sum_{T\subseteq \{1,...,m\},T\neq\emptyset}\psi_{S\cup T}\ \mathbb{P}\left[\prod_{j\in S }{A}(K-j)=1\right].
\end{split}
\tag{C.1.2}
\end{equation}

Next, we prove statement (ii).
Under the MSM (2), $Y^{\bar a}$ can be expressed as follows:
\begin{equation*}
   Y^{\bar a}
= \psi_0 + \sum_{j=1}^{K}\psi_j a(K-j) + \varepsilon,
\end{equation*}
where $\mathbb{E}[\varepsilon]=0$.
Specifically, $Y^{\bar{a}(K-m-1),\underline{a}(K-m)=1_m}$ and $Y^{\bar{a}(K-m-1),\underline{a}(K-m)=0_m}$ can be expressed as follows:
\begin{equation*}
   Y^{\bar{a}(K-m-1),\underline{a}(K-m)=1_m}
= \psi_0 + \sum_{j=1}^{m}\psi_j + \sum_{j=m+1}^{K}\psi_j a(K-j)+ \varepsilon,
\end{equation*}
\begin{equation*}
   Y^{\bar{a}(K-m-1),\underline{a}(K-m)=0_m}
= \psi_0 + \sum_{j=m+1}^{K}\psi_ja(K-j)+ \varepsilon.
\end{equation*}
Thus, under the MSM (2), the equation (\ref{eq: C1 prepare}) can be expressed as follows:

\begin{equation}
\label{eq: MSM HRMSM link main}
\begin{split}
    \theta^{(m)}
    &=\sum_{\bar{a}(K-m-1)\in\bar{\mathcal{A}}(K-m-1)}\left\{\psi_0 + \sum_{j=1}^{m}\psi_j + \sum_{j=m+1}^{K}\psi_j a(K-j) -\left(\psi_0 + \sum_{j=m+1}^{K}\psi_ja(K-j)\right) \right\}\\
    &\quad\quad\times \mathbb{P}[\bar{A}(K-m-1)=\bar{a}(K-m-1)]\\
    &=\sum_{j=1}^{m}\psi_j.
\end{split}
\tag{C.1.3}
\end{equation}

\end{proof}

\subsection{Proof of Lemma 2}
\label{subsec: lemma2}
\begin{proof}
First, we prove statement (i).
For convenience, we write $p_{a,\emptyset}^{(m)}\equiv1$ for $a\in\{0,1\}$.
Under (A1)--(A3) and the MSM (1), ${\theta}_{sw}^{(m)}$  can be expressed as follows:
\begin{equation}
    \label{eq: SW converge saturated}
    \begin{split}
        {\theta}_{sw}^{(m)}&=\sum_{\bar{a}(K-m-1)\in\bar{\mathcal{A}}(K-m-1)}\mathbb{E}[Y^{\bar{a}(K-m-1),\underline{a}(K-m)=1_m}]\\
        &\quad \quad \times \mathbb{P}[\bar{A}(K-m-1)=\bar{a}(K-m-1)\mid \underline{A}(K-m)=1_m]\\
        &\quad -\sum_{\bar{a}(K-m-1)\in\bar{\mathcal{A}}(K-m-1)}\mathbb{E}[Y^{\bar{a}(K-m-1),\underline{a}(K-m)=0_m}]\\
        &\quad \quad \quad \times \mathbb{P}[\bar{A}(K-m-1)=\bar{a}(K-m-1)\mid \underline{A}(K-m)=0_m] \because \text{(\ref{eq: SW converge})}\\
        &=\sum_{\bar{a}(K-m-1)\in\bar{\mathcal{A}}(K-m-1)}\left\{
\sum_{S\subseteq \{m+1,...,K\}}\left(\sum_{T\subseteq \{1,...,m\}}\psi_{S\cup T}\right)\prod_{j\in S}a(K-j) \right\}\\
        &\quad \quad \times \mathbb{P}[\bar{A}(K-m-1)=\bar{a}(K-m-1)\mid \underline{A}(K-m)=1_m]\\
        &\quad -\sum_{\bar{a}(K-m-1)\in\bar{\mathcal{A}}(K-m-1)}\left\{\sum_{S\subseteq \{m+1,...,K\}}\psi_S\prod_{j\in S }a(K-j) \right\}\\
        &\quad \quad \times \mathbb{P}[\bar{A}(K-m-1)=\bar{a}(K-m-1)\mid \underline{A}(K-m)=0_m]\because \text{the MSM (1)}\\
        &=\sum_{S\subseteq \{m+1,...,K\}}\sum_{T\subseteq \{1,...,m\}}\psi_{S\cup T}\ p_{1,S}^{(m)}-\sum_{S\subseteq \{m+1,...,K\}}\psi_{S}\ p_{0,S}^{(m)}\\
        &=\sum_{S\subseteq \{m+1,...,K\}}\sum_{T\subseteq \{1,...,m\},T\neq\emptyset}\psi_{S\cup T}\ p_{1,S}^{(m)}+\sum_{S\subseteq \{m+1,...,K\}}\psi_{S}\left(p_{1,S}^{(m)}-p_{0,S}^{(m)}\right)\\
        &=\sum_{T\subseteq \{1,...,m\},T\neq\emptyset}\psi_T+\sum_{S\subseteq \{m+1,...,K\},S\neq\emptyset}\sum_{T\subseteq \{1,...,m\},T\neq\emptyset}\psi_{S\cup T}\ p_{1,S}^{(m)}+\sum_{S\subseteq \{m+1,...,K\},S\neq \emptyset}\psi_{S}\left(p_{1,S}^{(m)}-p_{0,S}^{(m)}\right).
    \end{split}
    \tag{C.2.1}
\end{equation}
Under (A1)--(A3) and the MSM (1), by equations (\ref{eq: RSW converge}) and (\ref{eq: MSM HRMSM link main}), $\theta_{rsw}^{(m)}$ can be expressed as follows:
\begin{equation}
    \label{eq: RSW convergence saturated}\theta_{rsw}^{(m)}={\sum_{T\subseteq \{1,...,m\},T\neq\emptyset}}\psi_{T}\ +\sum_{S\subseteq \{m+1,...,K\},S\neq \emptyset}\sum_{T\subseteq \{1,...,m\},T\neq\emptyset}\psi_{S\cup T}\ \mathbb{P}\left[\prod_{j\in S }{A}(K-j)=1\right].
    \tag{C.2.2}
\end{equation}
Furthermore, under the MSM (1), $\theta^{(K)}$ can be expressed as follows:
\begin{equation}
    \label{eq: K convergence saturated}
    \begin{split}
    \theta^{(K)}&=\sum_{S\subseteq \{1,...,K\},S\neq\emptyset }\psi_S=\sum_{S\subseteq \{m+1,...,K\}}\sum_{T\subseteq \{1,...,m\},T\neq\emptyset}\psi_{S\cup T}+\sum_{S\subseteq \{m+1,...,K\},S\neq\emptyset }\psi_{S}\\
    &=\sum_{T\subseteq \{1,...,m\},T\neq\emptyset}\psi_T+\sum_{S\subseteq \{m+1,...,K\},S\neq\emptyset}\sum_{T\subseteq \{1,...,m\},T\neq\emptyset}\psi_{S\cup T}+\sum_{S\subseteq \{m+1,...,K\},S\neq \emptyset}\psi_S.
    \end{split}
    \tag{C.2.3}
\end{equation}

By equations (\ref{eq: SW converge saturated}) and (\ref{eq: K convergence saturated}), the following equation holds:
\begin{equation*}
\label{eq: K-SW saturated}
    \theta^{(K)}-\theta_{sw}^{(m)}=\sum_{S\subseteq \{m+1,...,K\},S\neq\emptyset}\left[\sum_{T\subseteq \{1,...,m\},T\neq\emptyset}\psi_{S\cup T}\left(1-p_{1,S}^{(m)}\right)+\psi_{S}\left\{1-\left(p_{1,S}^{(m)}-p_{0,S}^{(m)}\right)\right\}\right].
\end{equation*}
By equations (\ref{eq: RSW convergence saturated}) and (\ref{eq: K convergence saturated}), the following equation holds:
\begin{equation*}
\label{eq: K-RSW saturated}
    \theta^{(K)}-\theta_{rsw}^{(m)}=\sum_{S\subseteq \{m+1,...,K\},S\neq\emptyset}\left[\sum_{T\subseteq \{1,...,m\},T\neq\emptyset}\psi_{S\cup T}\left(1-\mathbb{P}\left[\prod_{j\in S }{A}(K-j)=1\right]\right)+\psi_S\right].
\end{equation*}
By equations (\ref{eq: SW converge saturated}) and (\ref{eq: RSW convergence saturated}), the following equation holds:
\begin{equation*}
    \theta_{sw}^{(m)}-\theta_{rsw}^{(m)}=\sum_{S\subseteq \{m+1,...,K\},S\neq\emptyset}\left[\sum_{T\subseteq \{1,...,m\},T\neq\emptyset}\psi_{S\cup T}\left(p_{1,S}^{(m)}- \mathbb{P}\left[\prod_{j\in S }{A}(K-j)=1\right]\right)+\psi_{S}\left\{p_{1,S}^{(m)}-p_{0,S}^{(m)}\right\}\right].
\end{equation*}

Next, we prove statement (ii).
Under (A1)--(A3) and the MSM (2), ${\theta}_{sw}^{(m)}$  can be expressed as follows:
\begin{equation}
    \label{eq: SW converge main}
    \begin{split}
        {\theta}_{sw}^{(m)}&=\sum_{\bar{a}(K-m-1)\in\bar{\mathcal{A}}(K-m-1)}\mathbb{E}[Y^{\bar{a}(K-m-1),\underline{a}(K-m)=1_m}]\\
        &\quad \quad \times \mathbb{P}[\bar{A}(K-m-1)=\bar{a}(K-m-1)\mid \underline{A}(K-m)=1_m]\\
        &\quad -\sum_{\bar{a}(K-m-1)\in\bar{\mathcal{A}}(K-m-1)}\mathbb{E}[Y^{\bar{a}(K-m-1),\underline{a}(K-m)=0_m}]\\
        &\quad \quad \quad \times \mathbb{P}[\bar{A}(K-m-1)=\bar{a}(K-m-1)\mid \underline{A}(K-m)=0_m] \because \text{(\ref{eq: SW converge})}\\
        &=\sum_{\bar{a}(K-m-1)\in\bar{\mathcal{A}}(K-m-1)}\left\{\psi_0 + \sum_{j=1}^{m}\psi_j + \sum_{j=m+1}^{K}\psi_j a(K-j) \right\}\\
        &\quad \quad \times \mathbb{P}[\bar{A}(K-m-1)=\bar{a}(K-m-1)\mid \underline{A}(K-m)=1_m]\\
        &\quad -\sum_{\bar{a}(K-m-1)\in\bar{\mathcal{A}}(K-m-1)}\left\{\psi_0 + \sum_{j=m+1}^{K}\psi_j a(K-j) \right\}\\
        &\quad \quad \times \mathbb{P}[\bar{A}(K-m-1)=\bar{a}(K-m-1)\mid \underline{A}(K-m)=0_m]\because \text{the MSM (2)}\\
        &=\sum_{j=1}^{m}\psi_j +\sum_{j=m+1}^{K}\psi_j q_j^{(m)}.
    \end{split}
    \tag{C.2.4}
\end{equation}
Under (A1)--(A3) and the MSM (2), by equations (\ref{eq: RSW converge}) and (\ref{eq: MSM HRMSM link main}), $\theta_{rsw}^{(m)}$ can be expressed as follows:
\begin{equation}
    \label{eq: RSW convergence main}\theta_{rsw}^{(m)}=\sum_{j=1}^{m}\psi_j.
    \tag{C.2.5}
\end{equation}
Furthermore, under the MSM (2), $\theta^{(K)}$ can be expressed as follows:
\begin{equation}
    \label{eq: K convergence main}\theta^{(K)}=\sum_{j=1}^{K}\psi_j=\sum_{j=1}^{m}\psi_j +\sum_{j=m+1}^{K}\psi_j.
    \tag{C.2.6}
\end{equation}
By equations (\ref{eq: SW converge main}) and (\ref{eq: K convergence main}), the following equation holds:
\begin{equation*}
    \theta^{(K)}-\theta_{sw}^{(m)}=\sum_{j=m+1}^{K}\psi_j \left\{1-q_j^{(m)}\right\}.
\end{equation*}
By equations (\ref{eq: RSW convergence main}) and (\ref{eq: K convergence main}), the following equation holds:
\begin{equation*}
    \theta^{(K)}-\theta_{rsw}^{(m)}=\sum_{j=m+1}^{K}\psi_j.
\end{equation*}
By equations (\ref{eq: SW converge main}) and (\ref{eq: RSW convergence main}), the following equation holds:
\begin{equation*}
    \theta_{sw}^{(m)}-\theta_{rsw}^{(m)}=\sum_{j=m+1}^{K}\psi_jq_j^{(m)}.
\end{equation*}
\end{proof}

\subsection{Proof of Theorem 1}
\label{subsec: proof2}
\begin{proof}
To begin with, using the same logic as the Appendix of \cite{MSMtest}, we prove that $\hat{\theta}_{sw}^{(m)}-\hat{\theta}_{rsw}^{(m)}$ is RAL.
Since both $\hat{\theta}_{sw}^{(m)}$ and $\hat{\theta}_{rsw}^{(m)}$ are RAL, the following equation holds:
\begin{equation}
\label{eq: RAL}
\begin{split}
    \sqrt{n}\{(\hat{\theta}_{sw}^{(m)}-\hat{\theta}_{rsw}^{(m)})-({\theta}_{sw}^{(m)}-{\theta}_{rsw}^{(m)})\}&=\sqrt{n}(\hat{\theta}_{sw}^{(m)}-{\theta}_{sw}^{(m)})-\sqrt{n}(\hat{\theta}_{rsw}^{(m)}-{\theta}_{rsw}^{(m)})\\
    &=\frac{1}{\sqrt{n}}\sum_{i=1}^{n}(\varphi_{sw,i}^{(m)}-\varphi_{rsw,i}^{(m)})+o_p(1),
\end{split}
\tag{C.3.1}
\end{equation}
where $\varphi_{w,i}^{(m)}$ is the influence function of the estimator $\hat{\theta}_{w}^{(m)}$ with $\mathbb{E}[\varphi_{w,i}^{(m)}]=0$ and $\mathbb{V}[\varphi_{w,i}^{(m)}]<\infty$ for $w\in\{sw, rsw\}$, and $o_p(1)$ is a term that converges in probability to zero as $n$ goes to infinity.
Since $\varphi_{w,i}^{(m)}$ is an element of the Hilbert space $\mathcal{H}$ with mean zero and finite variance, with covariance inner product, for $w\in\{sw, rsw\}$, $\mathbb{E}[\varphi_{sw,i}^{(m)}-\varphi_{rsw,i}^{(m)}]=0$ and $\mathbb{V}[\varphi_{sw,i}^{(m)}-\varphi_{rsw,i}^{(m)}]<\infty$.

That is, the following statement holds:
\begin{equation}
\label{eq: asynorm}
    \sqrt{n}\{(\hat{\theta}_{sw}^{(m)}-\hat{\theta}_{rsw}^{(m)})-({\theta}_{sw}^{(m)}-{\theta}_{rsw}^{(m)})\}\xrightarrow[]{d}N(0,\mathbb{V}[\varphi_{sw,i}^{(m)}-\varphi_{rsw,i}^{(m)}]),\tag{C.3.2}
\end{equation}
by central limit theorem.
Thus, the following statement holds:
\begin{equation*}
    \frac{\sqrt{n}\{(\hat{\theta}_{sw}^{(m)}-\hat{\theta}_{rsw}^{(m)})-({\theta}_{sw}^{(m)}-{\theta}_{rsw}^{(m)})\}}{\sqrt{\mathbb{V}[\varphi_{sw,i}^{(m)}-\varphi_{rsw,i}^{(m)}]}}\xrightarrow[]{d}N(0,1),
\end{equation*}
and thus
\begin{equation*}
    \frac{{n}\{(\hat{\theta}_{sw}^{(m)}-\hat{\theta}_{rsw}^{(m)})-({\theta}_{sw}^{(m)}-{\theta}_{rsw}^{(m)})\}^2}{{\mathbb{V}[\varphi_{sw,i}^{(m)}-\varphi_{rsw,i}^{(m)}]}}\xrightarrow[]{d}\chi^2(1).
\end{equation*}
Note that the following statement also holds:
\begin{equation*}
    \frac{\mathbb{V}[\varphi_{sw,i}^{(m)}-\varphi_{rsw,i}^{(m)}]}{n{\mathbb{V}[\hat{\theta}_{sw}^{(m)}-\hat{\theta}_{rsw}^{(m)}]}}\xrightarrow[]{p}1.
\end{equation*}
Thus, from Slutsky's theorem, the following statement holds:
\begin{equation*}
\begin{split}
    &\frac{\{(\hat{\theta}_{sw}^{(m)}-\hat{\theta}_{rsw}^{(m)})-({\theta}_{sw}^{(m)}-{\theta}_{rsw}^{(m)})\}^2}{{\mathbb{V}[\hat{\theta}_{sw}^{(m)}-\hat{\theta}_{rsw}^{(m)}]}}\\
    &=\frac{\mathbb{V}[\varphi_{sw,i}^{(m)}-\varphi_{rsw,i}^{(m)}]}{n{\mathbb{V}[\hat{\theta}_{sw}^{(m)}-\hat{\theta}_{rsw}^{(m)}]}}\times\frac{n\{(\hat{\theta}_{sw}^{(m)}-\hat{\theta}_{rsw}^{(m)})-({\theta}_{sw}^{(m)}-{\theta}_{rsw}^{(m)})\}^2}{{\mathbb{V}[\varphi_{sw,i}^{(m)}-\varphi_{rsw,i}^{(m)}]}}\xrightarrow[]{d}\chi^2(1).
\end{split}
\end{equation*}
Under $\hat{\mathbb{V}}[\hat{\theta}_{sw}^{(m)}-\hat{\theta}_{rsw}^{(m)}]\xrightarrow[]{p}{\mathbb{V}}[\hat{\theta}_{sw}^{(m)}-\hat{\theta}_{rsw}^{(m)}]$, the following statement also holds:
\begin{equation*}
\label{eq: chisq}
    \frac{\{(\hat{\theta}_{sw}^{(m)}-\hat{\theta}_{rsw}^{(m)})-({\theta}_{sw}^{(m)}-{\theta}_{rsw}^{(m)})\}^2}{\hat{\mathbb{V}}[\hat{\theta}_{sw}^{(m)}-\hat{\theta}_{rsw}^{(m)}]}\xrightarrow[]{d}\chi^2(1),
\end{equation*}
i.e., $D^{(m)}\xrightarrow[]{d}F_{D^{(m)}}$.
Then, \(\displaystyle
\lim_{n\rightarrow \infty}\mathbb{P}[h_{\alpha}(D^{(m)})=1]=1-F_{D^{(m)}}\left(\chi^2_\alpha(1)\right)\) holds.

Especially, under $H_0^{(m)}$, $D^{(m)}\xrightarrow[]{d}\chi^2(1)$ holds. Thus, $\lim_{n\rightarrow \infty}\mathbb{P}[h_\alpha(D^{(m)})=1\mid H_0^{(m)}]=\alpha$ holds.
Therefore, the following inequality holds:
\begin{equation*}
    \lim_{n\rightarrow \infty}\mathbb{P}[\tilde{m}_\alpha>m^*]=\lim_{n\rightarrow \infty}\mathbb{P}\left[\prod_{m=1}^{m^*}h_{\alpha}(D^{(m)})=1\ \middle| \ H_0^{(m^*)}\right]
     \leq\lim_{n\rightarrow \infty}\mathbb{P}\left[h_{\alpha}(D^{(m^*)})=1\ \middle|\  H_0^{(m^*)}\right]=\alpha.
\end{equation*}
\end{proof}

\subsection{Proof of Theorem 2}
\label{subsec: proof4}
\begin{proof}
By equation (\ref{eq: PSW converge}), under (A1)--(A3) and (A7), ${\theta}_{psw}^{(m)}$ can be expressed as follows:
\begin{equation}
    \label{eq: PSW converge A7}
    \begin{split}
        {\theta}_{psw}^{(m)}&=\sum_{\bar{a}(K-m-1)\in\bar{\mathcal{A}}(K-m-1)}\mathbb{E}[Y^{\bar{a}(K-m-1),\underline{a}(K-m)=1_m}]\\
        &\quad \quad \times \mathbb{P}[\bar{A}(K-m-1)=\bar{a}(K-m-1)\mid \underline{A}(K-m)=1_m]\\
        &\quad -\sum_{\bar{a}(K-m-1)\in\bar{\mathcal{A}}(K-m-1)}\mathbb{E}[Y^{\bar{a}(K-m-1),\underline{a}(K-m)=0_m}]\\
        &\quad \quad \quad \times \mathbb{P}[\bar{A}(K-m-1)=\bar{a}(K-m-1)\mid \underline{A}(K-m)=0_m].
    \end{split}
    \tag{C.4.1}
\end{equation}
By equations (\ref{eq: SW converge}) and (\ref{eq: PSW converge A7}), ${\theta}_{psw}^{(m)}={\theta}_{sw}^{(m)}$ holds.
\end{proof}

\subsection{Proof of Theorem 3}
\label{subsec: proof5}
\begin{proof}
By direct calculation, the following equation holds:
\begin{equation*}
    \mathbb{E}\left[\prod_{k=K-m}^{K-1}I(A(k)=a) W_{w}^{(m)}\right]=\mathbb{P}\left[\underline{A}(K-m)=a_{m}\right],
\end{equation*}
for $w\in\{sw,rsw,psw\}$ and $a\in\{0,1\}$. 
Also by direct calculation, under $\mu_{a,w}^{(m)}=\mathbb{E}[Y^{\bar{a}=a_K}]$, the following equation holds:
\begin{equation*}
    \mathbb{E}\left[\prod_{k=K-m}^{K-1}I(A(k)=a)\{W_{w}^{(m)}(Y-\mu_{a,w}^{(m)})\}^2\right]=\mathbb{E}\left[\prod_{k=K-m}^{K-1}I(A(k)=a) \{W_{w}^{(m)} (Y-\mathbb{E}[Y^{\bar{a}=a_K}])\}^2\right],
\end{equation*}
for $w\in\{sw,rsw,psw\}$ and $a\in\{0,1\}$.
Thus, ${asyvar_{w}^{(m)}}$ can be expressed as follows:
\begin{equation*}
\begin{split}
    {asyvar_{w}^{(m)}}=&\frac{\mathbb{E}\left[\prod_{k=K-m}^{K-1}I(A(k)=1) \{W_{w}^{(m)} (Y-\mathbb{E}[Y^{\bar{a}=1_K}])\}^2\right]}{\mathbb{P}\left[\underline{A}(K-m)=1_{m}\right]^2}\\
    &+\frac{\mathbb{E}\left[\prod_{k=K-m}^{K-1}I(A(k)=0) \{W_{w}^{(m)}  (Y-\mathbb{E}[Y^{\bar{a}=0_K}])\}^2\right]}{\mathbb{P}\left[\underline{A}(K-m)=0_{m}\right]^2},
\end{split}
\end{equation*}
for $w\in\{sw,rsw,psw\}$.

On the numerator of $asyvar_{sw}^{(m)}$, The following equation holds:
\begin{equation*}
\begin{split}
    &\mathbb{E}\left[\prod_{k=K-m}^{K-1}I(A(k)=a) \{W_{sw}^{(m)} (Y-\mathbb{E}[Y^{\bar{a}=a_K}])\}^2\right]\\
    &=\mathbb{E}\left[\{W_{sw}/W_{psw}^{(m)}\}^2\prod_{k=K-m}^{K-1}I(A(k)=a)\{W_{psw}^{(m)} (Y-\mathbb{E}[Y^{\bar{a}=a_K}])\}^2 \right]\\
    &=\mathbb{E}\left[\{W_{sw}/W_{psw}^{(m)}\}^2\right]\mathbb{E}\left[\prod_{k=K-m}^{K-1}I(A(k)=a)\{W_{psw}^{(m)} (Y-\mathbb{E}[Y^{\bar{a}=a_K}])\}^2\right]\\
    &\quad +\mathbb{COV}\left[\{W_{sw}/W_{psw}^{(m)}\}^2,\prod_{k=K-m}^{K-1}I(A(k)=a)\{W_{psw}^{(m)} (Y-\mathbb{E}[Y^{\bar{a}=a_K}])\}^2\right],
\end{split}
\end{equation*}
for $a\in\{0,1\}$.
Thus, the following equation holds:
\begin{equation*}
    asyvar_{sw}^{(m)}=\mathbb{E}\left[\{W_{sw}/W_{psw}^{(m)}\}^2\right] asyvar_{psw}^{(m)}+c_1.
\end{equation*}
Since $\mathbb{E}\left[W_{sw}/W_{psw}^{(m)}\right]=1$, (i) $asyvar_{sw}^{(m)}=\{1+\mathbb{V}[W_{sw}/W_{psw}^{(m)}]\} asyvar_{psw}^{(m)}+c_1$ holds.

(ii) $asyvar_{rsw}^{(m)}=\{1+\mathbb{V}[W_{rsw}^{(m)}/W_{psw}^{(m)}]\} asyvar_{psw}^{(m)}+c_2$ can be shown by the same procedure.
\end{proof}

\subsection{Proof of Theorem 4}
\label{subsec: proof6}
\begin{proof}
By direct calculation, the structural causal model (5) can also be expressed as follows:
\begin{equation}
  \label{eq: SCMV change}
  \begin{split}
   L(k)&=g_{L(k)}\left(\{\varepsilon_{L(t)}\mid 0\leq t\leq k\},\{\varepsilon_{A(t)}\mid 0\leq t\leq k-1\}\right) ,\quad 0\leq k\leq K-1, \\
   A(k)&=g_{A(k)}\left(\{\varepsilon_{L(t)}\mid 0\leq t\leq k\},\{\varepsilon_{A(t)}\mid 0\leq t\leq k\}\right) ,\quad 0\leq k\leq K-1, \\
   Y &= g_{Y}\left(\{\varepsilon_{L(t)}\mid 0\leq t\leq K-1\},\{\varepsilon_{A(t)}\mid 0\leq t\leq K-1\},\varepsilon_{Y}\right),
  \end{split}
  \tag{C.6.1}
\end{equation}
where $g_{L(0)}(\cdot),\ldots,g_{L(K-1)}(\cdot),g_{A(0)}(\cdot)\ldots, g_{A(K-1)}(\cdot), g_{Y}(\cdot)$ are corresponding functions.
Thus, $A(k)$ depends only on $\{\varepsilon_{L(t)}\mid 0\leq t\leq k\}$ and $\{\varepsilon_{A(t)}\mid 0\leq t\leq k\}$, for $0\leq k\leq K-1$.

Under the structural causal model (5), the structural causal model after the intervention $\bar{A}=\bar{a}$ can be expressed as follows:
 \begin{equation*}
  \label{eq: SCMV after int}
  \begin{split}
   L(k)&=f_{L(k)}\left(\bar{L}(k-1),\bar{a}(k-1),\varepsilon_{L(k)}\right) ,\quad 0\leq k\leq K-1, \\
   A(k)&=a(k) ,\quad 0\leq k\leq K-1, \\
   Y &= f_{Y}\left(\bar{L}(K-1),\bar{a}(K-1),\varepsilon_{Y}\right).
  \end{split}
\end{equation*}
Thus, under (A1), the following equation holds:
\begin{equation}
   \label{eq: SCMV Ya}
   \begin{split}
    Y^{\bar{a}} &= f_{Y}\left(\bar{L}(K-1),\bar{a}(K-1),\varepsilon_{Y}\right)\\
    &=f_{Y}\left(\{f_{L(k)}\left(\bar{L}(k-1),\bar{a}(k-1),\varepsilon_{L(k)}\right)\mid 0\leq k\leq K-1\},\bar{a}(K-1),\varepsilon_{Y}\right).
   \end{split}
   \tag{C.6.2}
\end{equation}

Now we prove that (A7)' holds under (A8) and (A9).
If (\ref{eq: SCMV Ya}) does not depend on $\{\varepsilon_{L(k)}\mid 1\leq k\leq K-m\}$, i.e., the following equation holds:
\begin{equation}
\label{eq: A7' proof}
\begin{split}
     Y^{\bar{a}}&=g_0\left(\bar{a},\varepsilon_{L(0)}, \{\varepsilon_{L(k)}\mid K-m+1\leq k\leq K-1\},\varepsilon_{Y}\right),
\end{split}
\tag{C.6.3}
\end{equation}
where $g_0(\cdot)$ is a corresponding function, then (A7)' holds because $\bar{A}(K-m-1)$ only depends on $\{\varepsilon_{L(t)}\mid 0\leq t\leq K-m-1\}$ and $\{\varepsilon_{A(t)}\mid 0\leq t\leq K-m-1\}$ by (\ref{eq: SCMV change}).
Thus, it is enough to show that equation (\ref{eq: A7' proof}) holds under (A8) and (A9).
Now assume that equation (\ref{eq: A7' proof}) does not hold, i.e., equation (\ref{eq: SCMV Ya}) depends on at least one of the elements of $\{\varepsilon_{L(k)}\mid 1\leq k\leq K-m\}$.
Combining this assumption with (A8), there must exist the directed path from $A(k-1)$ to $Y$ through $L(k)$ and not through $\underline{A}(k)$ for at least one $k\leq K-m$.
This implies that (A9) does not hold.
Take the contraposition, equation (\ref{eq: A7' proof}) holds under (A8) and (A9).

Next, we prove that (A7) holds under (A8)--(A10).
We have already shown that (\ref{eq: A7' proof}) holds under (A8) and (A9).
If we additionally assume (A10), then the following equation holds:
\begin{equation}
\label{eq: A7 proof}
\begin{split}
     Y^{\bar{a}}&=g_1\left(\bar{a}, \{\varepsilon_{L(k)}\mid K-m+1\leq k\leq K-1\},\varepsilon_{Y}\right),
\end{split}
\tag{C.6.4}
\end{equation}
where $g_1(\cdot)$ is a corresponding function, and then (A7) holds.
\end{proof}

\newpage
\section{Additional simulation results}
\subsection{Results of the second simulation}

\setcounter{figure}{0}
\setcounter{table}{0}

\spacingset{1.1}
\begin{figure}[H]
    \centering
    \includegraphics[width=1\linewidth]{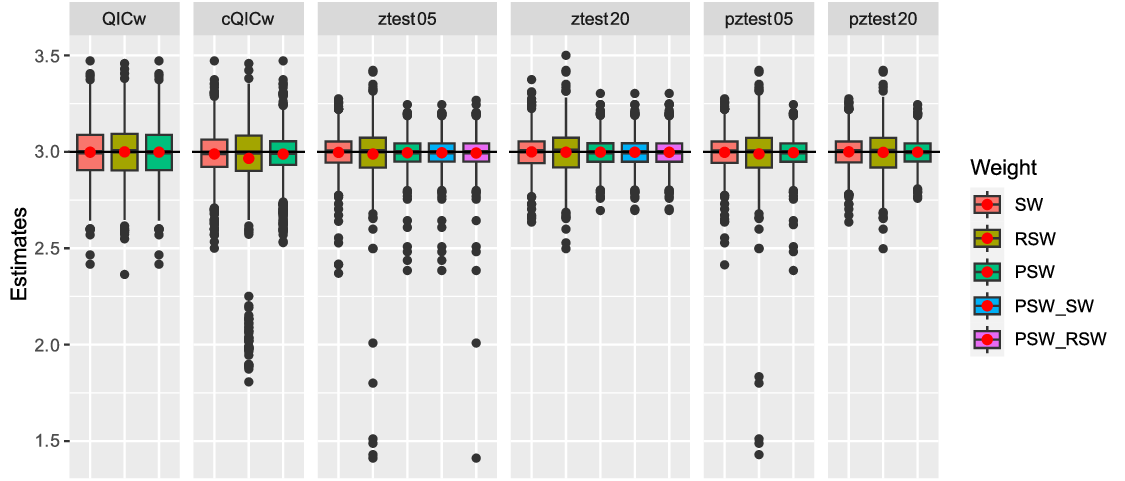}
    \caption{
    Box-plots of estimates of $\theta^{(K)}$ over 1000 runs of the second simulation with $(\alpha_0,\alpha_1, \alpha_2, \pi_1, \delta_0, \delta_1, \delta_2, \delta_3)=(0,0,1,4,0,1,2,0)$.
    The horizontal line is drawn at true value $\theta^{(K)}=3$.
    Twenty-two methods for estimating $\theta^{(K)}$ with combinations of selection methods and IP-weights are compared.
    Six gray blocks represent selection methods, where QICw, cQICw, ztest05, ztest20, pztest05, pztest20 is $\tilde{m}_{\text{QICw}}$, $\tilde{m}_{\text{cQICw}}$, $\tilde{m}_{0.05}$, $\tilde{m}_{0.20}$, $\hat{m}_{0.05}$, $\hat{m}_{0.20}$, respectively.
    For $m\in \{\tilde{m}_\text{QICw}, \tilde{m}_\text{cQICw}, \tilde{m}_{0.05}, \tilde{m}_{0.20}, \hat{m}_{0.05}, \hat{m}_{0.20}\}$, SW, RSW, PSW is $\hat{\theta}_{sw,main}^{(m)}$, $\hat{\theta}_{rsw,main}^{(m)}$, $\hat{\theta}_{psw,main}^{(m)}$, respectively.
    For $m\in \{\tilde{m}_{0.05}, \tilde{m}_{0.20}\}$, PSW\_SW, PSW\_RSW is $\hat{\theta}_{sw/psw,main}^{(m)}$, $\hat{\theta}_{rsw/psw,main}^{(m)}$, respectively.}
\end{figure}

\begin{table}[H]
\caption{(a) Selection probability of each $m\in \{1,2,3,4\}$ and (b) Estimation performance for $\theta^{(K)}$ over 1000 runs of the second simulation with $(\alpha_0,\alpha_1, \alpha_2, \pi_1, \delta_0, \delta_1, \delta_2, \delta_3)=(0,0,1,4,0,1,2,0)$.
In (a), six methods for selecting $m^*$ are compared, where QICw, cQICw, ztest05, ztest20, pztest05, pztest20 is $\tilde{m}_{\text{QICw}}$, $\tilde{m}_{\text{cQICw}}$, $\tilde{m}_{0.05}$, $\tilde{m}_{0.20}$, $\hat{m}_{0.05}$, $\hat{m}_{0.20}$, respectively.
Bold letter represents the selection probability of true $m^*=2$.
In (b), twenty-two methods for estimating $\theta^{(K)}$ with combinations of selection methods and IP-weights are compared. 
For $m\in \{\tilde{m}_\text{QICw}, \tilde{m}_\text{cQICw}, \tilde{m}_{0.05}, \tilde{m}_{0.20}, \hat{m}_{0.05}, \hat{m}_{0.20}\}$, SW, RSW, PSW is $\hat{\theta}_{sw,main}^{(m)}$, $\hat{\theta}_{rsw,main}^{(m)}$, $\hat{\theta}_{psw,main}^{(m)}$, respectively.
For $m\in \{\tilde{m}_{0.05}, \tilde{m}_{0.20}\}$, PSW\_SW, PSW\_RSW is $\hat{\theta}_{sw/psw,main}^{(m)}$, $\hat{\theta}_{rsw/psw,main}^{(m)}$, respectively.
Bias is the average of the estimates over 1000 simulations minus the true value $\theta^{(K)}=3$.
SE is the Monte Carlo standard error over 1000 simulations.
RMSE is the root mean squared error of the estimates over 1000 simulations.
CP is the proportion out of 1000 simulations for which the 95 percent confidence interval using the naïve sandwich variance estimator, that does not take into account uncertainty due to estimating IP-weights and selecting MSMs, includes the true value $\theta^{(K)}=3$.
\\
}
\centering
\resizebox{0.95\textwidth}{!}{%
\begin{tabular}{lcccclcccc}
\hline
\multirow{2}{*}{Selection method} & \multicolumn{4}{c}{(a) Selection probability}                                                                  & \multirow{2}{*}{Weight} & \multicolumn{4}{c}{(b) Estimation performance} \\ \cline{2-5} \cline{7-10} 
                                    & $m=1$                   & \textbf{$m=2$}                    & $m=3$                    & $m=4$                    &                         & Bias         & SE          & RMSE      & CP \\ \hline
\multirow{3}{*}{QICw}               & \multirow{3}{*}{0.000} & \multirow{3}{*}{\textbf{0.000}} & \multirow{3}{*}{0.026} & \multirow{3}{*}{0.974} & SW                      & -0.002       & 0.140       & 0.140      & 0.961         \\
                                    &                        &                                 &                        &                        & RSW                     & 0.000        & 0.139       & 0.139      & 0.961         \\
                                    &                        &                                 &                        &                        & PSW                     & -0.002       & 0.139       & 0.139      & 0.962         \\ \hline
\multirow{3}{*}{cQICw}              & \multirow{3}{*}{0.035} & \multirow{3}{*}{\textbf{0.486}} & \multirow{3}{*}{0.180} & \multirow{3}{*}{0.299} & SW                      & -0.011       & 0.126       & 0.126      & 0.914         \\
                                    &                        &                                 &                        &                        & RSW                     & -0.034       & 0.219       & 0.222      & 0.927         \\
                                    &                        &                                 &                        &                        & PSW                     & -0.012       & 0.118       & 0.119      & 0.919         \\ \hline
\multirow{5}{*}{ztest05}            & \multirow{5}{*}{0.006} & \multirow{5}{*}{\textbf{0.994}} & \multirow{5}{*}{0.000} & \multirow{5}{*}{0.000} & SW                      & -0.003       & 0.096       & 0.096      & 0.944         \\
                                    &                        &                                 &                        &                        & RSW                     & -0.011       & 0.162       & 0.163      & 0.954         \\
                                    &                        &                                 &                        &                        & PSW                     & -0.005       & 0.079       & 0.079      & 0.951         \\
                                    &                        &                                 &                        &                        & PSW\_SW                 & -0.005       & 0.079       & 0.079      & 0.951         \\
                                    &                        &                                 &                        &                        & PSW\_RSW                & -0.006       & 0.097       & 0.097      & 0.950         \\ \hline
\multirow{5}{*}{ztest20}            & \multirow{5}{*}{0.000} & \multirow{5}{*}{\textbf{0.951}} & \multirow{5}{*}{0.048} & \multirow{5}{*}{0.001} & SW                      & 0.000        & 0.094       & 0.094      & 0.941         \\
                                    &                        &                                 &                        &                        & RSW                     & -0.002       & 0.120       & 0.120      & 0.967         \\
                                    &                        &                                 &                        &                        & PSW                     & -0.002       & 0.073       & 0.073      & 0.953         \\
                                    &                        &                                 &                        &                        & PSW\_SW                 & -0.002       & 0.074       & 0.074      & 0.951         \\
                                    &                        &                                 &                        &                        & PSW\_RSW                & -0.002       & 0.074       & 0.074      & 0.950         \\ \hline
\multirow{3}{*}{pztest05}           & \multirow{3}{*}{0.005} & \multirow{3}{*}{\textbf{0.994}} & \multirow{3}{*}{0.001} & \multirow{3}{*}{0.000} & SW                      & -0.003       & 0.093       & 0.093      & 0.945         \\
                                    &                        &                                 &                        &                        & RSW                     & -0.011       & 0.156       & 0.156      & 0.956         \\
                                    &                        &                                 &                        &                        & PSW                     & -0.004       & 0.077       & 0.077      & 0.952         \\ \hline
\multirow{3}{*}{pztest20}           & \multirow{3}{*}{0.000} & \multirow{3}{*}{\textbf{0.986}} & \multirow{3}{*}{0.014} & \multirow{3}{*}{0.000} & SW                      & 0.000        & 0.088       & 0.088      & 0.946         \\
                                    &                        &                                 &                        &                        & RSW                     & -0.003       & 0.118       & 0.118      & 0.969         \\
                                    &                        &                                 &                        &                        & PSW                     & -0.002       & 0.070       & 0.070      & 0.954         \\ \hline
\end{tabular}%
}
\end{table}

\spacingset{2.0}

\subsection{Results of the third simulation}
\label{subsec: sim3}

\spacingset{1.1}
\label{subsec: sim3result}
\begin{figure}[H]
    \centering
    \includegraphics[width=1\linewidth]{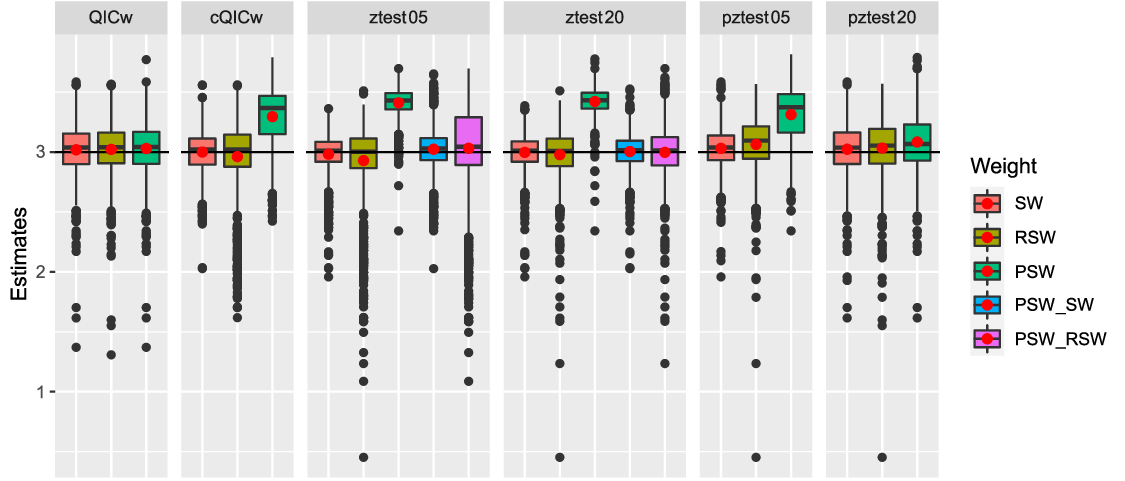}
    \caption{
    Box-plots of estimates of $\theta^{(K)}$ over 1000 runs of the third simulation with $(\alpha_0,\alpha_1, \alpha_2, \pi_1, \delta_0, \delta_1, \delta_2, \delta_3)=(0.5,0,1,4,0.5,1,2,0)$.
    The horizontal line is drawn at true value $\theta^{(K)}=3$.
    Twenty-two methods for estimating $\theta^{(K)}$ with combinations of selection methods and IP-weights are compared.
    Six gray blocks represent selection methods, where QICw, cQICw, ztest05, ztest20, pztest05, pztest20 is $\tilde{m}_{\text{QICw}}$, $\tilde{m}_{\text{cQICw}}$, $\tilde{m}_{0.05}$, $\tilde{m}_{0.20}$, $\hat{m}_{0.05}$, $\hat{m}_{0.20}$, respectively.
    For $m\in \{\tilde{m}_\text{QICw}, \tilde{m}_\text{cQICw}, \tilde{m}_{0.05}, \tilde{m}_{0.20}, \hat{m}_{0.05}, \hat{m}_{0.20}\}$, SW, RSW, PSW is $\hat{\theta}_{sw,main}^{(m)}$, $\hat{\theta}_{rsw,main}^{(m)}$, $\hat{\theta}_{psw,main}^{(m)}$, respectively.
    For $m\in \{\tilde{m}_{0.05}, \tilde{m}_{0.20}\}$, PSW\_SW, PSW\_RSW is $\hat{\theta}_{sw/psw,main}^{(m)}$, $\hat{\theta}_{rsw/psw,main}^{(m)}$, respectively.}
\end{figure}

\begin{table}[H]
\caption{(a) Selection probability of each $m\in \{1,2,3,4\}$ and (b) Estimation performance for $\theta^{(K)}$ over 1000 runs of the third simulation with $(\alpha_0,\alpha_1, \alpha_2, \pi_1, \delta_0, \delta_1, \delta_2, \delta_3)=(0.5,0,1,4,0.5,1,2,0)$.
In (a), six methods for selecting $m^*$ are compared, where QICw, cQICw, ztest05, ztest20, pztest05, pztest20 is $\tilde{m}_{\text{QICw}}$, $\tilde{m}_{\text{cQICw}}$, $\tilde{m}_{0.05}$, $\tilde{m}_{0.20}$, $\hat{m}_{0.05}$, $\hat{m}_{0.20}$, respectively.
Bold letter represents the selection probability of true $m^*=2$.
In (b), twenty-two methods for estimating $\theta^{(K)}$ with combinations of selection methods and IP-weights are compared. 
For $m\in \{\tilde{m}_\text{QICw}, \tilde{m}_\text{cQICw}, \tilde{m}_{0.05}, \tilde{m}_{0.20}, \hat{m}_{0.05}, \hat{m}_{0.20}\}$, SW, RSW, PSW is $\hat{\theta}_{sw,main}^{(m)}$, $\hat{\theta}_{rsw,main}^{(m)}$, $\hat{\theta}_{psw,main}^{(m)}$, respectively.
For $m\in \{\tilde{m}_{0.05}, \tilde{m}_{0.20}\}$, PSW\_SW, PSW\_RSW is $\hat{\theta}_{sw/psw,main}^{(m)}$, $\hat{\theta}_{rsw/psw,main}^{(m)}$, respectively.
Bias is the average of the estimates over 1000 simulations minus the true value $\theta^{(K)}=3$.
SE is the Monte Carlo standard error over 1000 simulations.
RMSE is the root mean squared error of the estimates over 1000 simulations.
CP is the proportion out of 1000 simulations for which the 95 percent confidence interval using the naïve sandwich variance estimator, that does not take into account uncertainty due to estimating IP-weights and selecting MSMs, includes the true value $\theta^{(K)}=3$.
\\
}
\centering
\resizebox{0.95\textwidth}{!}{%
\begin{tabular}{lcccclcccc}
\hline
\multirow{2}{*}{Selection method} & \multicolumn{4}{c}{(a) Selection probability}                                                                  & \multirow{2}{*}{Weight} & \multicolumn{4}{c}{(b) Estimation performance} \\ \cline{2-5} \cline{7-10} 
                                    & $m=1$   & \textbf{$m=2$} & $m=3$   & $m=4$   &        & Bias   & SE     & RMSE  & CP   \\ \hline
\multirow{3}{*}{QICw}              & \multirow{3}{*}{0.000}   & \multirow{3}{*}{\textbf{0.002}} & \multirow{3}{*}{0.022}   & \multirow{3}{*}{0.976}   & SW     & {0.019}
  & {0.217}  & {0.218} & {0.919} \\
                                   &         &                &         &         & RSW    & 0.022  & 0.224  & 0.225 & 0.916 \\
                                   &         &                &         &         & PSW    & 0.029  & 0.226  & 0.228 & 0.897 \\ \hline
\multirow{3}{*}{cQICw}             & \multirow{3}{*}{0.069}   & \multirow{3}{*}{\textbf{0.449}} & \multirow{3}{*}{0.171}   & \multirow{3}{*}{0.311}   & SW     & 0.000  & 0.187  & 0.187 & 0.872 \\
                                   &         &                &         &         & RSW    & -0.038 & 0.304  & 0.307 & 0.866 \\
                                   &         &                &         &         & PSW    & 0.296  & 0.240  & 0.381 & 0.348 \\ \hline
\multirow{5}{*}{ztest05}           & \multirow{5}{*}{0.078}   & \multirow{5}{*}{\textbf{0.918}} & \multirow{5}{*}{0.004}  & \multirow{5}{*}{0.000}   & SW     & -0.018 & 0.179  & 0.180 & 0.888 \\
                                   &         &                &         &         & RSW    & -0.071 & 0.326  & 0.333 & 0.887 \\
                                   &         &                &         &         & PSW    & 0.411  & 0.130  & 0.431 & 0.078 \\
                                   &         &                &         &         & PSW\_SW& 0.024  & 0.195  & 0.197 & 0.832 \\
                                   &         &                &         &         & PSW\_RSW& 0.031 & 0.376  & 0.378 & 0.655 \\ \hline
\multirow{5}{*}{ztest20}           & \multirow{5}{*}{0.014}   & \multirow{5}{*}{\textbf{0.934}} & \multirow{5}{*}{0.050}   & \multirow{5}{*}{0.002}   & SW     & -0.003 & 0.156  & 0.156 & 0.932 \\
                                   &         &                &         &         & RSW    & -0.022 & 0.226  & 0.227 & 0.963 \\
                                   &         &                &         &         & PSW    & 0.422  & 0.119  & 0.438 & 0.064 \\
                                   &         &                &         &         & PSW\_SW& 0.004  & 0.154  & 0.154 & 0.931 \\
                                   &         &                &         &         & PSW\_RSW& -0.002 & 0.224 & 0.224 & 0.930 \\ \hline
\multirow{3}{*}{pztest05}          & \multirow{3}{*}{0.003}   & \multirow{3}{*}{\textbf{0.312}} & \multirow{3}{*}{0.341}   & \multirow{3}{*}{0.344}   & SW     & 0.031  & 0.181  & 0.184 & 0.921 \\
                                   &         &                &         &         & RSW    & 0.064  & 0.241  & 0.249 & 0.889 \\
                                   &         &                &         &         & PSW    & 0.313  & 0.233  & 0.390 & 0.397 \\ \hline
\multirow{3}{*}{pztest20}          & \multirow{3}{*}{0.001}   & \multirow{3}{*}{\textbf{0.053}} & \multirow{3}{*}{0.088}   & \multirow{3}{*}{0.858}   & SW     & 0.023  & 0.215  & 0.216 & 0.921 \\
                                   &         &                &         &         & RSW    & 0.033  & 0.256  & 0.258 & 0.902 \\
                                   &         &                &         &         & PSW    & 0.084  & 0.245  & 0.259 & 0.834 \\ \hline
\end{tabular}%
}
\end{table}

\spacingset{2.0}

\subsection{Results of the third simulation with adjusting $L(0)$}
In this section, we make a modification to $\hat{\theta}_{w,main}^{(m)}$ in Section \ref{subsec: sim3}.
Specifically, we condition $L(0)$ on the outcome regression model and the numerator of the IP-weights.

\spacingset{1.1}
\begin{figure}[H]
    \centering
    \includegraphics[width=0.65\linewidth]{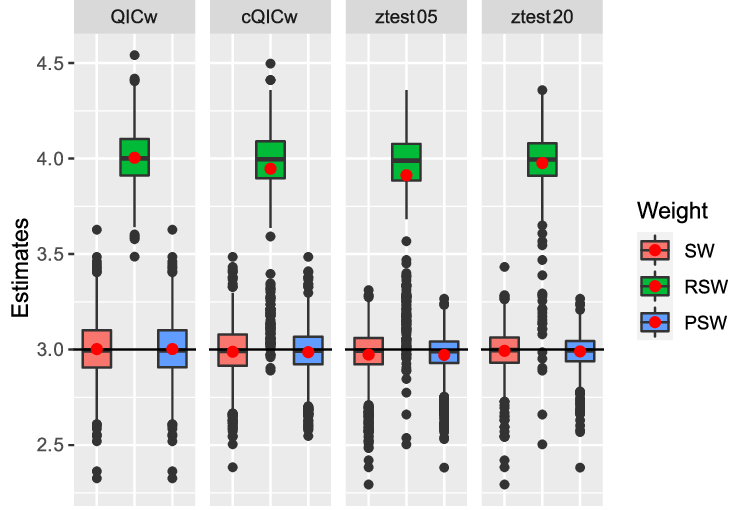}
    \caption{
    Box-plots of estimates of $\theta^{(K)}$ over 1000 runs of the third simulation with $(\alpha_0,\alpha_1, \alpha_2, \pi_1, \delta_0, \delta_1, \delta_2, \delta_3)=(0.5,0,1,4,0.5,1,2,0)$.
    The horizontal line is drawn at true value $\theta^{(K)}=3$.
    Twelve methods for estimating $\theta^{(K)}$ with combinations of selection methods and IP-weights are compared.
    Four gray blocks represent selection methods, where QICw, cQICw, ztest05, ztest20 is $\tilde{m}_{\text{QICw}}$, $\tilde{m}_{\text{cQICw}}$, $\tilde{m}_{0.05}$, $\tilde{m}_{0.20}$, respectively.
    For $m\in \{\tilde{m}_\text{QICw}, \tilde{m}_\text{cQICw}, \tilde{m}_{0.05}, \tilde{m}_{0.20}\}$, SW, RSW, PSW is $\hat{\theta}_{sw,main}^{(m)}$, $\hat{\theta}_{rsw,main}^{(m)}$, $\hat{\theta}_{psw,main}^{(m)}$, respectively.}
\end{figure}

\begin{table}[H]
\caption{(a) Selection probability of each $m\in \{1,2,3,4\}$ and (b) Estimation performance for $\theta^{(K)}$ over 1000 runs of the third simulation with $(\alpha_0,\alpha_1, \alpha_2, \pi_1, \delta_0, \delta_1, \delta_2, \delta_3)=(0.5,0,1,4,0.5,1,2,0)$.
In (a), four methods for selecting $m^*$ are compared, where QICw, cQICw, ztest05, ztest20 is $\tilde{m}_{\text{QICw}}$, $\tilde{m}_{\text{cQICw}}$, $\tilde{m}_{0.05}$, $\tilde{m}_{0.20}$, respectively.
Bold letter represents the selection probability of true $m^*=2$.
In (b), twelve methods for estimating $\theta^{(K)}$ with combinations of selection methods and IP-weights are compared. 
For $m\in \{\tilde{m}_\text{QICw}, \tilde{m}_\text{cQICw}, \tilde{m}_{0.05}, \tilde{m}_{0.20}\}$, SW, RSW, PSW is $\hat{\theta}_{sw,main}^{(m)}$, $\hat{\theta}_{rsw,main}^{(m)}$, $\hat{\theta}_{psw,main}^{(m)}$, respectively.
Bias is the average of the estimates over 1000 simulations minus the true value $\theta^{(K)}=3$.
SE is the Monte Carlo standard error over 1000 simulations.
RMSE is the root mean squared error of the estimates over 1000 simulations.
CP is the proportion out of 1000 simulations for which the 95 percent confidence interval using the naïve sandwich variance estimator, that does not take into account uncertainty due to estimating IP-weights and selecting MSMs, includes the true value $\theta^{(K)}=3$.
\\
}
\centering
\resizebox{0.95\textwidth}{!}{%
\begin{tabular}{lcccclcccc}
\hline
\multirow{2}{*}{Selection method} & \multicolumn{4}{c}{(a) Selection probability}                                                                  & \multirow{2}{*}{Weight} & \multicolumn{4}{c}{(b) Estimation performance} \\ \cline{2-5} \cline{7-10} 
                                     & $m=1$                   & \textbf{$m=2$}                    & $m=3$                    & $m=4$                    &                         & Bias         & SE          & RMSE      & CP \\ \hline
\multirow{3}{*}{QICw}            & \multirow{3}{*}{0.000} & \multirow{3}{*}{\textbf{0.002}} & \multirow{3}{*}{0.022} & \multirow{3}{*}{0.976} & SW                      & 0.003        & 0.153       & 0.153      & 0.951         \\
                                    &                        &                                 &                        &                        & RSW                     & 1.004        & 0.138       & 1.014      & 0.004         \\
                                    &                        &                                 &                        &                        & PSW                     & 0.003        & 0.153       & 0.153      & 0.951         \\
                                 \hline
\multirow{3}{*}{cQICw}            & \multirow{3}{*}{0.069} & \multirow{3}{*}{\textbf{0.449}} & \multirow{3}{*}{0.171} & \multirow{3}{*}{0.311} & SW                      & -0.011       & 0.141       & 0.141      & 0.891         \\
                                    &                        &                                 &                        &                        & RSW                     & 0.947        & 0.255       & 0.981      & 0.060         \\
                                    &                        &                                 &                        &                        & PSW                     & -0.013       & 0.132       & 0.133      & 0.891         \\
                                   \hline
\multirow{3}{*}{ztest05}           & \multirow{3}{*}{0.103} & \multirow{3}{*}{\textbf{0.893}} & \multirow{3}{*}{0.004} & \multirow{3}{*}{0.000} & SW                      & -0.026       & 0.138       & 0.140      & 0.869         \\
                                    &                        &                                 &                        &                        & RSW                     & 0.913        & 0.293       & 0.959      & 0.074         \\
                                    &                        &                                 &                        &                        & PSW                     & -0.028       & 0.115       & 0.118      & 0.867         \\ \hline
\multirow{3}{*}{ztest20}           & \multirow{3}{*}{0.022} & \multirow{3}{*}{\textbf{0.928}} & \multirow{3}{*}{0.048} & \multirow{3}{*}{0.002} & SW                      & -0.006       & 0.112       & 0.112      & 0.934         \\
                                    &                        &                                 &                        &                        & RSW                     & 0.976        & 0.179       & 0.993      & 0.021         \\
                                    &                        &                                 &                        &                        & PSW                     & -0.009       & 0.091       & 0.091      & 0.945         \\ \hline
\end{tabular}%
}
\end{table}

\subsection{Results of the fourth simulation}

\spacingset{1.1}
\begin{figure}[H]
    \centering
    \includegraphics[width=\linewidth]{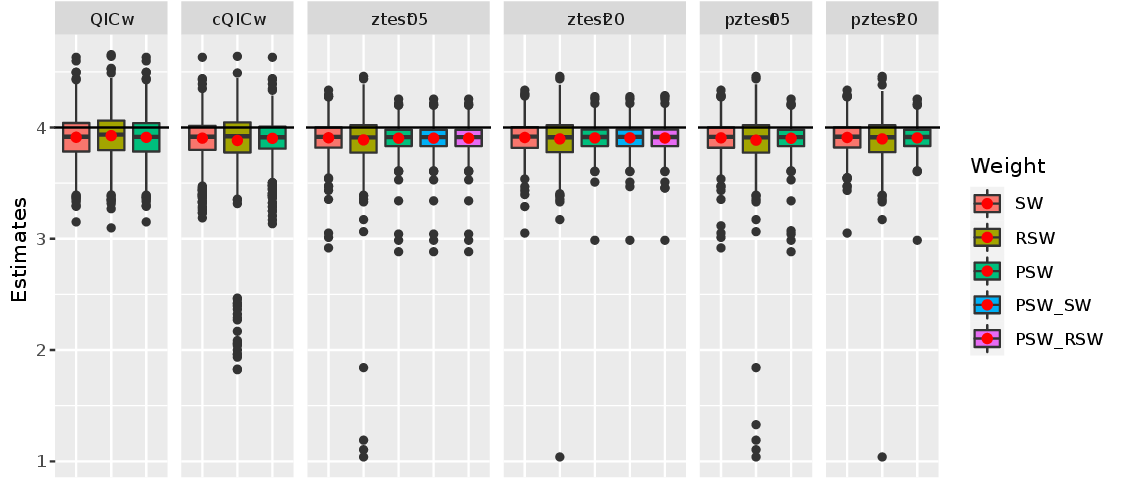}
    \caption{
    Box-plots of estimates of $\theta^{(K)}$ over 1000 runs of the fourth simulation with $(\alpha_0,\alpha_1, \alpha_2, \pi_1, \delta_0, \delta_1, \delta_2, \delta_3)=(0,0,1,4,0,1,2,1)$.
    The horizontal line is drawn at true value $\theta^{(K)}=4$.
    Twenty-two methods for estimating $\theta^{(K)}$ with combinations of selection methods and IP-weights are compared.
    Six gray blocks represent selection methods, where QICw, cQICw, ztest05, ztest20, pztest05, pztest20 is $\tilde{m}_{{\text{QICw}}}$, $\tilde{m}_{{\text{cQICw}}}$, $\tilde{m}_{0.05}$, $\tilde{m}_{0.20}$, $\hat{m}_{0.05}$, $\hat{m}_{0.20}$, respectively.
    For $m\in \{\tilde{m}_{\text{QICw}}, \tilde{m}_{\text{cQICw}}, \tilde{m}_{0.05}, \tilde{m}_{0.20}, \hat{m}_{0.05}, \hat{m}_{0.20}\}$, SW, RSW, PSW is $\hat{\theta}_{sw}^{(m)}$, $\hat{\theta}_{rsw}^{(m)}$, $\hat{\theta}_{psw}^{(m)}$, respectively.
    For $m\in \{\tilde{m}_{0.05}, \tilde{m}_{0.20}\}$, PSW\_SW, PSW\_RSW is $\hat{\theta}_{sw/psw}^{(m)}$, $\hat{\theta}_{rsw/psw}^{(m)}$, respectively.}
\end{figure}

\begin{table}[H]
\caption{(a) Selection probability of each $m\in \{1,2,3,4\}$ and (b) Estimation performance for $\theta^{(K)}$ over 1000 runs of the fourth simulation with $(\alpha_0,\alpha_1, \alpha_2, \pi_1, \delta_0, \delta_1, \delta_2, \delta_3)=(0,0,1,4,0,1,2,1)$.
In (a), four methods for selecting $m^*$ are compared, where QICw, cQICw, ztest05, ztest20 is $\tilde{m}_{\text{QICw}}$, $\tilde{m}_{\text{cQICw}}$, $\tilde{m}_{0.05}$, $\tilde{m}_{0.20}$, respectively.
Bold letter represents the selection probability of true $m^*=2$.
In (b), twelve methods for estimating $\theta^{(K)}$ with combinations of selection methods and IP-weights are compared. 
For $m\in \{\tilde{m}_\text{QICw}, \tilde{m}_\text{cQICw}, \tilde{m}_{0.05}, \tilde{m}_{0.20}\}$, SW, RSW, PSW is $\hat{\theta}_{sw,main}^{(m)}$, $\hat{\theta}_{rsw,main}^{(m)}$, $\hat{\theta}_{psw,main}^{(m)}$, respectively.
Bias is the average of the estimates over 1000 simulations minus the true value $\theta^{(K)}=4$.
SE is the Monte Carlo standard error over 1000 simulations.
RMSE is the root mean squared error of the estimates over 1000 simulations.
CP is the proportion out of 1000 simulations for which the 95 percent confidence interval using the naïve sandwich variance estimator, that does not take into account uncertainty due to estimating IP-weights and selecting MSMs, includes the true value $\theta^{(K)}=4$.
\\
}
\centering
\resizebox{0.95\textwidth}{!}{%
\begin{tabular}{lcccclcccc}
\hline
\multirow{2}{*}{Selection method} & \multicolumn{4}{c}{(a) Selection probability} & 
\multirow{2}{*}{Weight} & \multicolumn{4}{c}{(b) Estimation performance} \\ 
\cline{2-5} \cline{7-10}
 & $m=1$ & \textbf{$m=2$} & $m=3$ & $m=4$ &  & Bias & SE & RMSE & CP \\ 
\hline
\multirow{3}{*}{QICw} 
 & \multirow{3}{*}{0.000} & \multirow{3}{*}{\textbf{0.000}} & \multirow{3}{*}{0.022} & \multirow{3}{*}{0.978} 
 & SW     & -0.088 & 0.200 & 0.219 & 0.936 \\
 &  &  &  &  & RSW    & -0.073 & 0.207 & 0.219 & 0.949 \\
 &  &  &  &  & PSW    & -0.088 & 0.199 & 0.217 & 0.936 \\ 
\hline
\multirow{3}{*}{cQICw}
 & \multirow{3}{*}{0.020} & \multirow{3}{*}{\textbf{0.438}} & \multirow{3}{*}{0.190} & \multirow{3}{*}{0.352}
 & SW     & -0.096 & 0.182 & 0.205 & 0.905 \\
 &  &  &  &  & RSW    & -0.117 & 0.312 & 0.333 & 0.922 \\
 &  &  &  &  & PSW    & -0.097 & 0.170 & 0.195 & 0.891 \\ 
\hline
\multirow{5}{*}{ztest05}
 & \multirow{5}{*}{0.004} & \multirow{5}{*}{\textbf{0.995}} & \multirow{5}{*}{0.001} & \multirow{5}{*}{0.000}
 & SW     & -0.092 & 0.144 & 0.170 & 0.910 \\
 &  &  &  &  & RSW    & -0.111 & 0.247 & 0.270 & 0.924 \\
 &  &  &  &  & PSW    & -0.096 & 0.120 & 0.155 & 0.867 \\
 &  &  &  &  & PSW\_SW & -0.096 & 0.120 & 0.155 & 0.867 \\
 &  &  &  &  & PSW\_RSW & -0.096 & 0.120 & 0.155 & 0.867 \\ 
\hline
\multirow{5}{*}{ztest20}
 & \multirow{5}{*}{0.001} & \multirow{5}{*}{\textbf{0.955}} & \multirow{5}{*}{0.044} & \multirow{5}{*}{0.000}
 & SW     & -0.090 & 0.143 & 0.167 & 0.906 \\
 &  &  &  &  & RSW    & -0.103 & 0.200 & 0.226 & 0.939 \\
 &  &  &  &  & PSW    & -0.094 & 0.114 & 0.148 & 0.866 \\
 &  &  &  &  & PSW\_SW & -0.093 & 0.115 & 0.148 & 0.867 \\
 &  &  &  &  & PSW\_RSW & -0.094 & 0.116 & 0.148 & 0.866 \\ 
\hline
\multirow{3}{*}{pztest05}
 & \multirow{3}{*}{0.005} & \multirow{3}{*}{\textbf{0.994}} & \multirow{3}{*}{0.001} & \multirow{3}{*}{0.000}
 & SW     & -0.093 & 0.146 & 0.173 & 0.909 \\
 &  &  &  &  & RSW    & -0.114 & 0.259 & 0.283 & 0.923 \\
 &  &  &  &  & PSW    & -0.097 & 0.123 & 0.155 & 0.866 \\ 
\hline
\multirow{3}{*}{pztest20}
 & \multirow{3}{*}{0.001} & \multirow{3}{*}{\textbf{0.982}} & \multirow{3}{*}{0.017} & \multirow{3}{*}{0.000}
 & SW     & -0.090 & 0.137 & 0.164 & 0.912 \\
 &  &  &  &  & RSW    & -0.103 & 0.198 & 0.224 & 0.935 \\
 &  &  &  &  & PSW    & -0.093 & 0.111 & 0.145 & 0.870 \\ 
\hline
\end{tabular}
}
\end{table}

\subsection{Results of the fifth simulation}

\spacingset{1.1}
\begin{figure}[H]
    \centering
    \includegraphics[width=\linewidth]{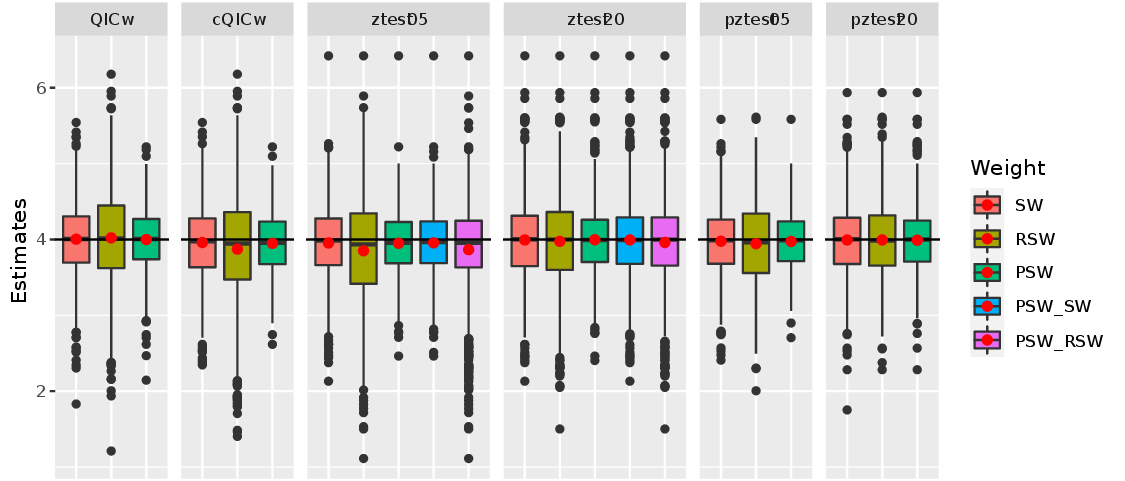}
    \caption{
    Box-plots of estimates of $\theta^{(K)}$ over 1000 runs of the fifth simulation with $(\alpha_0,\alpha_1, \alpha_2, \pi_1, \delta_0, \delta_1, \delta_2, \delta_3)=(0,0,1,4,0,1,2,1)$.
    The horizontal line is drawn at true value $\theta^{(K)}=4$.
    Twenty-two methods for estimating $\theta^{(K)}$ with combinations of selection methods and IP-weights are compared.
    Six gray blocks represent selection methods, where QICw, cQICw, ztest05, ztest20, pztest05, pztest20 is $\tilde{m}_{{\text{QICw}}}$, $\tilde{m}_{{\text{cQICw}}}$, $\tilde{m}_{0.05}$, $\tilde{m}_{0.20}$, $\hat{m}_{0.05}$, $\hat{m}_{0.20}$, respectively.
    For $m\in \{\tilde{m}_{\text{QICw}}, \tilde{m}_{\text{cQICw}}, \tilde{m}_{0.05}, \tilde{m}_{0.20}, \hat{m}_{0.05}, \hat{m}_{0.20}\}$, SW, RSW, PSW is $\hat{\theta}_{sw}^{(m)}$, $\hat{\theta}_{rsw}^{(m)}$, $\hat{\theta}_{psw}^{(m)}$, respectively.
    For $m\in \{\tilde{m}_{0.05}, \tilde{m}_{0.20}\}$, PSW\_SW, PSW\_RSW is $\hat{\theta}_{sw/psw}^{(m)}$, $\hat{\theta}_{rsw/psw}^{(m)}$, respectively.}
\end{figure}

\begin{table}[H]
\caption{(a) Selection probability of each $m\in \{1,2,3,4\}$ and (b) Estimation performance for $\theta^{(K)}$ over 1000 runs of the fifth simulation with $(\alpha_0,\alpha_1, \alpha_2, \pi_1, \delta_0, \delta_1, \delta_2, \delta_3)=(0,0,1,4,0,1,2,1)$.
In (a), four methods for selecting $m^*$ are compared, where QICw, cQICw, ztest05, ztest20 is $\tilde{m}_{\text{QICw}}$, $\tilde{m}_{\text{cQICw}}$, $\tilde{m}_{0.05}$, $\tilde{m}_{0.20}$, respectively.
Bold letter represents the selection probability of true $m^*=2$.
In (b), twelve methods for estimating $\theta^{(K)}$ with combinations of selection methods and IP-weights are compared. 
For $m\in \{\tilde{m}_\text{QICw}, \tilde{m}_\text{cQICw}, \tilde{m}_{0.05}, \tilde{m}_{0.20}\}$, SW, RSW, PSW is $\hat{\theta}_{sw,main}^{(m)}$, $\hat{\theta}_{rsw,main}^{(m)}$, $\hat{\theta}_{psw,main}^{(m)}$, respectively.
Bias is the average of the estimates over 1000 simulations minus the true value $\theta^{(K)}=4$.
SE is the Monte Carlo standard error over 1000 simulations.
RMSE is the root mean squared error of the estimates over 1000 simulations.
CP is the proportion out of 1000 simulations for which the 95 percent confidence interval using the naïve sandwich variance estimator, that does not take into account uncertainty due to estimating IP-weights and selecting MSMs, includes the true value $\theta^{(K)}=4$.
\\
}
\centering
\resizebox{0.95\textwidth}{!}{%
\begin{tabular}{lcccclcccc}
\hline
\multirow{2}{*}{Selection method} & \multicolumn{4}{c}{(a) Selection probability} & 
\multirow{2}{*}{Weight} & \multicolumn{4}{c}{(b) Estimation performance} \\ 
\cline{2-5} \cline{7-10}
 & $m=1$ & \textbf{$m=2$} & $m=3$ & $m=4$ &  & Bias & SE & RMSE & CP \\ 
\hline
\multirow{3}{*}{QICw} 
 & \multirow{3}{*}{0.000} & \multirow{3}{*}{\textbf{0.529}} & \multirow{3}{*}{0.470} & \multirow{3}{*}{0.001} 
 & SW     &  0.004 & 0.481 & 0.481 & 0.922 \\
 &  &  &  &  & RSW    &  0.021 & 0.628 & 0.628 & 0.880 \\
 &  &  &  &  & PSW    & -0.002 & 0.399 & 0.399 & 0.932 \\ 
\hline
\multirow{3}{*}{cQICw}
 & \multirow{3}{*}{0.119} & \multirow{3}{*}{\textbf{0.517}} & \multirow{3}{*}{0.363} & \multirow{3}{*}{0.001}
 & SW     & -0.040 & 0.493 & 0.494 & 0.896 \\
 &  &  &  &  & RSW    & -0.127 & 0.746 & 0.757 & 0.794 \\
 &  &  &  &  & PSW    & -0.052 & 0.410 & 0.414 & 0.884 \\ 
\hline
\multirow{5}{*}{ztest05}
 & \multirow{5}{*}{0.183} & \multirow{5}{*}{\textbf{0.747}} & \multirow{5}{*}{0.065} & \multirow{5}{*}{0.005}
 & SW     & -0.046 & 0.497 & 0.499 & 0.861 \\
 &  &  &  &  & RSW    & -0.149 & 0.716 & 0.732 & 0.774 \\
 &  &  &  &  & PSW    & -0.050 & 0.400 & 0.402 & 0.880 \\
 &  &  &  &  & PSW\_SW & -0.045 & 0.419 & 0.422 & 0.864 \\
 &  &  &  &  & PSW\_RSW & -0.134 & 0.622 & 0.636 & 0.795 \\ 
\hline
\multirow{5}{*}{ztest20}
 & \multirow{5}{*}{0.074} & \multirow{5}{*}{\textbf{0.677}} & \multirow{5}{*}{0.187} & \multirow{5}{*}{0.062}
 & SW     & -0.006 & 0.533 & 0.533 & 0.880 \\
 &  &  &  &  & RSW    & -0.025 & 0.620 & 0.620 & 0.868 \\
 &  &  &  &  & PSW    & -0.004 & 0.452 & 0.452 & 0.890 \\
 &  &  &  &  & PSW\_SW & -0.006 & 0.509 & 0.509 & 0.863 \\
 &  &  &  &  & PSW\_RSW & -0.040 & 0.571 & 0.572 & 0.846 \\ 
\hline
\multirow{3}{*}{pztest05}
 & \multirow{3}{*}{0.089} & \multirow{3}{*}{\textbf{0.832}} & \multirow{3}{*}{0.071} & \multirow{3}{*}{0.008}
 & SW     & -0.024 & 0.431 & 0.431 & 0.917 \\
 &  &  &  &  & RSW    & -0.056 & 0.554 & 0.558 & 0.885 \\
 &  &  &  &  & PSW    & -0.026 & 0.362 & 0.362 & 0.917 \\ 
\hline
\multirow{3}{*}{pztest20}
 & \multirow{3}{*}{0.028} & \multirow{3}{*}{\textbf{0.722}} & \multirow{3}{*}{0.180} & \multirow{3}{*}{0.070}
 & SW     & -0.007 & 0.474 & 0.473 & 0.920 \\
 &  &  &  &  & RSW    & -0.006 & 0.492 & 0.492 & 0.937 \\
 &  &  &  &  & PSW    & -0.010 & 0.409 & 0.409 & 0.921 \\ 
\hline
\end{tabular}
}
\end{table}

\subsection{Results of the sixth simulation}

\spacingset{1.1}
\begin{figure}[H]
    \centering
    \includegraphics[width=\linewidth]{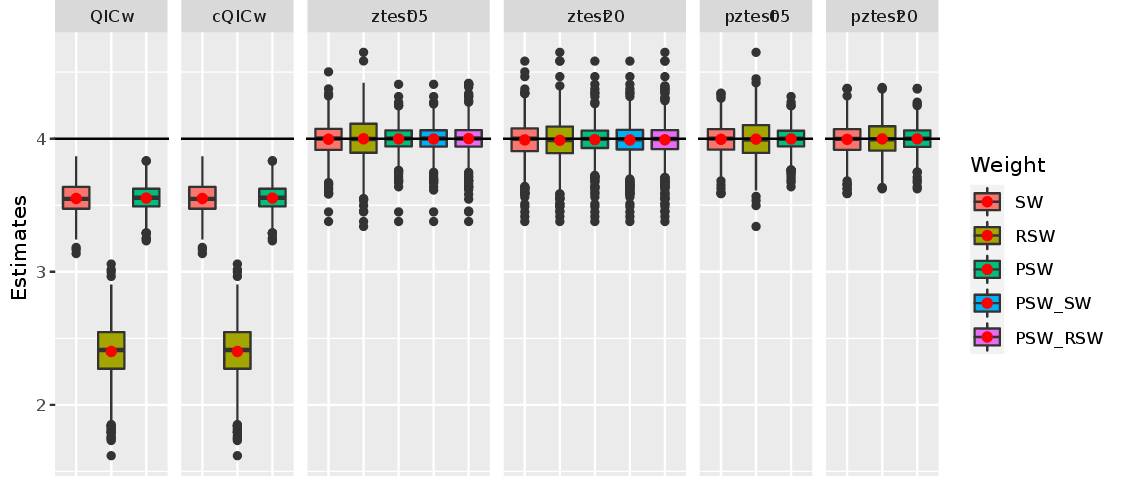}
    \caption{
    Box-plots of estimates of $\theta^{(K)}$ over 1000 runs of the sixth simulation with $(\alpha_0,\alpha_1, \alpha_2, \pi_1, \delta_0, \delta_1, \delta_2, \delta_3)=(0,0,1,40,0,1,2,1)$.
    The horizontal line is drawn at true value $\theta^{(K)}=4$.
    Twenty-two methods for estimating $\theta^{(K)}$ with combinations of selection methods and IP-weights are compared.
    Six gray blocks represent selection methods, where QICw, cQICw, ztest05, ztest20, pztest05, pztest20 is $\tilde{m}_{{\text{QICw}}}$, $\tilde{m}_{{\text{cQICw}}}$, $\tilde{m}_{0.05}$, $\tilde{m}_{0.20}$, $\hat{m}_{0.05}$, $\hat{m}_{0.20}$, respectively.
    For $m\in \{\tilde{m}_{\text{QICw}}, \tilde{m}_{\text{cQICw}}, \tilde{m}_{0.05}, \tilde{m}_{0.20}, \hat{m}_{0.05}, \hat{m}_{0.20}\}$, SW, RSW, PSW is $\hat{\theta}_{sw}^{(m)}$, $\hat{\theta}_{rsw}^{(m)}$, $\hat{\theta}_{psw}^{(m)}$, respectively.
    For $m\in \{\tilde{m}_{0.05}, \tilde{m}_{0.20}\}$, PSW\_SW, PSW\_RSW is $\hat{\theta}_{sw/psw}^{(m)}$, $\hat{\theta}_{rsw/psw}^{(m)}$, respectively.}
\end{figure}

\begin{table}[H]
\caption{(a) Selection probability of each $m\in \{1,2,3,4\}$ and (b) Estimation performance for $\theta^{(K)}$ over 1000 runs of the sixth simulation with $(\alpha_0,\alpha_1, \alpha_2, \pi_1, \delta_0, \delta_1, \delta_2, \delta_3)=(0,0,1,40,0,1,2,1)$.
In (a), four methods for selecting $m^*$ are compared, where QICw, cQICw, ztest05, ztest20 is $\tilde{m}_{\text{QICw}}$, $\tilde{m}_{\text{cQICw}}$, $\tilde{m}_{0.05}$, $\tilde{m}_{0.20}$, respectively.
Bold letter represents the selection probability of true $m^*=2$.
In (b), twelve methods for estimating $\theta^{(K)}$ with combinations of selection methods and IP-weights are compared. 
For $m\in \{\tilde{m}_\text{QICw}, \tilde{m}_\text{cQICw}, \tilde{m}_{0.05}, \tilde{m}_{0.20}\}$, SW, RSW, PSW is $\hat{\theta}_{sw,main}^{(m)}$, $\hat{\theta}_{rsw,main}^{(m)}$, $\hat{\theta}_{psw,main}^{(m)}$, respectively.
Bias is the average of the estimates over 1000 simulations minus the true value $\theta^{(K)}=4$.
SE is the Monte Carlo standard error over 1000 simulations.
RMSE is the root mean squared error of the estimates over 1000 simulations.
CP is the proportion out of 1000 simulations for which the 95 percent confidence interval using the naïve sandwich variance estimator, that does not take into account uncertainty due to estimating IP-weights and selecting MSMs, includes the true value $\theta^{(K)}=4$.
\\
}
\centering
\resizebox{0.95\textwidth}{!}{%
\begin{tabular}{lcccclcccc}
\hline
\multirow{2}{*}{Selection method} & \multicolumn{4}{c}{(a) Selection probability} & 
\multirow{2}{*}{Weight} & \multicolumn{4}{c}{(b) Estimation performance} \\ 
\cline{2-5} \cline{7-10}
 & $m=1$ & \textbf{$m=2$} & $m=3$ & $m=4$ &  & Bias & SE & RMSE & CP \\ 
\hline
\multirow{3}{*}{QICw} 
 & \multirow{3}{*}{1.000} & \multirow{3}{*}{\textbf{0.000}} & \multirow{3}{*}{0.000} & \multirow{3}{*}{0.000} 
 & SW     & -0.449 & 0.117 & 0.464 & 0.029 \\
 &  &  &  &  & RSW    & -1.600 & 0.220 & 1.614 & 0.000 \\
 &  &  &  &  & PSW    & -0.446 & 0.098 & 0.457 & 0.000 \\ 
\hline
\multirow{3}{*}{cQICw}
 & \multirow{3}{*}{1.000} & \multirow{3}{*}{\textbf{0.000}} & \multirow{3}{*}{0.000} & \multirow{3}{*}{0.000}
 & SW     & -0.449 & 0.117 & 0.464 & 0.029 \\
 &  &  &  &  & RSW    & -1.600 & 0.220 & 1.614 & 0.000 \\
 &  &  &  &  & PSW    & -0.446 & 0.098 & 0.457 & 0.000 \\ 
\hline
\multirow{5}{*}{ztest05}
 & \multirow{5}{*}{0.000} & \multirow{5}{*}{\textbf{0.938}} & \multirow{5}{*}{0.058} & \multirow{5}{*}{0.004}
 & SW     & -0.004 & 0.122 & 0.122 & 0.944 \\
 &  &  &  &  & RSW    & -0.000 & 0.166 & 0.167 & 0.961 \\
 &  &  &  &  & PSW    & -0.001 & 0.096 & 0.095 & 0.950 \\
 &  &  &  &  & PSW\_SW & -0.002 & 0.101 & 0.100 & 0.943 \\
 &  &  &  &  & PSW\_RSW &  0.001 & 0.110 & 0.110 & 0.939 \\ 
\hline
\multirow{5}{*}{ztest20}
 & \multirow{5}{*}{0.000} & \multirow{5}{*}{\textbf{0.790}} & \multirow{5}{*}{0.169} & \multirow{5}{*}{0.041}
 & SW     & -0.009 & 0.138 & 0.138 & 0.932 \\
 &  &  &  &  & RSW    & -0.012 & 0.159 & 0.158 & 0.968 \\
 &  &  &  &  & PSW    & -0.007 & 0.113 & 0.114 & 0.941 \\
 &  &  &  &  & PSW\_SW & -0.009 & 0.128 & 0.126 & 0.923 \\
 &  &  &  &  & PSW\_RSW & -0.008 & 0.133 & 0.134 & 0.931 \\ 
\hline
\multirow{3}{*}{pztest05}
 & \multirow{3}{*}{0.000} & \multirow{3}{*}{\textbf{0.950}} & \multirow{3}{*}{0.049} & \multirow{3}{*}{0.001}
 & SW     & -0.004 & 0.113 & 0.114 & 0.955 \\
 &  &  &  &  & RSW    & -0.003 & 0.156 & 0.155 & 0.975 \\
 &  &  &  &  & PSW    & -0.001 & 0.089 & 0.089 & 0.958 \\ 
\hline
\multirow{3}{*}{pztest20}
 & \multirow{3}{*}{0.000} & \multirow{3}{*}{\textbf{0.791}} & \multirow{3}{*}{0.180} & \multirow{3}{*}{0.029}
 & SW     & -0.005 & 0.118 & 0.118 & 0.954 \\
 &  &  &  &  & RSW    & -0.001 & 0.133 & 0.134 & 0.995 \\
 &  &  &  &  & PSW    & -0.001 & 0.094 & 0.095 & 0.958 \\ 
\hline
\end{tabular}%
}
\end{table}

\newpage
\spacingset{2.0}

\setcounter{table}{0}
\setcounter{figure}{0}

\section{Additional data analysis}

\subsection{Estimation results when varying $m$}

\spacingset{1.1}

\begin{figure}[H]
    \centering
    \includegraphics[width=\linewidth]{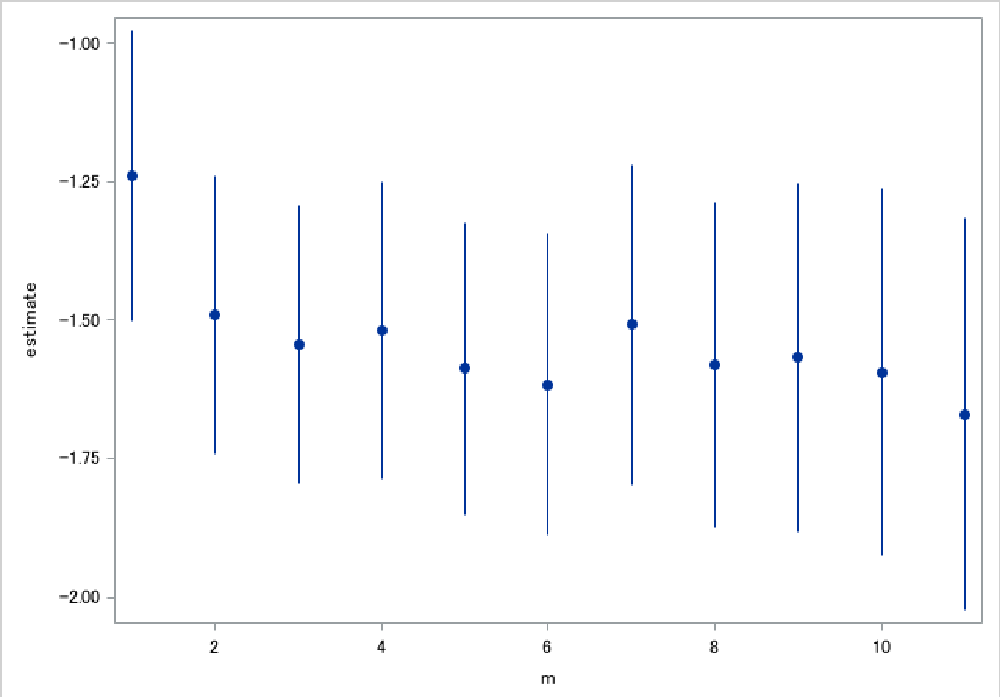}
    \caption{The SW estimates of $\theta^{(K)}$ using hemodialysis patients' data for $m=1,...,11$. The x-axis represents $m$. For each $m$, the dot represents the point estimate $\hat\theta_{sw}^{(m)}$ and the line represents the 95 percent confidence interval $[\hat\theta_{sw}^{(m)}-1.96\times SE^{(m)},\hat\theta_{sw}^{(m)}+1.96\times SE^{(m)}]$, where $ SE^{(m)}$ is the estimated standard error calculated by the naïve sandwich variance estimator.}
\end{figure}

\begin{figure}[H]
    \centering
    \includegraphics[width=\linewidth]{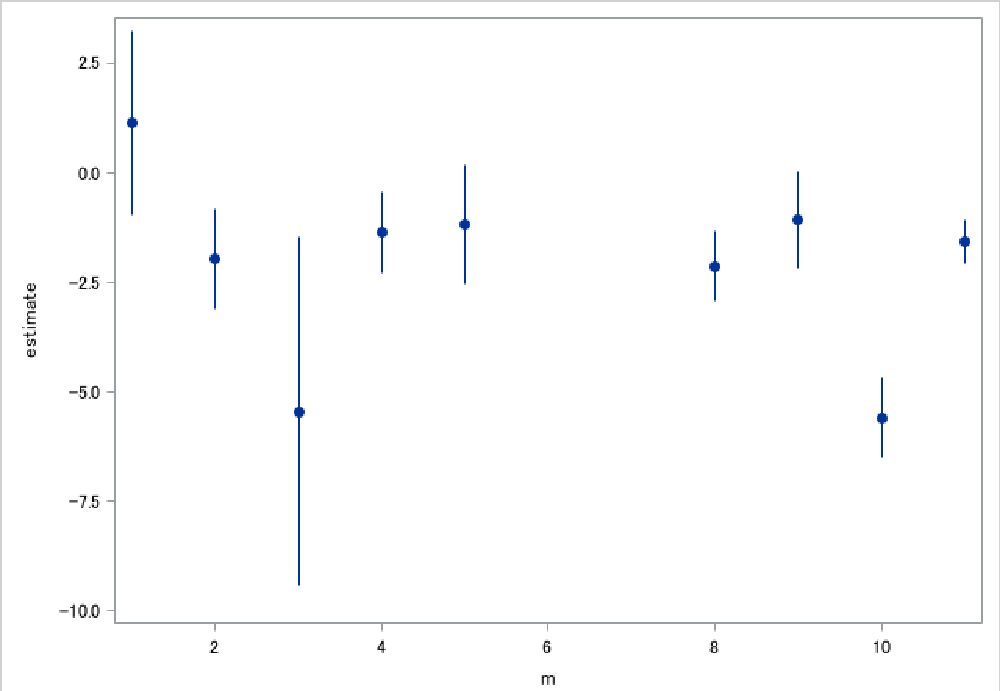}
    \caption{The RSW estimates of $\theta^{(K)}$ using hemodialysis patients' data for $m=1,...,11$. The x-axis represents $m$. For each $m$, the dot represents the point estimate $\hat\theta_{rsw}^{(m)}$ and the line represents the 95 percent confidence interval $[\hat\theta_{rsw}^{(m)}-1.96\times SE^{(m)},\hat\theta_{rsw}^{(m)}+1.96\times SE^{(m)}]$, where $ SE^{(m)}$ is the estimated standard error calculated by the naïve sandwich variance estimator. Estimates could not be calculated for $m=6$ and $7$.}
\end{figure}

\begin{figure}[H]
    \centering
    \includegraphics[width=\linewidth]{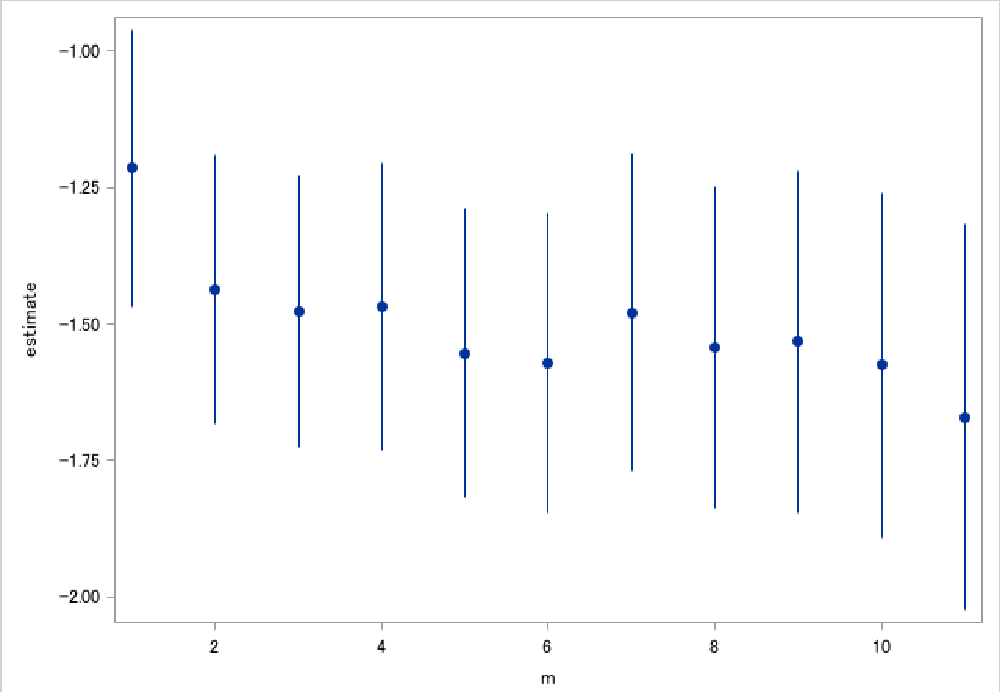}
    \caption{The PSW estimates of $\theta^{(K)}$ using hemodialysis patients' data for $m=1,...,11$. The x-axis represents $m$. For each $m$, the dot represents the point estimate $\hat\theta_{psw}^{(m)}$ and the line represents the 95 percent confidence interval $[\hat\theta_{psw}^{(m)}-1.96\times SE^{(m)},\hat\theta_{psw}^{(m)}+1.96\times SE^{(m)}]$, where $ SE^{(m)}$ is the estimated standard error calculated by the naïve sandwich variance estimator.}
\end{figure}

\subsection{Checking (A4)'}

\spacingset{1.1}

\begin{figure}[H]
    \centering
    \includegraphics[width=0.5\linewidth]{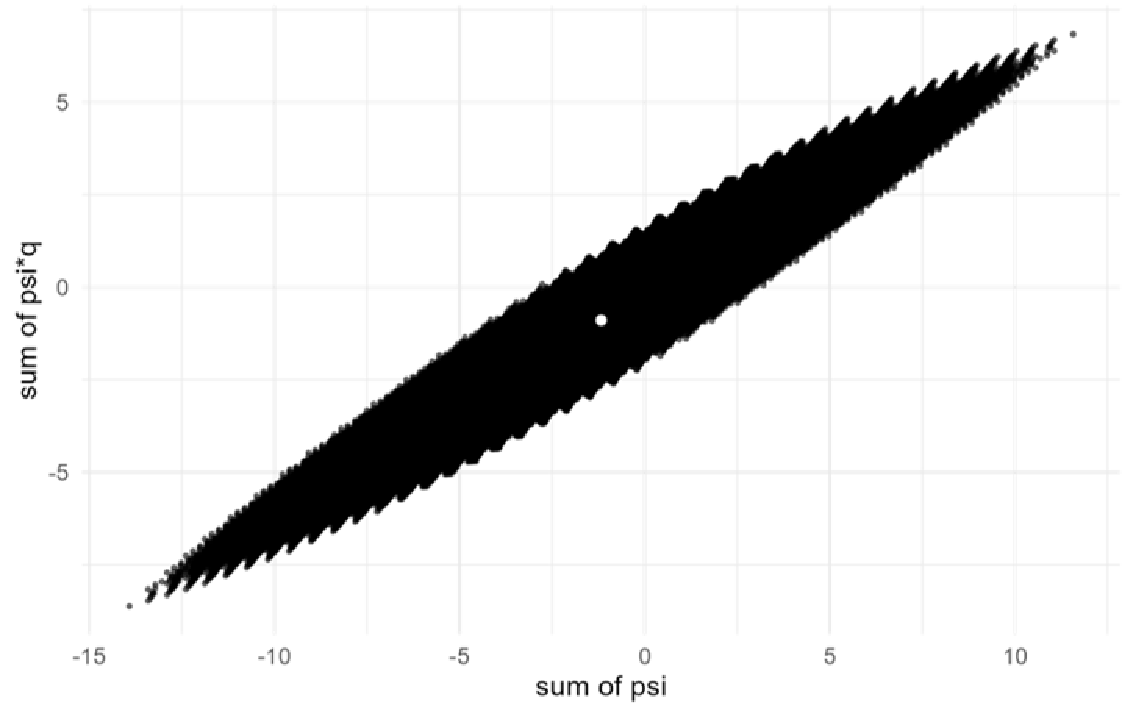}
    \caption{The x-axis represents $\sum_{j=1}^{11}\psi_j$ and the y-axis represents $\sum_{j=1}^{11}\psi_jq_j^{(1)}$ when $(\psi_1,...,\psi_{11})$ are varied within the confidence interval shown in Table E.1 and substitute for $(q_1^{(1)},...,q_{11}^{(1)})$ the values from Table E.2.
    The dot is the value substituting the point estimates of $(\psi_1,...,\psi_{11})$ shown in Table E.1.}
\end{figure}

\spacingset{1.1}
\begin{figure}[H]
    \centering
    \includegraphics[width=0.3\linewidth]{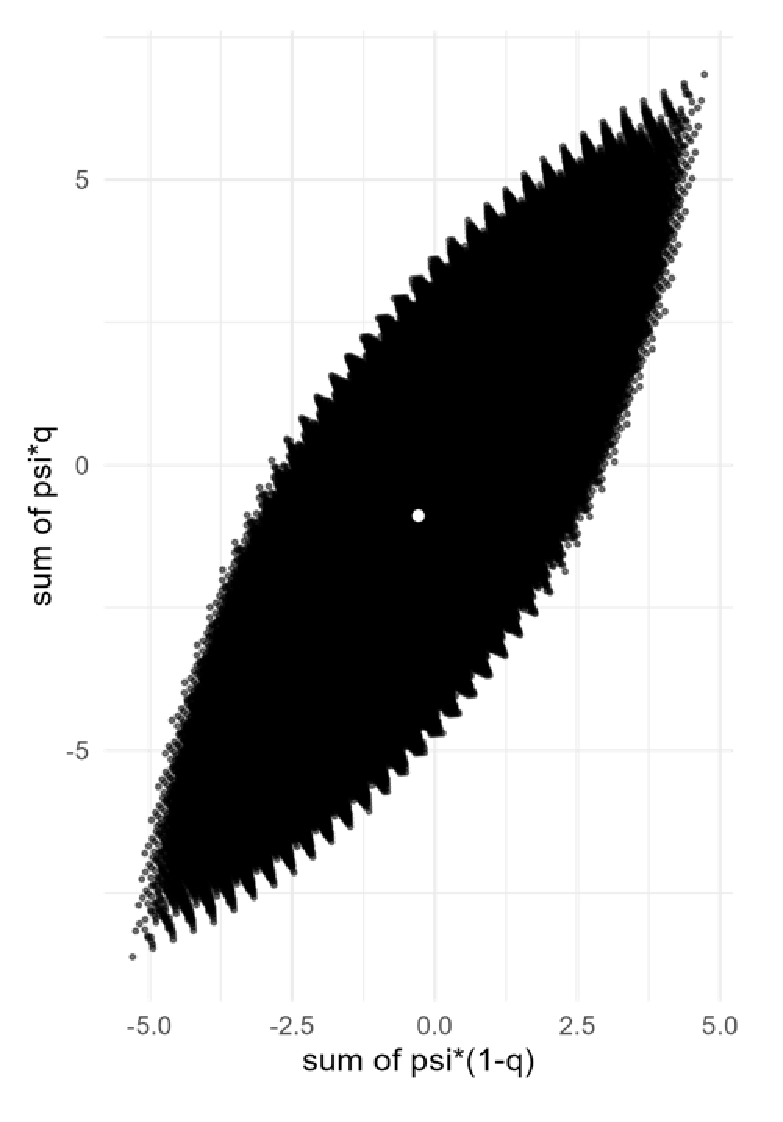}
    \caption{
    The x-axis represents $\sum_{j=1}^{11}\psi_j(1-q_j^{(1)})$ and the y-axis represents $\sum_{j=1}^{11}\psi_jq_j^{(1)}$ when $(\psi_1,...,\psi_{11})$ are varied within the confidence interval shown in Table E.1 and substitute for $(q_1^{(1)},...,q_{11}^{(1)})$ the values from Table E.2.
    The dot is the value substituting the point estimates of $(\psi_1,...,\psi_{11})$ shown in Table E.1.}
\end{figure}

\subsection{Checking (A5)'}
\spacingset{1.1}
\begin{table}[H]
\centering
\caption{Estimation results for $(\psi_1,...,\psi_{11})$ in the main effect MSM with $m=11$ using PSW. 
ES is the point estimate, and SE is the estimated standard error calculated by the naïve sandwich variance estimator.
LCL is the 95 percent lower confidence limit, i.e., $\text{ES}-1.96\times\text{SE}$, and UCL is the 95 percent upper confidence limit, i.e., $\text{ES}+1.96\times\text{SE}$.\\ 
}
\begin{tabular}{lrrrr}
\toprule
 & ES & SE & LCL & UCL \\
\midrule
$\psi_1$   & 0.807  & 0.401 & 0.021 & 1.593 \\
$\psi_2$   & -0.351 & 0.519 & -1.367 & 0.666 \\
$\psi_3$   & -0.970 & 0.716 & -2.373 & 0.433 \\
$\psi_4$   &  0.984 & 0.899 & -0.778 & 2.745 \\
$\psi_5$   & -0.545 & 0.710 & -1.936 & 0.847 \\
$\psi_6$   & -1.537 & 0.647 & -2.805 & -0.270 \\
$\psi_7$   &  2.175 & 0.623 &  0.954 & 3.395 \\
$\psi_8$   & -0.932 & 0.681 & -2.266 & 0.402 \\
$\psi_9$   &  0.222 & 0.608 & -0.974 & 1.414 \\
$\psi_{10}$ &  0.133 & 0.581 & -1.005 & 1.271 \\
$\psi_{11}$ & -0.351 & 0.519 & -1.367 & 0.666 \\
\bottomrule
\end{tabular}
\end{table}

\subsection{Checking (A6)'}
\spacingset{1.1}
\begin{table}[H]
\centering
\caption{$q_j^{(m)}$ in hemodialysis patients' data for $j=m+1,...,12$ and $m=1,...,11$. $q_j^{(m)}$ is calculated using the empirical distribution.  
\\ 
}
\resizebox{0.9\textwidth}{!}{%
\begin{tabular}{@{}lcccccccccccc@{}}
\toprule
 & $q_2^{(m)}$ & $q_3^{(m)}$ & $q_4^{(m)}$ & $q_5^{(m)}$ & $q_6^{(m)}$ & $q_7^{(m)}$ & $q_8^{(m)}$ & $q_9^{(m)}$ & $q_{10}^{(m)}$ & $q_{11}^{(m)}$ & $q_{12}^{(m)}$ \\ 
\midrule
$m=1$  & 0.894 & 0.831 & 0.749 & 0.690 & 0.586 & 0.554 & 0.513 & 0.450 & 0.382 & 0.300 & 0.152 \\
$m=2$  &       & 0.929 & 0.837 & 0.766 & 0.650 & 0.614 & 0.569 & 0.498 & 0.422 & 0.326 & 0.171 \\
$m=3$  &       &       & 0.896 & 0.820 & 0.694 & 0.651 & 0.602 & 0.526 & 0.444 & 0.340 & 0.184 \\
$m=4$  &       &       &       & 0.909 & 0.769 & 0.714 & 0.660 & 0.575 & 0.483 & 0.368 & 0.199 \\
$m=5$  &       &       &       &       & 0.846 & 0.786 & 0.726 & 0.632 & 0.531 & 0.405 & 0.219 \\
$m=6$  &       &       &       &       &       & 0.929 & 0.842 & 0.740 & 0.621 & 0.471 & 0.259 \\
$m=7$  &       &       &       &       &       &       & 0.898 & 0.788 & 0.660 & 0.499 & 0.271 \\
$m=8$  &       &       &       &       &       &       &       & 0.868 & 0.734 & 0.555 & 0.301 \\
$m=9$  &       &       &       &       &       &       &       &       & 0.847 & 0.640 & 0.347 \\
$m=10$ &       &       &       &       &       &       &       &       &       & 0.756 & 0.410 \\
$m=11$ &       &       &       &       &       &       &       &       &       &       & 0.542 \\
\bottomrule
\end{tabular}%
}
\end{table}

\subsection{Association between $A(11)$ and $A(10)$}

\begin{table}[H]
\centering
\caption{$2\times2$ contingency table of $A(11)$ and $A(10)$.\\
}
\begin{tabular}{lcc}
\hline
 & $A(10)=0$ & $A(10)=1$ \\ \hline
$A(11)=0$ & 4412 & 7 \\
$A(11)=1$ & 23 & 198 \\ \hline
\end{tabular}
\end{table}

\end{document}